\documentclass{article}
\usepackage{arxiv}
\usepackage[utf8]{inputenc} 
\usepackage[T1]{fontenc}    
\usepackage[hidelinks]{hyperref}
\hypersetup{
    colorlinks=true,    
    citecolor=blue,     
    urlcolor=black,     
    linkcolor=red       
}
\usepackage{hyperref}
\usepackage[symbol]{footmisc}

\usepackage{url}            
\usepackage{booktabs}       
\usepackage{amsfonts}       
\usepackage{amssymb}        
\usepackage{nicefrac}       
\usepackage{microtype}      
\usepackage{lipsum}		
\usepackage{graphicx}
\usepackage[numbers,super]{natbib}
\usepackage{multibib}
\newcites{S}{References}
\usepackage{authblk}  
\usepackage{doi}
\usepackage{float}
\usepackage{placeins}
\usepackage{amsmath}
\usepackage{amsbsy}
\usepackage{soul}
\usepackage{subcaption}
\usepackage{lineno}
\usepackage{xcolor}
\usepackage{dsfont}
\usepackage{algorithm}
\usepackage{algpseudocode}
\usepackage{mathtools}
\usepackage{pgfplots}
\pgfplotsset{compat=1.18}
\usepgfplotslibrary{groupplots}
\usetikzlibrary{arrows.meta,positioning}
\usepackage{setspace}

\usepackage{tcolorbox}
\tcbuselibrary{skins,breakable,listings}
\usepackage{xcolor}
\usepackage{listings}
\usepackage{microtype}

\definecolor{promptbg}{HTML}{FAFAFA}
\definecolor{promptframe}{HTML}{D0D0D0}
\definecolor{roleSystem}{HTML}{0B5394}   
\definecolor{roleUser}{HTML}{38761D}     
\definecolor{roleAsst}{HTML}{7F6000}     
\definecolor{promptgray}{HTML}{444444}

\lstdefinestyle{promptstyle}{
  basicstyle=\ttfamily\small,
  columns=fullflexible,
  breaklines=true,
  breakatwhitespace=false,
  keepspaces=true,
  showstringspaces=false,
  tabsize=2,
  upquote=true,
}

\newtcolorbox{promptbox}[2][]{%
  enhanced,
  breakable,
  colback=promptbg,
  colframe=promptframe,
  boxrule=0.6pt,
  arc=2pt,
  left=6pt,right=6pt,top=6pt,bottom=6pt,
  fonttitle=\bfseries,
  title={#2},
  listing only,
  listing options={style=promptstyle},
  #1
}

\newcommand{\RoleSystem}{\textcolor{roleSystem}{\bfseries [SYSTEM]}}
\newcommand{\RoleUser}{\textcolor{roleUser}{\bfseries [USER]}}

\soulregister\eqref7
\soulregister\cref7
\soulregister\cite7
\soulregister\citep7
\usepackage{enumitem}
\usepackage{cleveref}
\crefname{figure}{Fig.}{Figs.}
\Crefname{figure}{Fig.}{Figs.}
\crefname{equation}{Eq.}{Eqs.}
\Crefname{equation}{Eq.}{Eqs.}
\crefname{table}{Table}{Tables}
\Crefname{table}{Table}{Tables}
\usepackage[labelfont=bf]{caption}
\newcommand{\vect}[1]{\boldsymbol{#1}}

\def\defi{\coloneqq}

\makeatletter
\let\standardaddcontentsline\addcontentsline
\newif\ifsupplementcontents
\supplementcontentsfalse
\renewcommand{\addcontentsline}[3]{%
  \ifsupplementcontents
    \standardaddcontentsline{#1}{#2}{#3}%
  \fi
}
\newcommand{\startsupplementcontents}{\global\supplementcontentstrue}
\newcommand{\supplementsectionlist}{\@starttoc{toc}}
\newcommand{\supplementfigurelist}{\@starttoc{lof}}
\newcommand{\supplementtablelist}{\@starttoc{lot}}
\makeatother

\title{\large Social Amplification Dominates Collective Hazard Response}

\date{} 					

\author[1]{Xiaolei Chu}
\author[1]{Guanren Zhou}
\author[2]{Marco Broccardo}
\author[3]{Didier Sornette\thanks{Corresponding author. Email: \href{mailto:dsornette@ethz.ch}{dsornette@ethz.ch}}}
\author[1]{Khalid M. Mosalam}
\author[1]{Ziqi Wang\thanks{Corresponding author. Email: \href{mailto:ziqiwang@berkeley.edu}{ziqiwang@berkeley.edu}}}

\affil[1]{Department of Civil and Environmental Engineering, University of California, Berkeley, United States}
\affil[2]{Department of Civil, Environmental and Mechanical Engineering, University of Trento, Italy}
\affil[3]{Institute of Risk Analysis, Prediction and Management, Southern University of Science and Technology, Shenzhen, China}

\begin{document}
\maketitle

\begin{abstract}
Large-scale hazards affect societies not only through direct physical impacts but also through emotions that spread across populations. Fueled by social amplification and networked communication, collective emotions often diverge markedly from underlying physical threats, pressuring policymakers toward suboptimal decisions that erode long-term societal resilience and misalign risk governance priorities. Yet when exactly these collective emotions mirror hazard severity and when they are warped by social dynamics remains poorly understood. We introduce a compact, interpretable model that couples hazard exposure with networked emotional contagion and identifies the transition from proportionate responses to an amplification regime sustained by negativity bias. Applying this framework to the COVID-19 pandemic in the United States, we integrate state-level epidemiological data with large-scale stress signals inferred from Twitter/X activity. Our analysis shows that social influence outweighed direct hazard forcing in over 80\% of U.S. states during the study period, and that amplified stress covaries with major economic indices. These findings reveal a measurable regularity in societal hazard response, enabling quantitative anticipation of collective emotional tipping points and supporting community resilience under large-scale hazards.
\end{abstract}

\section{Introduction}
When disasters strike, whether pandemics,  earthquakes, hurricanes, or technological failures, societies rarely respond in a coordinated and rational manner. Despite advances in hazard science and the development of extensive risk management frameworks, a chasm persists between this expertise and collective action. The core tenets of crisis management literature, emphasizing proactive preparation and structured organizational response, are well-documented, yet they are routinely overlooked in favor of reactive and often politicized decision-making. The COVID-19 pandemic made this discrepancy visible on a global scale: faced with a common biological threat, communities exhibited remarkably divergent policies, communication patterns, and emotional climates. This divergence was starkly illustrated by the widespread failure to implement existing pandemic preparedness plans. Despite having well-structured, pre-approved response frameworks in place, the initial, rapid reaction was often to sideline these protocols. Policy instead drifted, hastily improvised around nascent models and warped by the intense pressure of a 24-hour media cycle. These differences cannot be explained solely by epidemiological or resource variations. They point instead to deeper mechanisms through which emotion and social interaction shape collective behavior and resilience under global threat.

The growing literature on community resilience recognizes that recovery from disasters depends not only on physical and economic robustness but also on psychosocial stability~\citep{berkes2013community,koliou2020state}. Acute emotional surges, including fear, anger, denial, and exhaustion, are inevitable during disasters. However, when amplified by social networks, these transient reactions can evolve into disproportionately prolonged psychosocial impacts. Post-traumatic stress disorder (PTSD), anxiety, and depression are widely documented following major disasters~\citep{cullen2020mental, yabe2014psychological, schwartz2015impact}. Meta-analyses reveal an overall pooled prevalence of 22.6\% for post-pandemic PTSD, while the proportion of individuals meeting PTSD screening thresholds following hurricanes and earthquakes typically ranges from 10\% to 60\% \citep{yuan2021prevalence,galea2007exposure,alipour2020social}. Alarmingly, these effects can persist for years, with 12\% of PTSD patients from the 1999 İzmit earthquake retaining symptoms over a decade later~\citep{karamustafaliouglu2023ten}. Such enduring psychological disruptions, coupled with the echo chamber effects of social media, can trigger stress polarization and socioeconomic instability~\citep{del2016echo, kasperson1988social,hikichi2016can}. 

Beyond their direct societal costs, these amplified emotional dynamics also shape the decision environment faced by policymakers. In highly networked societies, leaders are embedded within the same information ecosystems that magnify public sentiment, and real-time emotional signals from media and online platforms can create strong feedback loops that compress decision-making timescales. Under these conditions, policy responses may become overly reactive to salient, emotionally charged narratives, favoring short-term visibility over long-term risk optimization. This can result in abrupt policy shifts, inconsistent strategies, and the diversion of resources away from less visible but structurally critical interventions, ultimately weakening societal resilience. Emotional resilience therefore demands quantitative, mechanistic understanding, not only to characterize psychosocial outcomes, but also to anticipate when collective emotions may decouple from underlying hazard severity and begin to steer governance in suboptimal directions. However, interpretable models that mechanistically link hazard exposure, social influence, and emotional contagion are currently lacking.

The key missing piece is a formal principle connecting micro-contagion to macro-responses: specifically, when does collective emotion remain proportionate to a hazard, and when does it diverge into a disproportionate and dominant societal force? Here we develop a quantitative model that treats collective emotion as an emergent phenomenon arising from two coupled forces: (i) the external field imposed by the hazard and (ii) emotional interactions transmitted through a social network. Individual emotional states fluctuate, but coupling through dense and heterogeneous connectivity can generate collective behavior analogous to that seen in many interacting systems. We formalize this intuition by representing each individual’s emotional state as influenced by neighbors and by hazard exposure, leading to a local Hamiltonian with two corresponding contributions. When emotional fluctuations are stochastic and local, and when collective dynamics evolve on slower timescales than individual interactions, the population can be approximated by a quasi-stationary distribution of Boltzmann form \citep{landau2013statistical}. This representation provides a direct bridge from individual contagion to population-level stability, and it frames abrupt shifts in sentiment as transitions between competing collective states.

Within this framework, we define the macroscopic stress prevalence—the fraction of individuals in heightened stress or arousal—as an order parameter that quantifies departure from an emotionally neutral population. This order parameter enables a compact description of collective polarization and permits low-dimensional characterization about regime change. The resulting mean-field Hamiltonian exposes a small set of governing parameters, including the relative strength of social influence and a negativity bias that makes negative states more contagious than positive ones. Phase diagrams then reveal a boundary separating proportional responses from tipping into majority high-stress states. 

The COVID-19 pandemic provides an unusually stringent test of these ideas: a common hazard experienced across many communities, yet accompanied by dramatic divergence in collective sentiment and behavior. We therefore examine state-level dynamics in the United States by integrating epidemiological statistics, mobility metrics, and stress signals derived from Twitter/X activity across all 50 states. This analysis reveals a consistent negativity bias across communities and shows that social influence outweighs direct hazard forcing in over 80\% of U.S. states. In this view, breakdowns in coordinated crisis response are not simply failures of information or resources; they are also consequences of a fragile balance between social structure, emotional contagion, and uncertainty. By making this balance measurable, the approach offers a quantitative basis for understanding the governing principles, anticipating tipping toward emotional polarization, and supporting community resilience under large-scale hazards.

\section{Main Results}

We present two main findings. First, our quantitative hazard–emotion model shows how social coupling transforms individual reactions into population-level stress, and reveals how negativity bias can shift communities from proportionate, hazard-tracking responses into dominant high-stress states. Second, applying the framework to U.S. state-level COVID-19 data shows that social amplification dominated collective stress responses in the vast majority of states, while collective emotional dynamics covaried with major economic indices, suggesting that socially amplified distress contributes to macroeconomic volatility.

\begin{figure}[t]
    \centering    \includegraphics[width=1.0\textwidth]{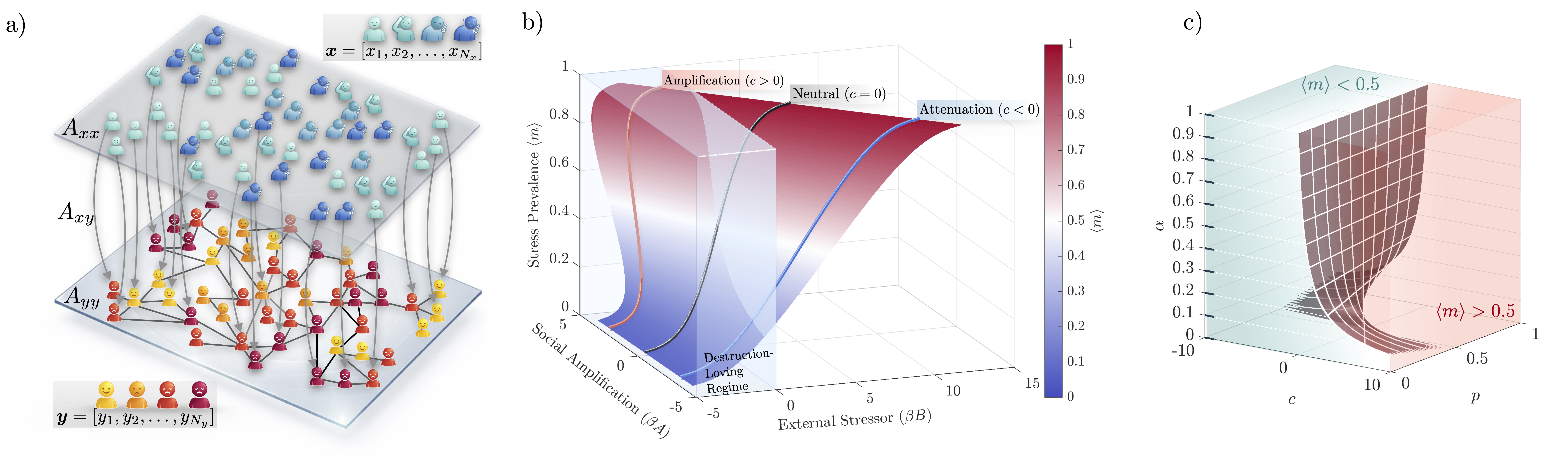}   
\caption{\textbf{Modeling hazard–emotion couplings.} 
\textbf{Panel (a):} Individuals’ emotional states are influenced by both tangible damage states and the emotions of others. Damage-to-emotion influence propagates through a bipartite hazard damage–human interaction network $\mathbf{A}_{\boldsymbol{xy}}$, while emotional interactions propagate on the social network $\mathbf{A}_{\boldsymbol{yy}}$. This study focuses on intra-layer emotional interactions within the social network and inter-layer coupling between hazard damage and emotions, treating hazard damage as prior information derived from domain-specific models and data. \textbf{Panel (b):} Mean stress prevalence $\langle m\rangle$ as a function of social amplification ($\beta A$) and external stressor ($\beta B$). For weak social amplification, responses are proportional. For strong amplification, responses become nonlinear and explosive. The cross-sections illustrate different social regimes: negativity bias ($c>0$) steepens the response, while positivity bias ($c<0$) attenuates it. A transition from a destruction-loving state ($B<0$) to a normal empathy state ($B>0$) can trigger a sharp, first-order phase transition in the amplification regime ($c>0$), while no such transition occurs for $c\leq0$. \textbf{Panel (c):} Limit-state surface $\langle m\rangle=0.5$ as a function of negativity bias $c$, physical damage prevalence $p$, and social interaction strength $\alpha$. For a community with positivity bias ($c<0$), the limit state is reached only when the physical damage is substantial ($p>0.5$). In contrast, in a negativity-biased community ($c>0$), even small physical damage can push the system toward the limit state. The interaction strength $\alpha$ modulates this balance: for small $\alpha$, physical damage dominates, whereas for large $\alpha$, the influence of $c$ becomes dominant.
}
\label{fig:multilayer_complex_network_demo}
\end{figure}

\subsection{Formulation of the Probabilistic Model for the Emotional Responses to Hazards}

Hazards can cause physical damage to both people and infrastructure, potentially triggering collective anxiety in communities. To model this process, we consider two sets of discrete vectors: (i) damage states of infrastructures and/or individuals, denoted by $\vect x = [x_1, x_2, \dots, x_{N_x}] \in\{0, 1, \dots, L\}^{N_x}$, and (ii) emotional states of individuals, denoted by $\vect y = [y_1, y_2, \dots, y_{N_y}] \in \{0, 1, \dots, L\}^{N_y}$. We let $0$ represent negligible damage or emotional impact, and $L$ the maximum damage or emotional impact. Since deterministic modeling is infeasible, for a given time window of interest, we seek a probabilistic description of the emotional states given the hazard damage states. As illustrated in \cref{fig:multilayer_complex_network_demo}(a), we construct a parsimonious local Hamiltonian for an individual interacting with their environment, defined as an effective energy function proportional to the negative logarithm of the stationary probability distribution over emotional states:
\begin{equation}
\label{eq:local_H}
\mathcal{H}_i = \alpha \sum_{\mathbf{A}_{\boldsymbol{yy}}(j,i)=1}\left(\left(y_i-y_j\right)^2 - c\,y_i\,y_j\right)
+(1-\alpha)\sum_{\mathbf{A}_{\boldsymbol{xy}}(j,i)=1}\left(x_j-y_i\right)^2,\quad i=1,2,\dots,N_y,
\end{equation}
where $\alpha \in [0,1]$ controls the relative importance of social interactions versus hazard damage, $c \in \mathbb{R}$ captures the asymmetry between low- and high-stress states, a phenomenon well established in social psychology~\citep{baumeister2001bad,rozin2001negativity},  $\mathbf{A}_{\vect{yy}}$ is the adjacency matrix of a social interaction network, and $\mathbf{A}_{\vect{xy}}$ is the biadjacency matrix of a bipartite hazard damage-human interaction network. We make the following remarks on the model:

\begin{itemize}
    \item \textbf{Main assumption}: An individual's emotional state is driven by social interactions and the severity of hazard damage, with $\alpha$ controlling the relative importance of these two drivers.
    
    \item \textbf{The first summation}: The emotional interaction is characterized by the sum of an empathy term $(y_j - y_i)^2$ and an amplification-attenuation term $-cy_i y_j$. If $c = 0$, the ground state of minimum energy favors $y_i$ to align with the average of $\{y_j\}$ that influence the individual. If $c > 0$, emotional amplification is triggered, as the ground state favors $y_i$ to be more stressful than the average of $\{y_j\}$, suggesting that stressful emotions are more contagious, known as the negativity bias; while $c < 0$ corresponds to the opposite: positivity bias. 
    
    \item \textbf{The second summation}: The direct emotional impact of hazard damage is characterized by the sum of the squared differences between the damage states $x_j$ and the emotional state $y_i$. As a result, the ground state favors an alignment between the physical damage and the emotional state. Amplification-attenuation terms are not introduced here because we assume that only human-to-human interactions can exhibit complex behaviors such as emotional amplification.
    
    \item \textbf{Networks}: Two directed networks are encoded in the model: the social interaction network, characterized by an $N_y \times N_y$ adjacency matrix $\mathbf A_{\vect{yy}}$, and the hazard damage-human interaction network, characterized by an $N_x \times N_y$ biadjacency matrix $\mathbf A_{\vect{xy}}$. We define $\mathbf A_{\vect{yy}}(j,i) = 1$ if the emotional state $y_j$ of the $j$-th individual affects the emotional state of the $i$-th individual, and $\mathbf A_{\vect{yy}}(j,i) = 0$ otherwise. Similarly, $\mathbf A_{\vect{xy}}(j,i) = 1$ if the damage state $x_j$ affects the emotional state of the $i$-th individual.
    
    \item \textbf{Parsimony and homogeneity}: 
     To retain interpretability and enable parameter inference, we adopt a homogeneous approximation in which $(\alpha, c)$ are treated as global parameters.     
\end{itemize}

Using the local Hamiltonian in \cref{eq:local_H}, we formulate a Gibbs simulation model (Algorithm~\ref{alg:Gibbs}) for emotional states. This model generalizes the Boltzmann distribution to systems with non-reciprocal interactions \cite{localHam1,localHam2}. The initial configuration $\vect y^{(0)}$ is set to the integer part of the average damage exposure, i.e., $y^{(0)}_i = \big\lfloor \frac{\sum_{\mathbf{A}_{\boldsymbol{xy}}(j,i)=1} x_j}{\sum_{j=1}^{N_x} \mathbf{A}_{\boldsymbol{xy}}(j,i)} \big\rceil$, where $\lfloor \cdot \rceil$ denotes round to the nearest integer.

\begin{algorithm}[H]
\caption{\textbf{Gibbs Simulation Model for Emotional States}}\label{alg:Gibbs}
\begin{itemize}[left=0pt]
    \item Given the current configuration $\vect y^{(k)}$, randomly select $i \in \{1,2,\dots,N_y\}$ and compute its local Hamiltonian $\mathcal{H}_i$. 
    \item Randomly perturb $y_i^{(k)}$ in $\vect y^{(k)}$ to obtain a proposal $\vect y'$, then compute its local Hamiltonian $\mathcal{H}_i'$ and $\Delta \mathcal{H}_i = \mathcal{H}_i' - \mathcal{H}_i$.
    \item Accept the proposal $\vect y^{(k+1)} \leftarrow \vect y'$, with probability $\min\{1, \exp(-\beta \Delta \mathcal{H}_i)\}$, where $\beta = 1/(k_BT)$. $T$ is an effective temperature that quantifies the intrinsic emotional volatility. Given that $T$ is a social analogue of temperature, we set the Boltzmann constant $k_B=1$. If $\vect y'$ is not accepted, $\vect y^{(k+1)} \leftarrow \vect y^{(k)}$. 
\end{itemize}
\end{algorithm}

For analytical tractability, we aggregate $\mathcal{H}_i$ to represent the statistical state of the system in Boltzmann form:
\begin{equation}\label{eq:global_H}
\begin{aligned}
  p_{\mathbf{Y}|\mathbf{X}}(\boldsymbol{y}|\boldsymbol{x})=&\frac{1}{Z}\exp{\left(-\beta\mathcal{H}\right)}\\
  =&\frac{1}{Z}\exp{\bigg(-\beta\Big(\frac{\alpha}{2}\sum_{\substack{1\leq i\leq N_y\\\mathbf{A}_{\boldsymbol{yy}}(j,i)=1}}((y_i-y_j)^2-c y_iy_j)+(1-\alpha)\sum_{\substack{1\leq i\leq N_y\\\mathbf{A}_{\boldsymbol{xy}}(j,i)=1}}(x_j-y_i)^2\Big)\bigg)}\,,       
\end{aligned}
\end{equation}
where $\mathcal{H}$ is the system Hamiltonian and $Z$ is the partition function. The Boltzmann distribution model is strictly consistent with the Gibbs simulation model when $\mathbf A_{\vect{yy}}$ is symmetric. In this context, the Hammersley-Clifford theorem\cite{bremaud2013markov} guarantees the form of \cref{eq:global_H}, and the Gibbs simulation model simply performs Gibbs sampling for the Boltzmann distribution. However, when $\mathbf{A}_{\vect{yy}}$ is asymmetric, the Boltzmann model  homogenizes the Gibbs simulation model by replacing non-reciprocal interactions with reciprocal ones and averaging directed interaction energies into undirected equivalents. Non-reciprocal couplings break the variational structure of the dynamics. As a result, the deterministic force comprises not only a potential gradient but also a non-conservative, solenoidal component. This drives complex transient behavior, characterized by burstiness, non-monotonicity, and the emergence of large, transient ``bubbles'' \cite{Sornettenonnormal2023},without changing the qualitative nature of the final equilibrium states. As an approximation of reality, \cref{eq:global_H} can be reasonable when emotional interactions reach a steady state that closely mimics thermodynamic equilibrium. This occurs when the emotional interactions relax rapidly during the time window of interest. In this work, we employ the Gibbs simulation model for numerical simulations and the Boltzmann distribution to derive analytical approximations.

To study collective stress within a population, we define the macroscopic quantity of interest, or order parameter, as the prevalence of hazard-induced stress: 
\begin{equation}
m(\boldsymbol{y}) = \frac{1}{N_y} \sum_{i=1}^{N_y} \mathds{1}(y_i)
\label{trhrbgr}
\end{equation}
where $\mathds{1}$ is a binary indicator function for hazard-induced emotional arousal. The order parameter $m$ is analogous to prevalence in epidemiology and magnetization in statistical physics. This coarse-grained description is adopted because our primary objective is not to characterize the full distribution of emotional intensities, but to quantify the collective prevalence of hazard-induced emotional arousal in the population. To this end, we distinguish emotionally neutral individuals ($y_i = 0$) from those exhibiting a non-negligible emotional response ($y_i > 0$). This binary coarse-graining is a deliberate modeling choice: at the macroscopic level, many collective outcomes of interest, such as social unrest, panic propagation, behavioral change, or demand for intervention, are driven primarily by whether individuals are emotionally aroused, rather than by the precise intensity of that arousal. Accordingly, this binary representation enables a tractable mean-field description of the emergence and prevalence of hazard-induced emotional arousal.

Adopting the binary states and following the Landau mean-field theory of phase transitions \cite{landau2013statistical} (see \emph{Methods and Materials} for details), the macroscopic system dynamics are governed by the Landau free energy density:
\begin{equation}
\label{eq:free_eng}
\begin{aligned}
    f(m) & = \beta\big((1-\alpha)\langle k_{\vect y}\rangle_{\mathbf{A}_{\boldsymbol{xy}}} - \frac{1}{2}\alpha c\langle k\rangle_{\mathbf{A}_{\boldsymbol{yy}}}\big) m^2-2 \beta(1-\alpha)\langle{x}\rangle_{\mathbf{A}_{\boldsymbol{xy}}} m + m \ln m + (1-m) \ln (1-m)\\
    &=-\beta Am^2-\beta Bm+m \ln m + (1-m) \ln (1-m)
\end{aligned}
\end{equation}
where the second line introduces two physically interpretable  composite parameters, $A$ and $B$, defined as follows:
\begin{subequations}\label{eq:AB}
    \begin{align}
        A & \defi -(1-\alpha)\langle k_{\vect y}\rangle_{\mathbf{A}_{\boldsymbol{xy} }}+\frac{1}{2}\alpha c\langle k\rangle_{\mathbf{A}_{\boldsymbol{yy}}}\,,\label{eq:A}\\\
        B & \defi 2 (1-\alpha) \langle x\rangle_{\mathbf{A}_{\boldsymbol{xy}}}\equiv 2 (1-\alpha) p \langle k_{\vect y}\rangle_{\mathbf{A}_{\boldsymbol{xy}}}\,,\label{eq:B}
    \end{align}
\end{subequations}
where $\langle k \rangle_{\mathbf{A}_{\boldsymbol{yy}}}$ represents the average in-degree of the social interaction network, $\langle k_{\vect y} \rangle_{\mathbf{A}_{\boldsymbol{xy}}}$ denotes the average in-degree of $\vect y$-nodes in the hazard damage-human interaction network, $\langle x \rangle_{\mathbf{A}_{\boldsymbol{xy}}}$ represents the average total damage states affecting each individual's emotional state, and $ p \defi \frac{\langle x \rangle_{\mathbf{A}_{\boldsymbol{xy}}}}{\langle k_{\vect{y}} \rangle_{\mathbf{A}_{\boldsymbol{xy}}}}$ is a damage rate. Parameter $A$ governs the collective dynamics of social network effects. Larger values indicate stronger social amplification, driven primarily by negativity bias ($c>0$) or dense social connectivity ($\langle k\rangle_{\mathbf{A}_{\boldsymbol{yy}}}\gg0$). Parameter $B$ quantifies the external forcing from hazard stressors; larger values arises mainly from high damage exposure ($\langle x_{\vect y}\rangle_{\mathbf{A}_{\boldsymbol{xy}}}$). \emph{Methods and Materials} presents the derivation of the mean-field model; \emph{Supplementary Information} S1–S3 further elaborates on the scope and properties of the effective Boltzmann approximation and the mean-field model.

\subsection{The Mechanism of Collective Emotional Response to Hazards}
\subsubsection{Fundamental Parameters}
The free energy described by \cref{eq:free_eng} identifies six fundamental parameters that govern the collective emotional response to hazards, as detailed below.
\begin{table}[h!]
    \centering
    \begin{tabular}{c p{\dimexpr\textwidth - 3cm\relax}}
        \toprule
        \textbf{Parameter} & \textbf{Interpretation} \\ \midrule        
        $\alpha $ & The relative importance of social interaction over hazard damage. If $ \alpha $ is close to 1, social interactions dominate the emotional response. \\ \midrule
        
        $c$  & Asymmetry between low- and high-stress emotional states: a positive $c$ suggests that high-stress states are more contagious, while a negative $c$ indicates the opposite. \\ \midrule
        
        $\langle k \rangle_{\mathbf{A}_{\boldsymbol{yy}}}$ & The average in-degree of the social interaction network, which quantifies the average number of connections per individual. \\ \midrule
        
        $\langle k_{\vect{y}} \rangle_{\mathbf{A}_{\boldsymbol{xy}}} $ & The average in-degree of $\vect y$-nodes in the hazard damage-human interaction network. This characterizes the average number of potential hazard exposures per individual. \\ \midrule
        
        $p \defi \frac{\langle x \rangle_{\mathbf{A}_{\boldsymbol{xy}}}}{\langle k_{\vect{y}} \rangle_{\mathbf{A}_{\boldsymbol{xy}}}} $ & 
        The average damage rate, defined as the mean damage level per hazard exposure affecting an individual. In binary damage systems ($x_j \in \{0,1\}$), $p = 1$ indicates that every potential hazard exposure leads to damage. \\ \midrule

        $T$ ~~~($\beta=1/T$) & Effective temperature, which quantifies the intrinsic variability in emotional state. Higher temperatures indicate greater emotional volatility, while lower temperatures indicate more stability.  \\ \bottomrule
    \end{tabular}
    \label{tab:parameters}
\end{table}

\subsubsection{Visualizing the Model}
The composite parameters $A$ and $B$ introduced in \cref{eq:free_eng} provide an effective means to image the dynamics of the originally six-dimensional mean-field model, as shown in \cref{fig:multilayer_complex_network_demo}(b). The figure maps the mean stress prevalence $\langle m \rangle$ as a function of $\beta A$ and $\beta B$, capturing the global landscape of  emotional responses. \cref{fig:multilayer_complex_network_demo}(b) further illustrates the role of the contagion asymmetry parameter \(c\). To reveal the full spectrum of model behaviors, we extend the parameter space to include negative values of $B$. The regime $B<0$ corresponds to a ``repulsive hazard-emotion coupling'' system, or a``destruction-loving'' state, which can be theoretically realized by replacing the attractive term \((x_i-y_i)^2\)  in the Hamiltonian with a repulsive term \((x_i+y_i)^2\). As demonstrated in \cref{fig:multilayer_complex_network_demo}(b), the composite parameter $A$ quantifies the social amplification of collective stress. Large values of $A$ correlate with explosive collective responses to external stressors. By analyzing the components of $A$ in \cref{eq:A}, we identify a critical divergence driven by the contagion asymmetry parameter $c$. In communities characterized by positivity bias ($c<0$), increasing social interaction effectively reduces $A$, thereby attenuating stress and fostering collective rationality. In contrast, for communities dominated by negativity bias ($c>0$), stronger social ties inflate $A$, driving the system toward collective irrationality manifested as hypersensitivity and explosive reactions to stressors.

\subsubsection{Limit-state Function for Emotional Polarization}
We define emotional polarization as the regime in which emotionally aroused individuals constitute a majority of the population, i.e., $\langle m \rangle>0.5$. This choice reflects a societal tipping point beyond which emotional arousal becomes collectively dominant and self-reinforcing, rather than a microscopic phase transition. The mean-field approximation of the limit-state function for emotional polarization is (Fig.~\ref{fig:multilayer_complex_network_demo}(c)):
\begin{equation}
\label{eq:limit_function}
G\left(\alpha, c,\langle k\rangle_{\mathbf{A}_{\boldsymbol{yy}}}, \langle k_{\vect y}\rangle_{\mathbf{A}_{\boldsymbol{xy}}}, p\right)= {\partial f(m) \over \partial m}\Big\vert_{m=1/2}=
-\frac{1}{2}\langle k\rangle_{\mathbf{A}_{\boldsymbol{yy}}}c+\langle k_{\vect y}\rangle_{\mathbf{A}_{\boldsymbol{xy}}}\left(1-2 p\right)\Big(\frac{1}{\alpha}-1\Big)\,.
\end{equation}
The sign of $G$ determines whether the free-energy minimum, and hence the equilibrium value of $m$, lies above or below the threshold $1/2$. As a result, $G$ serves as a criterion separating parameter regimes with minority versus majority emotional arousal.
Consistent with reliability theory, $G < 0$ corresponds to the undesirable outcome of $\langle m \rangle > 0.5$. This limit-state function is useful for discerning the trend of emotional polarization. In the realistic scenario where $\alpha\neq0,1$, \cref{eq:limit_function} suggests that $c$ and $p$ dominantly determine the sign of the function, and thus whether $\langle m \rangle > 0.5$. We make the following general observations on emotional polarization triggered by hazards:

\begin{itemize}
    \item For a community dominated by negativity bias ($c > 0$), there is a definite trend toward mass panic if the average damage rate is greater than $1/2$ ($p > 0.5$). If the average damage rate is not significant ($p<0.5$), the trend toward mass panic may still exist, determined by the competition between the effects of risk amplification (first term of \cref{eq:limit_function}) and hazard damage (second term). The competition is influenced by the connectivity of the social interaction network and the hazard damage-human interaction network.    
    \item For a community dominated by positivity bias or with no bias ($c \leq 0$), mass panic becomes possible only when the damage rate is substantial ($p > 0.5$).
\end{itemize}

These qualitative insights are made explicit and quantifiable by the limit-state function~$G$, which provides a closed-form criterion delineating parameter regimes associated with minority versus majority emotional arousal.

\subsection{Application to COVID-19 Datasets in the United States}
\subsubsection{Overview}
Beyond the severe casualties\cite{woolf2020excess,magesh2021disparities} and extensive economic\cite{chetty2024economic} and social disruptions\cite{levy2021social,andrade2022social}, COVID-19 has also caused ongoing mental health challenges \cite{cullen2020mental,talevi2020mental}. Understanding collective emotional responses during the COVID-19 pandemic is essential to refine risk management policies for future public health crises. In this section, we apply the proposed model to U.S. COVID-19 data for August 2020 and December 2021, for which sufficient state-level emotional and epidemiological data are available. We use the Twitter/X follow graph \cite{myers2014information} to map the topology of the social interaction network. Simultaneously, we construct the hazard–human interaction network using demographic contact patterns accounting for household, workplace, and school variations across states, sourced from the \href{https://www.un.org/development/desa/pd/data/household-size-and-composition}{United Nations Population Division 2022}. Finally, we conduct simulated control experiments to evaluate stress-mitigation strategies and investigate how these high-stress collective states impact economic indices. Full implementation details are provided in \emph{Methods and Materials} and \emph{Supplementary Information} S4--S8. 

\subsubsection{Distribution of $\alpha$ and $c$ Induced by COVID-19}
\begin{figure}[H]
    \centering
    \includegraphics[width=1.0\textwidth]{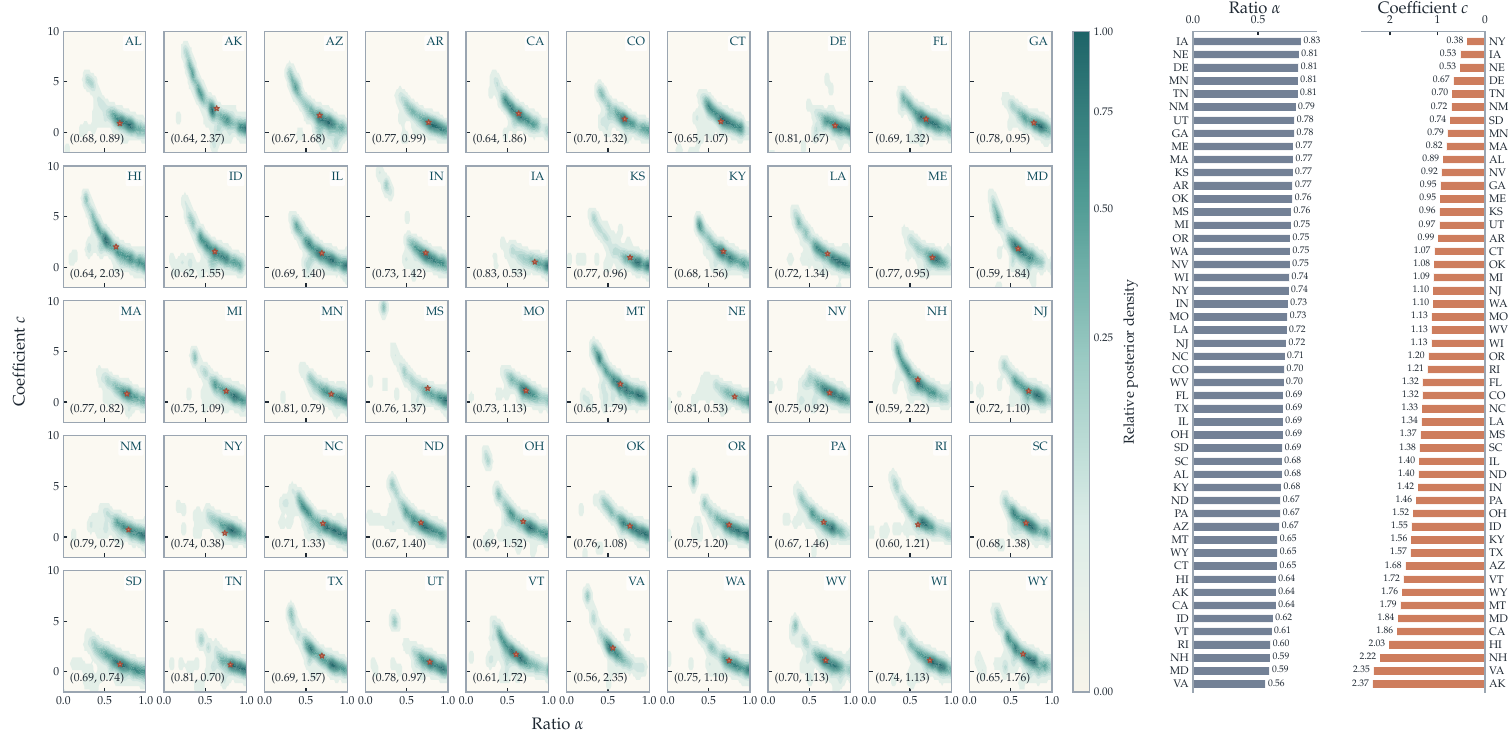}   
    \caption{\textbf{The joint posterior distributions of $\alpha$ and $c$ for each state in the United States.} The joint posterior densities of $\alpha$ and $c$ are obtained from Bayesian inference, accounting for both aleatory and epistemic uncertainties. Each subplot corresponds to a specific state, and the color scale shows the joint density normalized by the maximum density within that state panel; the color bar ranges from 0 to 1, where 1 denotes the panel-wise maximum posterior density and lower values denote lower relative density. The red star marks the posterior mean of ($\alpha$, $c$) for that state. The blue-gray bar plot labeled ``Ratio $\alpha$'' and the terracotta bar plot labeled ``Coefficient $c$'' on the right show the posterior mean values of $\alpha$ and $c$ for each state, sorted in descending and ascending order, respectively.}
    \label{fig:a_c_joint_multi_beta}
\end{figure}

We use the mean-field approximation and Bayesian parameter estimation to infer the posterior distribution of $(\alpha, c)$. \cref{fig:a_c_joint_multi_beta} shows the state-level posteriors, with colors in each panel normalized by the maximum density within that state. We assume that: (1) $\alpha$ and $c$ remain stable across the two observed periods, and (2) the effect of inter-state transport is negligible. The first assumption is reasonable because $\alpha$ and $c$ are intrinsic properties of a community and are unlikely to change significantly over a short timeframe. The second assumption is justified by the travel restrictions implemented during the COVID-19 pandemic. Recall that $\alpha$ represents the relative importance of social interaction over hazard damage, with $\alpha = 1$ indicating that social interactions dominate the emotional response. The parameter $c$ quantifies the asymmetry between low- and high-stress emotions: $c > 0$ reflects negativity bias and risk amplification, while $c < 0$ indicates positivity bias and risk attenuation. Parameter inference results reveal that all states have an average $\alpha > 0.5$, with some states (e.g., DE, IA, MN, NE) exhibiting high values exceeding $0.8$, highlighting the dominant influence of social interactions on the collective emotional response in the U.S. Additionally, all states show $c > 0$, confirming the prevalence of negativity bias. Significant negativity bias is observed in AK, HI, NH, and VA. 

Across the 50 U.S.\ states, we observe a negative correlation between the posterior mean of $\alpha$ and that of $c$. 
At the individual-state level, the joint posterior density of $(\alpha,c)$ typically forms a negatively sloped ridge, reflecting a partial identifiability and trade-off between the relative strength of social interactions and their emotional asymmetry. 
This arises naturally from the model structure, as collective emotional amplification depends primarily on the combined effect of $\alpha$ and $c$. Beyond this statistical dependence, the systematic state-level trend admits a plausible behavioral interpretation: in populations where emotional responses are strongly shaped by social interactions (large $\alpha$), stress propagation can occur even in the absence of strong negativity bias, whereas in populations with weaker social coupling (small $\alpha$), emotionally asymmetric interactions (large $c$) are required to generate comparable collective effects. 
These findings suggest a substitution between the strength of social influence and the degree of emotional bias, while acknowledging that posterior correlations may partly reflect model-inherent compensation effects.

\subsubsection{Limit-state Function-informed Control for Emotional Resilience}
The limit-state function introduced in \cref{eq:limit_function} suggests macroscopic control strategies to prevent collective distress by tuning its parameters. Specifically, given the current state of a community described by the limit-state function, we aim to increase its value (thus mitigating stress prevalence). Since $\alpha$ and $c$ are intrinsic properties unlikely to change over a short period, the tunable parameters include $\langle k\rangle_{\mathbf{A}_{\boldsymbol{yy}}}$ (e.g., reducing social media interactions by spending more time with family and neighbors), $\langle k_{\boldsymbol{y}}\rangle_{\mathbf{A}_{\boldsymbol{xy}}}$ (e.g., working from home and maintaining social distance), and $p$ (e.g., wearing masks or getting vaccinated).

\begin{figure}[h]
    \centering
    \includegraphics[width=1.0\textwidth]{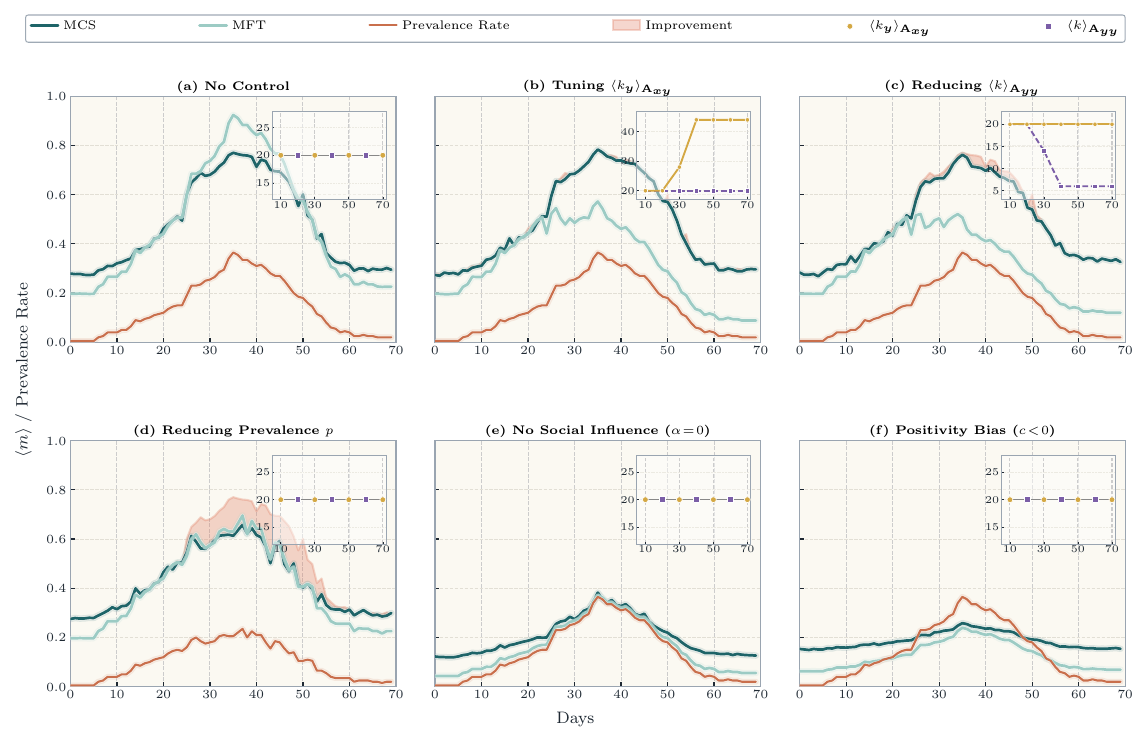}

\caption{\textbf{Numerical experiments on mitigating hazard-induced collective stress.} The figure displays the synthetic time evolution ($70$ days after the COVID-19 outbreak) of the stress prevalence under various control strategies. The strategies include: (a) no control, (b) tuning the average in-degree of $\boldsymbol{y}$-nodes in the hazard damage–human interaction network, (c) reducing the average degree of the social network, and (d) reducing the viral prevalence. Two additional cases are considered for comparison: (e) the \(\alpha=0\) case, in which the effects of social interactions are ignored, and (f) the positivity bias case (\(c<0\)), where social interactions suppress the high-stress state. In each panel, the dark green line represents the result of the Gibbs simulation model (Algorithm~\ref{alg:Gibbs}), while the light green line represents the prediction of the mean-field approximation. The orange line indicates the time evolution of the viral prevalence. The shaded patch represents the improvement compared to the no-control baseline. The inset in the upper-right corner of each panel shows the time evolution of the average network degrees $\langle k_{\boldsymbol{y}}\rangle_{\mathbf{A}_{\boldsymbol{xy}}}$ and $\langle k\rangle_{\mathbf{A}_{\boldsymbol{yy}}}$, illustrating which network parameter is being tuned by each control strategy. The results show that reducing the infectious rate is the most effective strategy. In the positivity bias case, the stress prevalence is even lower than the viral prevalence.}
    \label{fig:control_strategy}
\end{figure}

We conduct numerical control experiments with different parameter-tuning schemes to examine whether the limit-state function can guide control actions. Monte Carlo simulations (MCS in \cref{fig:control_strategy}) of 200 agents governed by the Gibbs model (Algorithm~\ref{alg:Gibbs}) are employed as a reference for the ``true'' dynamics. The initial hazard damage–human interaction network and social network both have average in-degree values of $\langle k_{\boldsymbol{y}}\rangle_{\mathbf{A}_{\boldsymbol{xy}}}=20$ and $\langle k\rangle_{\mathbf{A}_{\boldsymbol{yy}}}=20$, respectively. We set $\alpha = 0.67$ and $c = 0.70$, values that are representative for many U.S. states based on the results presented in the previous section.

We compare three hypothetical control schemes based on the limit-state function: (i) increasing $\langle k_{\boldsymbol{y}}\rangle_{\mathbf{A}_{\boldsymbol{xy}}}$ if \(p < 1/2\) and decreasing it if \(p \geq 1/2\); (ii) reducing $\langle k\rangle_{\mathbf{A}_{\boldsymbol{yy}}}$; and (iii) reducing the viral prevalence $p$. The trigger for taking control actions is a negative value of the limit-state function. When adjusting the network structure, we assume that the average in-degree cannot change by more than $2$, and that the reduction in the viral prevalence cannot exceed $40\%$\cite{howard2021evidence,cheng2020role}. We also run a baseline experiment with no control actions. Additionally, to understand the net contribution of social interactions on emotional responses, we conduct experiments with $\alpha = 0$, meaning that only the physical impact influences emotions. Moreover, to shed light on the potential emotional responses of a community dominated by positivity bias, we conduct experiments with \(c < 0\).

Note that these simulations primarily serve as a proof of concept. In practice, the costs, feasibility, and secondary consequences associated with tuning different parameters can vary substantially. In particular, the strategy of tuning $\langle k_{\boldsymbol{y}}\rangle_{\mathbf{A}_{\boldsymbol{xy}}}$ may appear counterintuitive or impractical: when there are more uninfected than infected individuals ($p < 1/2$), the limit-state function suggests that increasing physical contacts would reduce collective stress, relying on the reasoning that greater exposure to uninfected individuals may mitigate emotional arousal. However, this static interpretation neglects dynamic feedback effects, whereby increased physical contact can elevate viral prevalence over time. For illustrative purposes and with potential applicability to other hazard contexts where such feedbacks are weaker, we nonetheless include this control action in the numerical analysis.

\cref{fig:control_strategy} summarizes the outcomes under the different control strategies; full implementation details are provided in \emph{Supplementary Information} S7. The results indicate that reducing viral prevalence (i.e., lowering the average damage rate) is the most effective intervention in this case study, whereas adjustments to network structure yield more modest improvements. This ordering is consistent with the gradients of the limit-state function: the partial derivatives with respect to $\langle k_{\boldsymbol{y}}\rangle_{\mathbf{A}_{\boldsymbol{xy}}}$, $\langle k\rangle_{\mathbf{A}_{\boldsymbol{yy}}}$, and $p$ are on the order of $0.1$, $0.1$, and $10$, respectively. In the $\alpha = 0$ case, stress prevalence closely tracks viral prevalence. Under a positivity-bias scenario (simulated by setting $c = -0.70$), individuals tend to encourage others even at high damage levels, resulting in stress prevalence that remains noticeably lower than the viral prevalence. More broadly, these findings motivate future studies on hybrid control strategies that jointly tune epidemiological and social parameters while accounting for their costs and dynamic interactions. The proposed framework provides a principled foundation for developing and evaluating such cost-aware intervention strategies. 

\subsection{Impact of Stress Prevalence on Socio-Economic Indices}

To investigate the impact of stress prevalence on socio-economic performance, we select a set of indicators that capture key dimensions of state-level economic activity, labor market conditions, demographic characteristics, and asset market dynamics. The data covering April 2020 to March 2023 are obtained from the Federal Reserve Bank of St. Louis \cite{fred_database}. Specifically:

\begin{itemize}
    \item \textbf{Nominal and Real Gross State Product (NGSP\cite{fred_ngdp_us}, RGSP\cite{fred_gdpc1}):} These indicators provide a comprehensive measure of overall economic output, with NGSP reflecting market value in current dollars and RGSP adjusting for inflation to capture real growth.
    \item \textbf{Per Capita Personal Income (PCPI\cite{fred_personal_income}):} A direct measure of individual welfare and purchasing power, strongly linked to consumption behavior and standards of living.
    \item \textbf{Unemployment Rate (UR\cite{fred_unrate}):} A key barometer of labor market health and social stability, highly sensitive to economic shocks.
    \item \textbf{Population (POP\cite{fred_popthm}):} A structural determinant of both labor supply and aggregate demand, shaping long-term economic potential.
    \item \textbf{House Price Index (HPI\cite{fred_ussthpi}):} A proxy for the housing market and household wealth, exerting significant influence on consumption through wealth effects and expectations.
    \item \textbf{Coincident Economic Activity Index (CEAI\cite{fred_usphci}):} A composite indicator integrating employment, production, and income data, serving as a high-frequency ``thermometer'' of economic conditions.
\end{itemize}

The first-order Sobol index \cite{sobol1993sensitivity} quantifies the contribution of each input factor to the variance of the output. In this analysis, we treat stress prevalence and COVID-19 prevalence as input variables and socio-economic indices as outputs, in order to assess whether collective emotional states exert a measurable influence on socio-economic performance. Note that stress prevalence is inferred from the mean-field model due to the lack of state-level emotional data over the study period. As shown in \cref{fig:economic_impact} (see \emph{Supplementary Information} S8 for implementation details), during the early stage of the COVID-19 pandemic, when the viral prevalence is relatively low (prior to January 2022), variation in economic performance is dominated by the viral prevalence. As the viral prevalence increases (after January 2022), widespread high-stress emotional states begin to contribute substantially to economic variability, particularly for the Coincident Economic Activity Index (CEAI). These results suggest that as the hazard intensifies, collective emotional responses can influence economic outcomes, in some cases rivaling or exceeding the effects of physical damage. This finding underscores the importance of managing collective emotional dynamics during large-scale hazards to mitigate their broader socio-economic consequences.

\begin{figure}[h]
    \centering
    \includegraphics[width=1\textwidth]{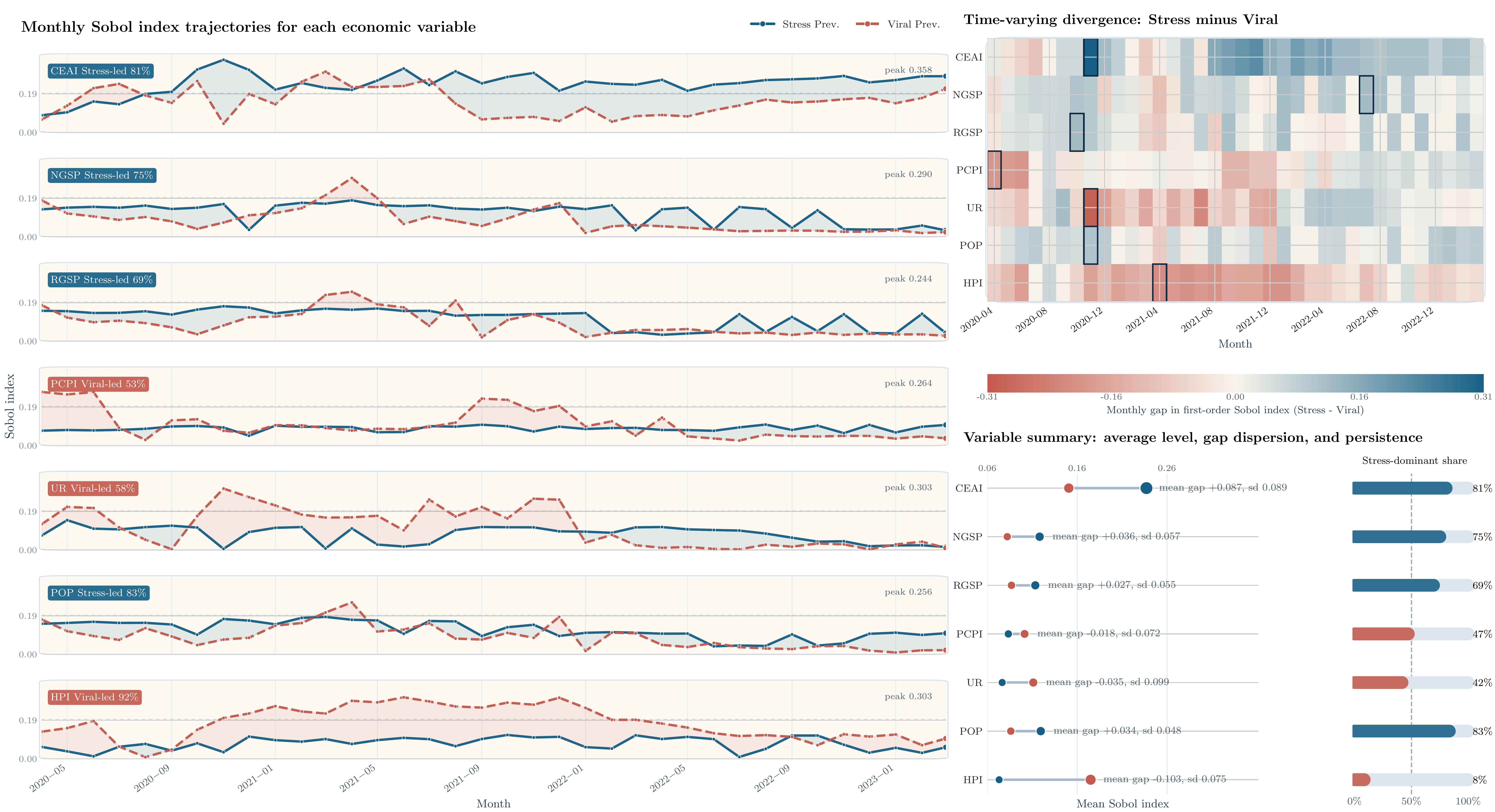}
    \caption{\textbf{Contributions of collective emotion and disease prevalence to macroeconomic variance.} Left: monthly first-order Sobol indices from April 2020 to March 2023 for inferred stress prevalence (blue solid line) and COVID-19 prevalence (orange dashed line). Blue and orange fills mark months in which the corresponding Sobol index is larger; the badge reports the share of months in which the indicated variable has the larger first-order Sobol index, and ``peak'' gives the maximum monthly first-order Sobol index in that strip. Upper-right: heatmap of the monthly gap in first-order Sobol index, defined as Stress minus Viral, where Stress and Viral denote the first-order Sobol indices of stress prevalence and COVID-19 prevalence, respectively; blue cells denote positive gaps, orange cells denote negative gaps, color intensity follows the color bar, and outlined cells mark the month with the largest absolute gap for each variable. Lower-right: variable-level summary of the mean Sobol indices of stress and viral prevalence (blue and orange dots, with dot area proportional to the corresponding mean Sobol index), the mean gap and its standard deviation (text), and the share of months in which stress has the larger first-order Sobol index (horizontal bar). CEAI: Coincident Economic Activity Index; NGSP: Nominal Gross State Product; RGSP: Real Gross State Product; PCPI: Per Capita Personal Income; UR: Unemployment Rate; POP: Population; HPI: House Price Index.}\label{fig:economic_impact}
\end{figure}

\section{Discussion}
Large-scale hazards do more than damage infrastructure, disrupt economies, and claim lives; they profoundly perturb the emotional landscape of societies, sometimes on a scale that exceeds the physical event itself. Our findings highlight the significance of this often-neglected dimension. We demonstrate that collective emotional responses are not merely reflections of hazard severity. Instead, they are co-produced by direct exposure and networked social transmission, allowing them to amplify through feedback loops. This perspective fundamentally shifts how we must approach recovery and resilience. Much of the existing institutional architecture for disaster preparedness assumes that public responses will remain broadly proportional and manageable. But when social amplification emerges, that premise breaks down. Plans conceived under conditions of rational coordination can be weakened, displaced, or abandoned once fear, uncertainty, and emotional contagion begin to organize collective behavior. The COVID-19 pandemic made this discrepancy globally visible, and similar departures from formal response frameworks are often observed after natural disasters. Resilience therefore cannot be reduced to the capacity to withstand physical shocks alone: recovery is shaped not only by material losses but also by collective emotional reactions and by the social processes that amplify them. In this light, the social amplification of hazard-induced emotions is not a secondary complication, but a central unresolved vulnerability in contemporary risk governance. By introducing a quantitative framework to analyze these dynamics, this study provides a critical foundation for addressing this gap.

We introduced a parsimonious statistical-physics-inspired framework in which individual emotional states evolve under the combined effects of hazard exposure and social interaction. A small set of interpretable parameters controls the relative importance of social influence and the asymmetry of emotional interactions, allowing negativity or positivity bias to be represented explicitly. A mean-field approximation then links these microscopic dynamics to a macroscopic measure of collective emotional arousal and identifies the conditions under which high-stress states become dominant. The model yields several insights. First, emotional asymmetry strongly shapes collective outcomes: under negativity bias, social interaction amplifies stress, whereas under neutral or positive bias it can instead damp emotional escalation. Second, collective emotional responses need not vary smoothly with hazard severity. Under sufficiently strong negativity bias, the system can undergo abrupt transitions and remain trapped in elevated-stress regimes even as damage declines, indicating that reducing physical harm alone may not restore proportionate collective responses. Third, widespread high-stress states can emerge even under relatively modest hazard damage when social amplification is strong enough. More broadly, these results suggest that collective distress is not simply a reflection of objective conditions, but an emergent property of coupled social dynamics that may also distort risk perception and decision-making.

These theoretical predictions are supported by empirical calibration to COVID-19 data across the 50 U.S. states. Using Bayesian inference, we estimated social-influence and emotional-bias parameters by combining epidemiological prevalence, sentiment signals extracted from Twitter/X posts using a fine-tuned BERTweet model, and proxies for demographic contact structure and online social connectivity. The inferred parameters indicate that negativity bias was prevalent in most states and that social amplification contributed substantially to collective emotional responses in the large majority of cases. In this setting, the mean-field limit-state analysis also helped identify directions of intervention, providing a basis for future work on strategies that jointly target hazard exposure and the social transmission of distress.

Overall, our results show that collective emotional responses to hazards can be understood through a small number of interpretable mechanisms within a unified quantitative framework. This has practical implications for resilience: societies may fail not only because hazards are severe, but also because social amplification drives reactions that become decoupled from the underlying threat. Extending this framework to other hazards, compound events and heterogeneous populations could help build a more complete science of how societies perceive, propagate and respond to extreme events.

\newpage
\section{Methods and Materials}
\subsection{Mean-Field Approximation}
The mean-field theory (MFT) coarse-grains a statistical physics system from the component/microscopic scale to the macroscopic scale described by the mean field $m$. Within the MFT framework, we assume that the variation of the emotional state $y_i$ for each individual around the mean field $m$ is small, i.e., $y_i = (y_i - m) + m = \delta y_i + m$, where $\delta y_i$ represents the small deviation from the mean. Under this assumption, the system Hamiltonian in \cref{eq:global_H} can be approximated by the mean-field Hamiltonian:
\begin{equation}
\label{eq:Hamiltonian_MF}
\begin{aligned}
\mathcal{H}_{\text{MF}}(m) &= \frac{\alpha}{2}\sum_{\substack{1\leq i\leq N_y\\\mathbf{A}_{\boldsymbol{yy}}(j,i)=1}}\left(-c m^2\right)+(1-\alpha)\sum_{\substack{1\leq i\leq N_y\\\mathbf{A}_{\boldsymbol{xy}}(j,i)=1}}\left(m^2-2 x_j m\right) \\
& = N_y\Big(-\frac{1}{2}\alpha c \langle k\rangle_{\mathbf{A}_{\boldsymbol{yy}}} m^2 + (1-\alpha)\langle k_{\boldsymbol{y}}\rangle_{\mathbf{A}_{\boldsymbol{xy}}} m^2 - 2(1-\alpha)\langle x\rangle_{\mathbf{A}_{\boldsymbol{xy}}} m \Big)\\
&\equiv N_y(-Am^2-Bm)\,.
\end{aligned}
\end{equation}
where constant terms (such as $x_j^2$) have been omitted as they cancel out in the normalization of the partition function. Using the mean-field system Hamiltonian, the partition function is approximated by:
\begin{equation}
\label{eq:partition_function_functional}
Z \approx Z_{\mathrm{MF}}=N_y\int_0^1 \mathrm{d}m \cdot \Omega\left(m\right) \exp{(-\beta \mathcal{H}_{\text{MF}}[{m}(\boldsymbol{y})])}\,. 
\end{equation}
where $\Omega\left(m\right)$ is 
the size of the phase space.

Let $N_1$ represent the number of spins with a value of $1$, and $N_0$ represent the number of spins with a value of $0$. We have:
\begin{equation}
\label{eq:phase_space}
\Omega(m) = \dfrac{N_y!}{N_1! N_0!}\,,
\end{equation}
where $N_1 + N_0 = N_y$ and $N_1 / N_y = m$. Using Stirling's formula, we obtain:
\begin{equation}
\label{eq:stirling_approx}
\begin{aligned}
    \lim_{N_y \rightarrow \infty} \ln{\Omega(m)} & = N_y \ln{N_y} - N_y - N_1 \ln{N_1} + N_1 - N_0 \ln{N_0} + N_0 \\
    & = N_y \left( -m \ln{m} - (1 - m) \ln{(1 - m)} \right)\,.
\end{aligned}
\end{equation}
Using Eq.~\eqref{eq:stirling_approx}, the partition function becomes:
\begin{equation}
\label{eq:partition_mft}
\begin{aligned}
    Z_{\text{MF}} & = N_y\int_0^1 \exp{(-\beta \mathcal{H}_{\text{MF}}(m) + N_y \left( -m \ln{m} - (1 - m) \ln{(1 - m)} \right))}\,\mathrm{d}m \\[3pt]
    & = N_y\int_0^1 \exp{\Big(-N_y\underbrace{\big(-\beta A m^2-\beta B m  + m \ln{m} + (1 - m) \ln{(1 - m)} \big)}_{\eqqcolon f(m)}\Big)}\,\mathrm{d}m\,.
\end{aligned}
\end{equation}
The saddle point approximation for (\ref{eq:partition_mft}) leads to the self-consistency condition for the stress prevalence:
\begin{equation}
\label{eq:self_cons}
\frac{\partial f(m)}{\partial m}\Big|_{m=\langle m\rangle}=-2\beta A\langle m\rangle-\beta B+\ln \frac{\langle m\rangle}{1-\langle m\rangle}=0\,.
\end{equation}
We use the mean-field formulas to study the mechanisms of collective emotional response. The accuracy of this mean-field approximation is evaluated in \emph{Supplementary Information} S3, while the analytical properties of the free-energy density are summarized in S2.

\subsection{Bayesian Model Inference}
\label{sec:BayModelInf}
We assume that the parameters $(\alpha, c)$ to be inferred are intrinsic properties of the community that remain stable over the time period during which the data were collected. Our goal is therefore to estimate the posterior distribution $f(\alpha, c | \{m^{(i)}\})$, conditional on observations of the stress prevalence $\{m^{(i)}\}$ recorded at different time intervals. Since we lack precise information on the temperature $T$—a hyperparameter related to the community’s inherent emotional variability—we integrate over it, resulting in:
\begin{equation}
\label{eq:posterior_Bayesian}
\begin{aligned}
f(\alpha, c | \{m^{(i)}\}) &\propto \int_{\Omega_T} f(\{m^{(i)}\} | \alpha, c, T) f(\alpha, c, T) \,dT \\
&= \prod_i \int_{\Omega_T} f(m^{(i)} | \alpha, c, T) f(\alpha, c) f(T) \,dT\,,
\end{aligned}
\end{equation}
where $T$ is assumed to be uniformly distributed over $\Omega_T \in [10,30]$. This range is chosen to keep inference in a non-degenerate regime of the mean-field free energy (\cref{eq:free_eng}): if $T$ is too large ($\beta\to 0$), the entropic term dominates and the equilibrium prevalence approaches $m\approx 0.5$; if $T$ is too small ($\beta\to\infty$), the entropic term becomes negligible and the equilibrium is governed almost entirely by the energetic coefficients $A$ and $B$. Hence, $[10,30]$ provides a practical balance for representing plausible emotional variability while avoiding these two extremes. The prior distribution $f(\alpha, c)$ is non-informative, which is also regarded as uniform. $f(m^{(i)} | \alpha, c, T)$ is the likelihood function. 

To construct the likelihood, we incorporate both aleatory and epistemic uncertainties in observing the stress prevalence from sentiment classification. For a given time interval where $m^{(i)}$ is estimated, let $N = N_0 + N_1 + N_2$ be the total number of collected posts, where $N_0$, $N_1$, and $N_2$ are the counts of posts classified as unaffected, affected, and irrelevant, respectively. By definition, $m^{(i)} = N_1 / (N_0 + N_1)$. The likelihood for observing $m^{(i)}$ can be expressed as:
{\small
\begin{equation}
\label{eq:Bayesian_inference}
\begin{aligned}
f(m^{(i)} | &\alpha, c, T) =\\
&\frac{1}{Z_{\text{MF}}(\alpha, c, T)}
\sum_{n_0=1}^N \sum_{n_1=0}^{N-n_0}
 \exp(-\beta\mathcal{H}_{\text{MF}}(\tfrac{n_1}{n_0 + n_1}; \alpha, c))\Omega(\tfrac{n_1}{n_0 + n_1})
\cdot 
\text{Multinomial}(N_0, N_1, N_2; N, p_c(n_0, n_1))\,.    
\end{aligned}
\end{equation}
}In \cref{eq:Bayesian_inference}, we distinguish between the observed counts $(N_0,N_1,N_2)$ produced by the sentiment classifier and the latent, unobserved counts $(n_0,n_1)$, which represent the true numbers of unaffected and affected posts in the population. The stress prevalence is estimated from the observed data as $m^{(i)} = N_1/(N_0+N_1)$; however, the mean-field model governs the distribution of the underlying, true emotional prevalence $m = n_1/(n_0+n_1)$. For this reason, the mean-field energy $\mathcal{H}_{\text{MF}}$ and the associated phase-space factor $\Omega(\cdot)$ are evaluated as functions of the latent variables $(n_0,n_1)$ rather than the observed counts $(N_0,N_1)$.

The likelihood construction therefore marginalizes over all admissible latent configurations $(n_0,n_1)$ that could have generated the observed counts $(N_0,N_1,N_2)$, explicitly accounting for epistemic uncertainty arising from classification errors. In this formulation, $n_0$ and $n_1$ denote the unknown true numbers of posts in the unaffected and affected classes, respectively, while $\Omega(\cdot)$ represents the size of the corresponding phase space, as defined in \cref{eq:phase_space}. The exponential term involving $\mathcal{H}_{\text{MF}}$ captures the aleatory variability inherent in the emotional dynamics over the time interval during which $m^{(i)}$ is measured, with additional model parameters treated as deterministic within that interval.

The multinomial term $\text{Multinomial}(N_0, N_1, N_2; N, p_c(n_0, n_1))$ provides the probabilistic link between the latent true counts and the observed classifier outputs, representing the probability of observing $(N_0,N_1,N_2)$ given $N$ total posts and a classification probability vector $p_c(n_0,n_1)$ that accounts for misclassification effects. We start the summation at $n_0 = 1$ to avoid the case $n_0 = n_1 = 0$, which would render $m^{(i)}$ undefined. 
 
 The probability vector $p_c(n_0, n_1)$, which accounts for classification errors, is given by:
\begin{equation}
p_c(n_0, n_1) = \frac{1}{N} [\,n_0, n_1, N-n_0-n_1\,]
\begin{bmatrix}
1-q & q/2 & q/2 \\
q/2 & 1-q & q/2 \\
q/2 & q/2 & 1-q
\end{bmatrix}\,,    
\end{equation}
where $q$ denotes the average classification error rate.
The probability vector $p_c(n_0,n_1)$ models classification errors through a symmetric confusion structure. The vector
$\frac{1}{N}[\,n_0, n_1, N-n_0-n_1\,]$
represents the true class proportions of unaffected, affected, and irrelevant posts, respectively. These true proportions are mapped to observed class probabilities by multiplication with a $3\times3$ confusion matrix, where $1-q$ is the probability of correct classification and $q$ is the average misclassification rate. Under the assumption of no systematic bias among incorrect labels, the misclassification probability $q$ is distributed uniformly across the two incorrect classes, yielding off-diagonal entries equal to $q/2$. This symmetric structure provides a parsimonious representation of classification uncertainty while preserving normalization and interpretability.

To accommodate the strong correlation between $\alpha$ and $c$, an affine-invariant Markov chain Monte Carlo sampler \cite{goodman2010ensemble} is employed to sample from the posterior distribution, \cref{eq:posterior_Bayesian}. Technical details of numerical implementation and robustness verification are reported in \emph{Supplementary Information} S4.

\subsection{Case Study on the COVID-19 Datasets}
\subsubsection{Overview}
To demonstrate the application of the proposed hazard-induced emotional response model, we choose the COVID-19 pandemic as a case study. The emotional response data extracted from social media posts and the epidemiological data from other sources do not have individual-to-individual correspondence. This implies that a fine-grained parameter estimation, such as using the Gibbs simulation or Boltzmann distribution model, is infeasible. To address this challenge, we rely on the mean-field approximation, which requires only macroscopic characterizations of the COVID-19 contact network and social interaction network, as well as aggregated epidemiological and emotional response data.

To obtain emotional responses to COVID-19, we fine-tuned a large language model, BERTweet\cite{nguyen2020bertweet}, for sentiment classification of Twitter/X posts across all U.S. states. Epidemiological data for the states are collected from open sources\cite{dong2020interactive}. The statistical property of the social interaction network is consistent with the Twitter following network, which aligns with the emotional response data. The hazard-emotion interaction network is consistent with the commuting pattern \cite{kerr2021controlling,colby2015projections}. 

\begin{figure}[h]
    \centering
       \includegraphics[width=1\textwidth]{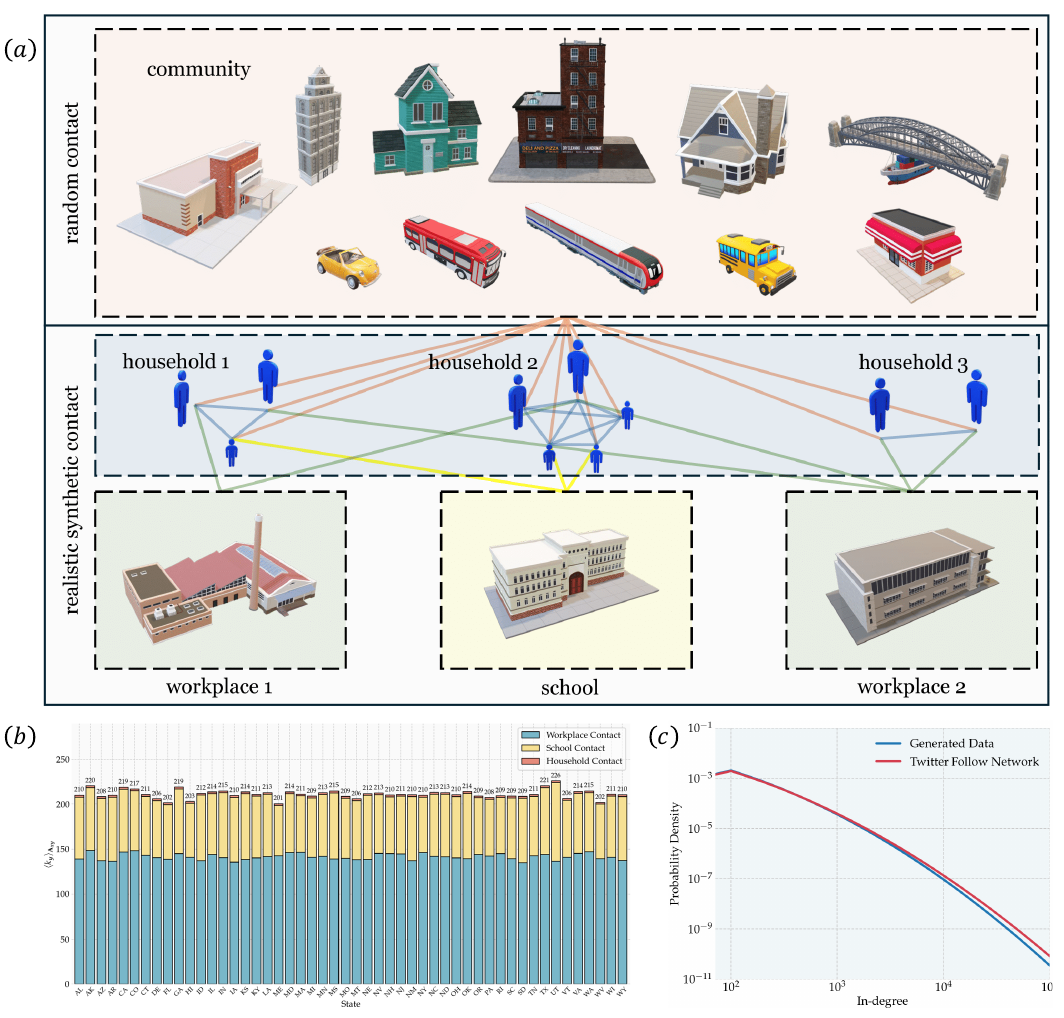}
\caption{\textbf{(a) Illustration of contact networks with multiple layers, including household, school, workplace, and community.} For household, school, and workplace contacts, we generate synthetic contact networks that align with census data. A random contact network is used for the community layer, where each individual establishes $n$ random contacts within the population. The value of $n$ follows a Poisson distribution with rate parameter $\lambda_c$, set to the expected number of community contacts ($\lambda_c=20$). 
\textbf{(b) Average in-degree of hazard damage–human interaction.} This panel illustrates the average in-degree for workplace, school, and household contacts. These values are extracted from the Covasim simulation \cite{kerr2021covasim} and are consistent with real-world data \cite{colby2015projections,huisman2009effects}. 
\textbf{(c) Average in-degree of the social interaction network.} The Twitter follow graph \cite{myers2014information} is used to represent the topology of the social interaction network, with an average in-degree of $\langle k \rangle_{\mathbf{A}_{\boldsymbol{yy}}} \approx 172$.}
    \label{fig:contact_network}
\end{figure}

\cref{fig:contact_network} illustrates properties of the two networks. The Twitter follow graph is used to represent the topology of the social interaction network, with an average in-degree of $\langle k \rangle_{\mathbf A_{\boldsymbol{yy}}} = 171.57$. 
Interestingly, this value is of the same order of magnitude as the Dunbar number, which posits a cognitive limit of approximately 150 stable social relationships that an individual can maintain \cite{dunbar1992neocortex,dunbar2010many}. 
While online social connections differ from offline social ties in strength and persistence, this proximity suggests that even in large-scale digital platforms, the effective number of socially influential connections may remain constrained by cognitive and attentional limits, consistent with empirical findings reported in related studies.

Using these elements, we calibrate the state-level parameters $\alpha$ and $c$ via the Bayesian model inference (see \cref{sec:BayModelInf}). We then conduct different control strategies informed by the limit-state function, such as reducing emotional contacts or lowering the viral prevalence rate, to examine the effectiveness of these strategies in mitigating collective stress. 

\subsubsection{Hazard Damage-Human Interaction Network and Social Interaction Network} 
\paragraph{Hazard damage-human interaction network.} 
The adjacency matrix of the hazard damage–human interaction network, $\mathbf{A}_{\boldsymbol{xy}}$, is extracted from contact records during the agent-based simulation. To ensure consistency with real-world data, the COVID-19 transmission simulation incorporates four different contact networks: household, school, workplace, and community contacts. For household, school, and workplace contacts, synthetic contact networks are generated based on census and survey data from Demographic and Health Surveys \cite{colby2015projections,huisman2009effects}, considering factors such as age distribution, household size, school enrollment, and employment rates. Community contact is modeled using a random contact network, where each person in the population can interact with any other individual. When constructing $\mathbf{A}_{\boldsymbol{xy}}$, only household, school, and workplace contacts are included. Community contacts are excluded because, in the agent-based simulation, the community contact network is modeled as a random network to represent incidental interactions. Since individuals in community contact do not necessarily know each other, the emotional state of individual $i$ is not influenced by the health condition of individual $j$ if they lack a direct connection. \cref{fig:contact_network}(a) illustrates the multi-layer contact networks. In mean-field theory, only the average in-degree $\langle k_{\boldsymbol{y}}\rangle_{\mathbf{A}_{\boldsymbol{xy}}}$ affects the results (values shown in \cref{fig:contact_network}(b)), meaning the detailed topology of $\mathbf{A}_{\boldsymbol{xy}}$ is not required during the calibration in \cref{fig:a_c_joint_multi_beta}. However, the full network topology of $\mathbf{A}_{\boldsymbol{xy}}$ is utilized in the Monte Carlo simulations to validate the mean-field approximation, as detailed in \emph{Supplementary Information} S3.

\paragraph{Social interaction network.}
There are various types of social interaction networks, including offline social networks (e.g., friend, family, professional, colleague, and acquaintance networks) and online social networks (e.g., Twitter/X, Facebook, and Instagram). In this study, we focus on the Twitter follow network, as it aligns with the emotional response data, which is also extracted from Twitter/X. The adjacency matrix of the social interaction network, $\mathbf{A}_{\boldsymbol{yy}}$, is directly extracted from the Twitter follow network, with its degree distribution shown in \cref{fig:contact_network}(c). In the mean-field approximation, only the average in-degree $\langle k\rangle_{\mathbf{A}_{\boldsymbol{yy}}}$ is required, so the detailed network topology of $\mathbf{A}_{\boldsymbol{yy}}$ is not necessary for calibration in \cref{fig:a_c_joint_multi_beta}. However, its full topology is incorporated into the Monte Carlo simulations for validation (\emph{Supplementary Information} S3).

\subsubsection{The Damage States}
\begin{figure}[tbp]
    \centering
    \includegraphics[width=\linewidth]{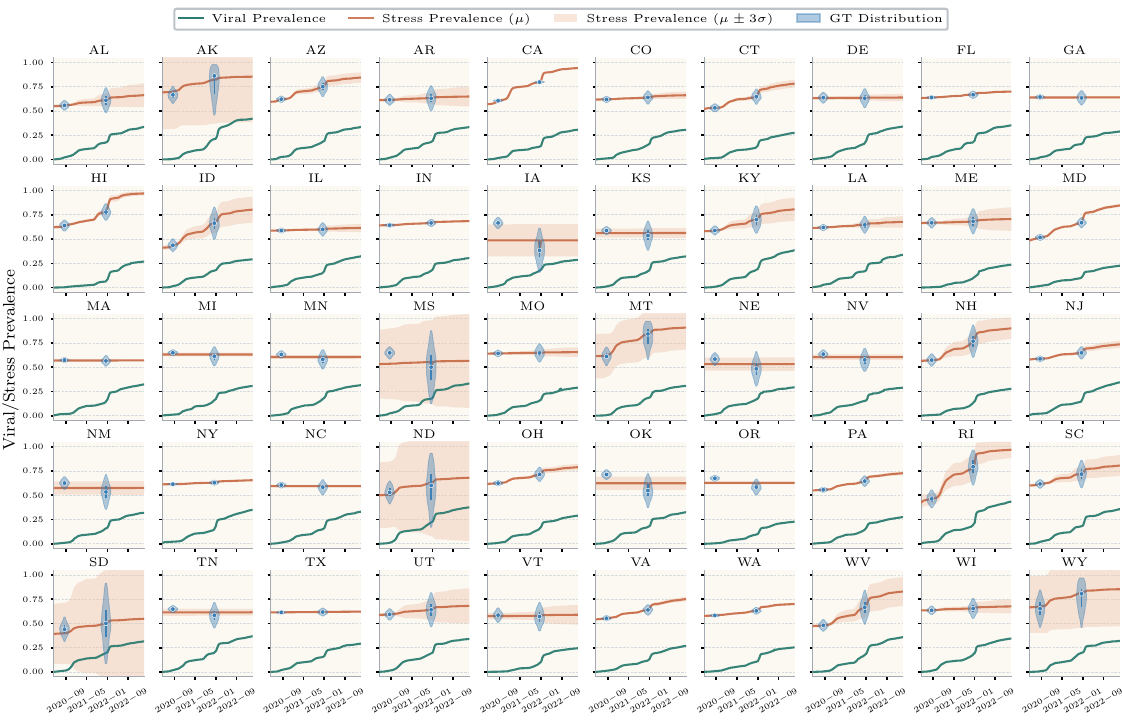}
    \caption{\textbf{Viral prevalence and stress prevalence for each state of the U.S. from April 2020 to March 2023.} In the viral prevalence, two abrupt increases are observed around December 2020 \cite{worldweekly} and January 2022 \cite{cele2022omicron}. Stress prevalence, except for the two observed months, is inferred using the mean-field model based on the posterior of $(\alpha, c)$. The green line denotes viral prevalence from the Center for Systems Science and Engineering at JHU \cite{dong2020interactive}. Violin plots show the distribution of stress prevalence across U.S. states in August 2020 (blue, left) and December 2021 (orange, right), where uncertainty arises from BERTweet classification.}
    \label{fig:data_description}
\end{figure}

We obtain the damage states (i.e., individuals' infectious conditions during COVID-19) from open sources \cite{dong2020interactive}, where the prevalence rate is used as the average damage rate $p$ (\cref{fig:data_description}). Since the Bayesian model inference relies on the mean-field approximation, the detailed health conditions of each individual are not required. For validating the mean-field approximation and synthetic control experiments, we use the Python package \textit{Covasim} \citep{kerr2021covasim} as an agent-based simulator to generate detailed pandemic dynamics. Covasim is based on the SEIR model \cite{aron1984seasonality,he2020seir}, where each agent is classified into one of the following states: susceptible, exposed (infected but not yet infectious), infectious, recovered, or dead. Infectious agents are further categorized based on symptom severity: asymptomatic, presymptomatic, mild, severe, or critical. By running Covasim, we obtain the health conditions of each individual, which are then used to construct the damage state vector $\vect{x}$. We assign a state value of $0$ to individuals who are susceptible, exposed, or recovered, and a value of $1$ to those who are infectious or deceased.

\subsubsection{Sentiment Analysis via Social Media Big Data Analytics}
\label{sec:TweetAnalysis}

To calibrate the parameters of the proposed model, we estimate the proportion of emotionally affected individuals ($m$) using social media data. We frame this as a text classification task where tweets are categorized as \textsc{Unaffected}, \textsc{Affected}, or \textsc{Other}.
We collected a dataset of tweets related to COVID-19 from Twitter/X using hashtags such as ``\#covid'' and ``\#omicron.'' After filtering for U.S.-based users and preprocessing, we constructed a supervised dataset of $2,460$ manually labeled tweets to train and validate our classifiers.

To determine the most effective classification strategy, we evaluated two primary approaches: (i) prompt engineering using large language models (LLMs) in zero-shot, few-shot, and chain-of-thought (CoT) settings; and (ii) fine-tuning pre-trained Transformer encoder-only models. Our experiments revealed that prompt engineering approaches yielded accuracies around 62\%, whereas fine-tuning a BERTweet model \citep{nguyen2020bertweet}—an encoder-only model trained on English tweets—achieved a significantly higher accuracy of 76\% (see \cref{tab:TextClassifier}). Consequently, we employed the fine-tuned BERTweet model for the final classification of the full dataset. The classification uncertainty is explicitly accounted for in the Bayesian observation model (\cref{eq:Bayesian_inference}). Implementation details regarding the data collection, preprocessing pipeline, model architecture, and experimental settings are provided in \emph{Supplementary Information} S5. Additional reliability evidence, including learning curves and confusion matrices for all classifier variants, is presented in S6.

\begin{table}[h]
    \centering
    \caption{\textbf{Performance scores of text classifier on validation set.}}
    \label{tab:TextClassifier}
    \begin{tabular}{l c c}
        \toprule
        \textbf{Model} & \textbf{Accuracy} & \textbf{Weighted F\textsubscript{1}} \\
        \midrule
        \textbf{Fine-tuned BERTweet}& $\mathbf{76.00}$\textbf{\%} & $\mathbf{0.761}$ \\
        Zero-shot (gpt-5.2) & $62.33\%$ & $0.629$ \\
        Few-shot (gpt-5.2) & $62.45\%$ & $0.628$ \\
        CoT (gpt-5.2) & $64.33\%$ & $0.652$ \\
        \bottomrule
    \end{tabular}
\end{table}

\newpage
\bibliographystyle{unsrtnat}
\bibliography{references} 

\newpage
\appendix
\setcounter{page}{1}
\renewcommand{\thepage}{S\arabic{page}}
\setcounter{figure}{0}
\setcounter{table}{0}
\setcounter{equation}{0}
\setcounter{algorithm}{0}
\setcounter{section}{0}
\setcounter{subsection}{0}
\setcounter{subsubsection}{0}
\renewcommand{\thefigure}{S\arabic{figure}}
\renewcommand{\thetable}{S\arabic{table}}
\renewcommand{\theequation}{S\arabic{equation}}
\renewcommand{\thealgorithm}{S\arabic{algorithm}}
\renewcommand{\theHalgorithm}{supp.\arabic{algorithm}}
\renewcommand{\thesection}{S\arabic{section}}
\renewcommand{\thesubsection}{S\arabic{section}.\arabic{subsection}}
\renewcommand{\thesubsubsection}{S\arabic{section}.\arabic{subsection}.\arabic{subsubsection}}
\renewcommand{\theHsection}{supp.\arabic{section}}
\renewcommand{\theHsubsection}{supp.\arabic{section}.\arabic{subsection}}
\renewcommand{\theHsubsubsection}{supp.\arabic{section}.\arabic{subsection}.\arabic{subsubsection}}
\startsupplementcontents

\section*{Supplementary Information}
\pdfbookmark[1]{Supplementary Information}{supplementary_information}
\begin{center}
{\large\bfseries Supplementary Information for}\\[3pt]
{\large\bfseries Social Amplification Dominates Collective Hazard Response}\\[2pt]
Xiaolei Chu$^{1}$, Guanren Zhou$^{1}$, Marco Broccardo$^{2}$, Didier Sornette$^{3}$, Khalid M. Mosalam$^{1}$, Ziqi Wang$^{1}$\\[4pt]
$^{1}$Department of Civil and Environmental Engineering, University of California, Berkeley, United States\\
$^{2}$Department of Civil, Environmental and Mechanical Engineering, University of Trento, Italy\\
$^{3}$Institute of Risk Analysis, Prediction and Management, Southern University of Science and Technology, China\\[4pt]
\texttt{Correspondence: dsornette@ethz.ch; ziqiwang@berkeley.edu}
\end{center}

\begin{center}
{\Large\bfseries\scshape Contents}
\end{center}
\noindent\rule{\textwidth}{0.4pt}
\vspace{4pt}
\begingroup
\setcounter{tocdepth}{1}
\setlength{\parindent}{0pt}
\setlength{\parskip}{2pt}
\noindent{\bfseries\scshape Sections}\par
\vspace{2pt}
\supplementsectionlist
\medskip

\noindent{\bfseries\scshape Figures}\par
\vspace{2pt}
\supplementfigurelist
\medskip

\noindent{\bfseries\scshape Tables}\par
\vspace{2pt}
\supplementtablelist
\endgroup
\vspace{2pt}
\noindent\rule{\textwidth}{0.4pt}
\bigskip

\newpage

\section{Extended Model Definition}
\label{sec:supp_extended_model}

This section provides a detailed description of the hazard-emotion interaction model used in the main text. The goal is to make the mathematical object, inference target, and interpretation boundary explicit before presenting detailed derivations and theoretical properties.

\subsection{Nomenclature}

To reduce notation switching between the main text and Supplementary Information (SI), core symbols are summarized in \cref{tab:supp_nomenclature_model,tab:supp_nomenclature_inference}.

\begin{table}[h]
\centering
\scriptsize
\caption[Nomenclature I: states, networks, and macroscopic observables]{\textbf{Nomenclature I: states, networks, and macroscopic observables.}}
\label{tab:supp_nomenclature_model}
\begin{tabular}{@{}p{2.1cm}p{3.1cm}p{7.2cm}p{2.6cm}@{}}
\toprule
\textbf{Symbol} & \textbf{Domain / Range} & \textbf{Meaning} & \textbf{First appearance} \\
\midrule
$N_x,N_y$ & $\mathbb{N}^+$ & Numbers of hazard-related units and emotional agents. & Main text, Sec. 2.1 \\
\midrule
$i,j$ & Index sets & Node indices in emotional and hazard networks. & Main text, \cref*{eq:local_H} \\
\midrule
$x_j$ & $\{0,1,\dots,L\}$ & Damage state of hazard node $j$ (binary in COVID case). & Main text, \cref*{eq:local_H} \\
\midrule
$y_i$ & $\{0,1,\dots,L\}$ & Emotional arousal state of individual $i$. & Main text, \cref*{eq:local_H} \\
\midrule
$\mathbf{X},\mathbf{Y}$ & $\{0,\dots,L\}^{N_x}$, $\{0,\dots,L\}^{N_y}$ & Vector forms of damage and emotion (also written as $\vect x,\vect y$ in the main text). & Main text, Sec. 2.1 \\
\midrule
$L$ & $\mathbb{N}^+$ & Maximum discrete state level for damage and emotion. & Main text, Sec. 2.1 \\
\midrule
$\mathds{1}(\cdot)$ & $\{0,1\}$ indicator & Indicator used in the prevalence definition. & Main text, \cref*{trhrbgr} \\
\midrule
$m(\mathbf{Y}),\langle m\rangle$ & $[0,1]$ & Stress prevalence and its equilibrium/mean value. & Main text, \cref*{trhrbgr,eq:self_cons} \\
\midrule
$\mathbf{A}_{\mathbf{yy}}$ & $\{0,1\}^{N_y\times N_y}$ & Directed social interaction adjacency matrix. & Main text, \cref*{eq:local_H} \\
\midrule
$\mathbf{A}_{\mathbf{xy}}$ & $\{0,1\}^{N_x\times N_y}$ & Hazard-to-human interaction (biadjacency) matrix. & Main text, \cref*{eq:local_H} \\
\midrule
$\mathbf{A}_{\mathbf{yy}}(j,i)$, $\mathbf{A}_{\mathbf{xy}}(j,i)$ & $\{0,1\}$ & Edge indicators (whether node $j$ influences/exposes individual $i$). & SI, Sec. S1.2 \\
\midrule
$k_i^{(\mathbf{yy})},k_i^{(\mathbf{xy})}$ & $\mathbb{N}_{\ge0}$ & Social and hazard-exposure in-degrees for individual $i$. & SI, Sec. S1.2 \\
\midrule
$\langle k\rangle_{\mathbf{A}_{\mathbf{yy}}}$, $\langle k_{\mathbf y}\rangle_{\mathbf{A}_{\mathbf{xy}}}$ & $\mathbb{R}_{\ge0}$ & Population-average in-degrees of social and hazard-exposure networks. & Main text, \cref*{eq:A} \\
\midrule
$\langle x\rangle_{\mathbf{A}_{\mathbf{xy}}},p$ & $\mathbb{R}_{\ge0}$, $[0,1]$ (binary setting) & Average accessible hazard intensity and damage rate $p=\langle x\rangle_{\mathbf{A}_{\mathbf{xy}}}/\langle k_{\mathbf y}\rangle_{\mathbf{A}_{\mathbf{xy}}}$. & Main text, \cref*{eq:B} \\
\bottomrule
\end{tabular}
\end{table}

\begin{table}[h]
\centering
\scriptsize
\caption[Nomenclature II: energetics, inference, and regime diagnostics]{\textbf{Nomenclature II: energetics, inference, and regime diagnostics.}}
\label{tab:supp_nomenclature_inference}
\begin{tabular}{@{}p{2.1cm}p{3.1cm}p{7.2cm}p{2.6cm}@{}}
\toprule
\textbf{Symbol} & \textbf{Domain / Range} & \textbf{Meaning} & \textbf{First appearance} \\
\midrule
$\alpha,c$ & $\alpha\in[0,1],\ c\in\mathbb{R}$ & Social-vs-hazard weighting and emotional asymmetry (negativity/positivity bias). & Main text, \cref*{eq:local_H} \\
\midrule
$\mathcal{H}_i,\mathcal{H}$ & $\mathbb{R}$ & Local and global Hamiltonians. & Main text, \cref*{eq:local_H,eq:global_H} \\
\midrule
$p_{\mathbf Y\mid\mathbf X}$ & Probability density/mass & Conditional model of emotions given hazard states. & Main text, \cref*{eq:global_H} \\
\midrule
$Z,\ Z(\mathbf{x}),\ Z_{\mathrm{MF}}$ & $\mathbb{R}_{>0}$ & Partition functions for exact/effective and mean-field forms. & Main text, \cref*{eq:global_H,eq:partition_mft}; SI, \cref{eq:supp_Zmf} \\
\midrule
$T,\beta$ & $T>0,\ \beta=1/T$ & Effective social-noise temperature and inverse temperature. & Main text, Alg. 1; \cref*{eq:global_H} \\
\midrule
$\Omega(m)$ & $\mathbb{R}_{>0}$ & Phase-space multiplicity at fixed prevalence $m$. & Main text, \cref*{eq:phase_space} \\
\midrule
$A,B$ & $\mathbb{R}$ and $\mathbb{R}_{\ge0}$ (physical domain) & Composite coefficients for social amplification and hazard forcing in free energy. & Main text, \cref*{eq:AB} \\
\midrule
$f(m)$ & $\mathbb{R}$ & Mean-field free-energy density. & Main text, \cref*{eq:free_eng} \\
\midrule
$G$ & $\mathbb{R}$ & Limit-state criterion for minority- vs majority-arousal regimes. & Main text, \cref*{eq:limit_function} \\
\midrule
$\theta$ & $(\alpha,c)$ & Community-level parameter vector in Bayesian inference. & SI, Sec. \ref{sec:supp_posterior_T} \\
\midrule
$\Omega_T$ & $[10,30]$ in this study & Prior support for temperature marginalization. & Main text, \cref*{eq:posterior_Bayesian} \\
\midrule
$N_0,N_1,N_2$ & $\mathbb{N}_{\ge0}$ & Observed classifier counts (unaffected, affected, irrelevant). & Main text, \cref*{eq:Bayesian_inference} \\
\midrule
$n_0,n_1,n_2$ & $\mathbb{N}_{\ge0}$ & Latent true class counts used in likelihood marginalization. & Main text, \cref*{eq:Bayesian_inference} \\
\midrule
$m^{(i)}$ & $[0,1]$ & Observed stress prevalence in window $i$. & Main text, \cref*{eq:posterior_Bayesian} \\
\midrule
$q$ & $[0,1]$ & Average sentiment-classifier misclassification rate in the observation model. & Main text, Bayesian observation model around \cref*{eq:Bayesian_inference} \\
\bottomrule
\end{tabular}
\end{table}

\subsection{State Variables and Sample Spaces}

Let $N_x$ denote the number of hazard-related units (e.g., individuals, facilities, or local exposure proxies) and $N_y$ denote the number of individuals whose emotional states are modeled. The model uses two discrete state vectors:
\begin{equation}
\mathbf X = (x_1,\dots,x_{N_x}) \in \{0,1,\dots,L\}^{N_x}, \qquad
\mathbf Y = (y_1,\dots,y_{N_y}) \in \{0,1,\dots,L\}^{N_y},
\end{equation}
where $x_j$ is hazard damage intensity and $y_i$ is emotional arousal intensity. In the COVID-19 case study, the binary damage setting $x_j\in\{0,1\}$ is adopted in practice, with $x_j=1$ indicating infectious/deceased status and $x_j=0$ otherwise (main text, \emph{Methods and Materials}).

At the macroscopic level, we focus on stress prevalence
\begin{equation}
m(\mathbf Y) \defi \frac{1}{N_y}\sum_{i=1}^{N_y}\mathds 1(y_i>0),
\end{equation}
namely the fraction of emotionally aroused individuals.

\subsection{Network Objects and Exposure Operators}

Two directed networks define the interaction structure:
\begin{equation}
\mathbf A_{\mathbf{yy}}\in\{0,1\}^{N_y\times N_y}, \qquad
\mathbf A_{\mathbf{xy}}\in\{0,1\}^{N_x\times N_y}.
\end{equation}
Here, $\mathbf A_{\mathbf{yy}}(j,i)=1$ means individual $j$ emotionally influences individual $i$, and $\mathbf A_{\mathbf{xy}}(j,i)=1$ means hazard node $j$ affects individual $i$.

Useful degree-like quantities are
\begin{equation}
k_i^{(\mathbf{yy})}=\sum_{j=1}^{N_y}\mathbf A_{\mathbf{yy}}(j,i),\qquad
k_i^{(\mathbf{xy})}=\sum_{j=1}^{N_x}\mathbf A_{\mathbf{xy}}(j,i),
\end{equation}
with population averages $\langle k\rangle_{\mathbf A_{\mathbf{yy}}}$ and $\langle k_{\mathbf y}\rangle_{\mathbf A_{\mathbf{xy}}}$ as defined in the main text.

The average hazard intensity per exposure is
\begin{equation}
p\defi \frac{\langle x\rangle_{\mathbf A_{\mathbf{xy}}}}{\langle k_{\mathbf y}\rangle_{\mathbf A_{\mathbf{xy}}}},
\end{equation}
which coincides with viral prevalence in the binary COVID-19 setup.

\subsection{Microscopic Energy and Update Dynamics}

Conditioned on $\mathbf X=\mathbf x$, the local Hamiltonian for individual $i$ is
\begin{equation}
\mathcal H_i =
\alpha\sum_{\mathbf A_{\mathbf{yy}}(j,i)=1}\left[(y_i-y_j)^2-cy_i y_j\right]
+(1-\alpha)\sum_{\mathbf A_{\mathbf{xy}}(j,i)=1}(x_j-y_i)^2,
\end{equation}
where $\alpha\in[0,1]$ weights social versus hazard forcing and $c\in\mathbb R$ controls emotional asymmetry ($c>0$: negativity bias; $c<0$: positivity bias / attenuation). This individual Hamiltonian captures the tension between social conformity and hazard response at the microscopic level.

Simulation uses asynchronous single-site Metropolis updates (Algorithm 1 in the main text):
\begin{equation}
\mathbb P(\mathbf y\to \mathbf y') = \min\{1,\exp(-\beta\Delta\mathcal H_i)\}, \qquad \beta=1/T.
\end{equation}
The temperature $T$ captures the variability of individual emotional responses.

\subsection{Quasi-Stationary Approximation and Scope}

The analytically tractable model uses a Boltzmann form
\begin{equation}
p_{\mathbf Y\mid \mathbf X}(\mathbf y\mid\mathbf x)\propto \exp\!\left[-\beta\mathcal H(\mathbf y\mid\mathbf x)\right],
\end{equation}
with global Hamiltonian $\mathcal H$ defined in \cref*{eq:global_H}. The Hammersley-Clifford theorem\citeS{si_bremaud2013markov} directly supports this Boltzmann/Markov-random-field form for reciprocal social couplings (symmetric $\mathbf A_{\mathbf{yy}}$). For asymmetric or mixed directed couplings, we use a homogenized effective-equilibrium approximation that captures equilibrium-like marginals under a quasi-stationary condition (fast emotional mixing relative to the observation window), as discussed in \cref{subsec:supp_directed_couplings}.

Hence, the model is designed for \emph{window-level inference and regime diagnosis}, rather than exact real-time non-equilibrium trajectory prediction.

\subsection{Assumptions and Interpretation Boundaries}

The key assumptions to construct this model and their implications are summarized below in \cref{tab:supp_assumptions}.
\begin{table}[h]
\centering
\caption[Core assumptions and interpretation boundaries of the hazard-emotion model]{\textbf{Core assumptions and interpretation boundaries of the hazard-emotion model.}}
\label{tab:supp_assumptions}
\begin{tabular}{p{3.2cm} p{5.7cm} p{5.5cm}}
\toprule
\textbf{Item} & \textbf{Assumption in this study} & \textbf{Implication / boundary} \\
\midrule
Time scale & Emotional interactions mix within each analysis window. & Supports quasi-stationary Boltzmann approximation; fast non-equilibrium transients are not fully resolved\citeS{bertini2015macroscopic,jarzynski1997nonequilibrium,crooks1999entropy,kamenev2023field}. \\
\midrule
Parameter homogeneity & $(\alpha,c)$ are community-level effective parameters per state. & Captures dominant aggregate behavior; within-community heterogeneity is not significant in the sense of mean field\citeS{si_landau2013statistical}. \\
\midrule
Network reduction in MFT & Only mean in-degrees enter mean-field formulas. & Enables closed-form analysis; network topologies only matter in the agent-level setting, which are proved to be not important for the macroscopic stress prevalence in \cref{sec:supp_mf_validity}. \\
\midrule
Macroscopic observable & Stress prevalence $m$ uses binary arousal indicator $y_i>0$. & Targets prevalence of emotional activation, not full intensity distribution. \\
\midrule
Observation model & Sentiment labels contain classification error. & Epistemic uncertainty is propagated through latent-count likelihood (main text, \cref*{eq:Bayesian_inference}). \\
\bottomrule
\end{tabular}
\end{table}

\newpage
\section{Theoretical Properties}
\label{sec:supp_theoretical_properties}

This section summarizes analytical properties implied by the free-energy density and the limit-state function. These properties formalize when the model predicts proportional hazard tracking versus social amplification and multistability.

\subsection{Curvature and Convexity Structure}

\begin{figure}[h]
    \centering
    \includegraphics[width=0.98\textwidth]{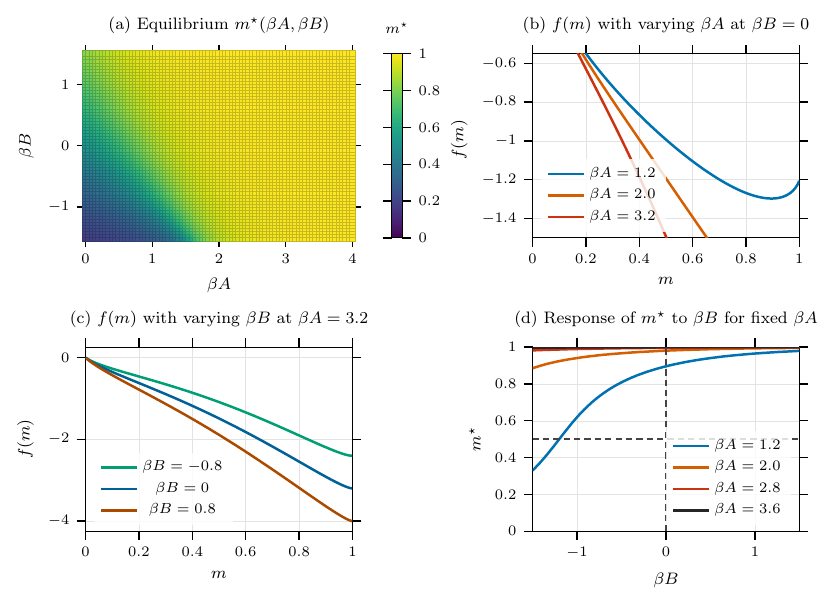}
    \caption[Free-energy geometry linking $\beta A$, $\beta B$, $m$, and $f(m)$]{\textbf{Free-energy geometry linking $\beta A$, $\beta B$, $m$, and $f(m)$.} (a) Equilibrium prevalence $m^{\star}$ over $(\beta A,\beta B)$. (b) Free-energy curves $f(m)$ for varying $\beta A$ at $\beta B=0$, showing the convex-to-nonconvex transition around $\beta A=2$. (c) Free-energy tilt induced by $\beta B$ at fixed $\beta A=3.2$. (d) Equilibrium response $m^{\star}(\beta B)$ for multiple fixed $\beta A$, illustrating stronger nonlinear susceptibility in the $\beta A>2$ regime.}
    \label{fig:supp_free_energy_landscape}
\end{figure}

From \cref*{eq:free_eng},
\begin{equation}
\label{eq:supp_second_derivative}
f''(m)= -2\beta A+\frac{1}{m(1-m)}.
\end{equation}
Since $\frac{1}{m(1-m)}\ge 4$ for $m\in(0,1)$, we have:
\begin{itemize}
    \item If $\beta A\le 2$, then $f''(m)\ge 0$ for all $m\in(0,1)$, so $f(m)$ is strictly convex and admits a unique equilibrium prevalence.
    \item If $\beta A>2$, non-convex regions can appear, enabling multiple local minima (metastability / hysteresis-like behavior) depending on $B$.
\end{itemize}
Thus, $\beta A=2$ is the curvature threshold for onset of amplification-driven nonlinearity in the mean-field landscape, quantitatively shown in \cref{fig:supp_free_energy_landscape}.

\subsection{Regime Interpretation of $A$ and $B$}

The two composite parameters have distinct structural roles:
\begin{equation}
A=-(1-\alpha)\langle k_{\mathbf y}\rangle_{\mathbf A_{\mathbf{xy}}}
+\frac{1}{2}\alpha c\langle k\rangle_{\mathbf A_{\mathbf{yy}}},
\qquad
B=2(1-\alpha)p\langle k_{\mathbf y}\rangle_{\mathbf A_{\mathbf{xy}}}\ge 0.
\end{equation}
\begin{itemize}
    \item $A$ controls endogenous social amplification strength.
    \item $B$ is exogenous hazard forcing in the feasible physical domain.
\end{itemize}
For $c>0$, increasing social connectivity $\langle k\rangle_{\mathbf A_{\mathbf{yy}}}$ increases $A$ and can push the system toward amplification. For $c\le 0$, the same connectivity increase tends to decrease or not increase amplification pressure.
These distinct roles are visualized in panel (a) of \cref{fig:supp_limit_state_boundary}.

\subsection{Explicit Phase Boundary from the Limit-State Function}

\begin{figure}[h]
    \centering
    \includegraphics[width=0.98\textwidth]{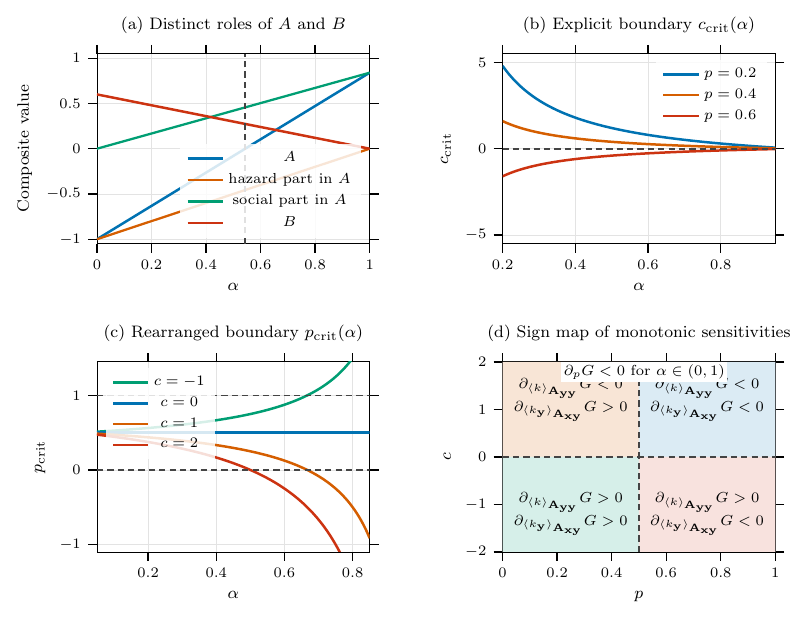}
    \caption[Regime interpretation, explicit phase boundaries, and monotonic sensitivities of the limit-state function]{\textbf{Regime interpretation, explicit phase boundaries, and monotonic sensitivities of the limit-state function.} (a) Decomposition of $A$ and $B$ under varying $\alpha$ with fixed $\langle k_{\mathbf y}\rangle_{\mathbf A_{\mathbf{xy}}}=1.0$, $\langle k\rangle_{\mathbf A_{\mathbf{yy}}}=1.2$, $c=1.4$, and $p=0.3$, showing that $A$ combines endogenous social amplification and hazard-alignment terms while $B$ acts as exogenous forcing. (b) Explicit boundary family $c_{\mathrm{crit}}(\alpha)$ from \cref{eq:supp_c_critical} for $p\in\{0.2,0.4,0.6\}$ with fixed ratio $\langle k_{\mathbf y}\rangle_{\mathbf A_{\mathbf{xy}}}/\langle k\rangle_{\mathbf A_{\mathbf{yy}}}=1$. (c) Rearranged threshold family $p_{\mathrm{crit}}(\alpha)$ from \cref{eq:supp_p_critical} for $c\in\{-1,0,1,2\}$ with fixed ratio $\langle k\rangle_{\mathbf A_{\mathbf{yy}}}/\langle k_{\mathbf y}\rangle_{\mathbf A_{\mathbf{xy}}}=1$; dashed lines indicate the feasible range $p\in[0,1]$. (d) Sign map of monotonic sensitivities in \cref*{eq:limit_function} over $(p,c)\in[0,1]\times[-2,2]$ for $\alpha\in(0,1)$, with sign changes at $p=1/2$ and $c=0$.}
    \label{fig:supp_limit_state_boundary}
\end{figure}

Setting \cref*{eq:limit_function} to zero yields an explicit boundary:
\begin{equation}
\label{eq:supp_c_critical}
c_{\mathrm{crit}}
=2\frac{\langle k_{\mathbf y}\rangle_{\mathbf A_{\mathbf{xy}}}}
{\langle k\rangle_{\mathbf A_{\mathbf{yy}}}}
(1-2p)\left(\frac{1}{\alpha}-1\right).
\end{equation}
Equivalent rearrangement gives the damage-rate threshold:
\begin{equation}
\label{eq:supp_p_critical}
p_{\mathrm{crit}}
=\frac{1}{2}
-\frac{1}{4}\frac{\alpha}{1-\alpha}
\frac{\langle k\rangle_{\mathbf A_{\mathbf{yy}}}}
{\langle k_{\mathbf y}\rangle_{\mathbf A_{\mathbf{xy}}}}\,c,
\end{equation}
which is valid when $\alpha\in(0,1)$. In this form, larger positivity attenuation ($c<0$) increases the damage needed to cross into majority-arousal regimes. Panels (b) and (c) of \cref{fig:supp_limit_state_boundary} visualize these boundary families.

\subsection{Monotonic Sensitivities of the Limit-State Function}

For short-horizon control where $(\alpha,c)$ are fixed intrinsic parameters, gradients of \cref*{eq:limit_function} are:
\begin{equation}
\frac{\partial G}{\partial p}
=-2\langle k_{\mathbf y}\rangle_{\mathbf A_{\mathbf{xy}}}\left(\frac{1}{\alpha}-1\right)<0,\quad
\frac{\partial G}{\partial \langle k\rangle_{\mathbf A_{\mathbf{yy}}}}
=-\frac{c}{2},\quad
\frac{\partial G}{\partial \langle k_{\mathbf y}\rangle_{\mathbf A_{\mathbf{xy}}}}
=(1-2p)\left(\frac{1}{\alpha}-1\right).
\end{equation}
Therefore:
\begin{itemize}
    \item Reducing $p$ always increases $G$ (mitigates majority-arousal risk) for $\alpha\in(0,1)$.
    \item Reducing social degree is beneficial when $c>0$, but can be counterproductive when $c<0$.
    \item The effect of changing $\langle k_{\mathbf y}\rangle_{\mathbf A_{\mathbf{xy}}}$ depends on whether $p$ is below or above $1/2$.
\end{itemize}
The corresponding sign structure is summarized in panel (d) of \cref{fig:supp_limit_state_boundary}.

\subsection{Singular Limits and Practical Diagnostics}

Two limits are useful for interpretation:
\begin{itemize}
    \item $\alpha\to 1$: the hazard-alignment contribution in the Hamiltonian is effectively suppressed, so emotional dynamics are governed predominantly by social contagion and the asymmetry parameter $c$. In this limit, individuals are perturbed mainly by others' emotional expressions rather than by direct hazard-alignment effects.
    \item $\alpha\to 0$: the social-interaction contribution in the Hamiltonian is effectively suppressed, so emotional states are governed primarily by hazard exposure and tend to align more directly with the underlying damage field.
\end{itemize}
Hence, in the singular limit $\alpha=0$, diagnostics should use the unscaled derivative $\partial f/\partial m\vert_{m=1/2}$ directly because the compact form of $G$ contains the factor $(1/\alpha-1)$ and is therefore undefined at $\alpha=0$. For $\alpha\in(0,1)$, $G$ still provides a compact closed-form boundary with identical sign information.

\subsection{Directed Couplings and Equilibrium Approximation}
\label{subsec:supp_directed_couplings}

To align with Hammersley-Clifford theorem \citeS{si_bremaud2013markov}, it is useful to distinguish three network structures:
\begin{itemize}
    \item \textbf{Reciprocal (symmetric) case}: for all $i\neq j$, $\mathbf A_{\mathbf{yy}}(j,i)=\mathbf A_{\mathbf{yy}}(i,j)$. This corresponds to an undirected pairwise Markov random field, where the Boltzmann factorization is theoretically consistent.
    \item \textbf{Fully asymmetric case}: for all $i\neq j$, mutual links do not coexist, i.e., $\mathbf A_{\mathbf{yy}}(j,i)\mathbf A_{\mathbf{yy}}(i,j)=0$. This is a pure leader-follower directed regime; strict detailed balance is generally violated, so the Boltzmann form is used as an effective approximation for window-level inference.
    \item \textbf{Mixed real-world case}: both reciprocal and one-way links coexist. Typical examples are celebrity--follower interactions (many one-way links) together with friend/family ties (often bidirectional links). This intermediate regime is exactly the setting where the effective-equilibrium approximation is most practically relevant.
\end{itemize}
Hence, when directed asymmetry is present, phase-boundary diagnostics remain informative for regime classification, while fine-grained transient path properties should be evaluated with the agent-based Gibbs simulation. In \cref{sec:supp_mf_validity}, we show that the mean-field phase diagram is robust to network structure, supporting the practical utility of the mean-field solution with the mixed real-world case.

\newpage
\section{Mean-field Validity}
\label{sec:supp_mf_validity}
\label{sec:validation_MFT}
We vary different parameters and compare the phase diagrams obtained from mean-field theory (MFT) with those from Monte Carlo simulations (MCS) to assess the accuracy of MFT. Here, MCS denotes the agent-level Gibbs simulation in \Cref*{alg:Gibbs}, i.e., asynchronous single-site Metropolis updates on the network. The MCS is performed with $4000$ nodes, while the MFT solution is obtained by finding the minimum of the free energy density, as given in \cref*{eq:free_eng}.

\begin{figure}[h]
    \centering
    \makebox[\textwidth][c]{\includegraphics[width=0.99\textwidth]{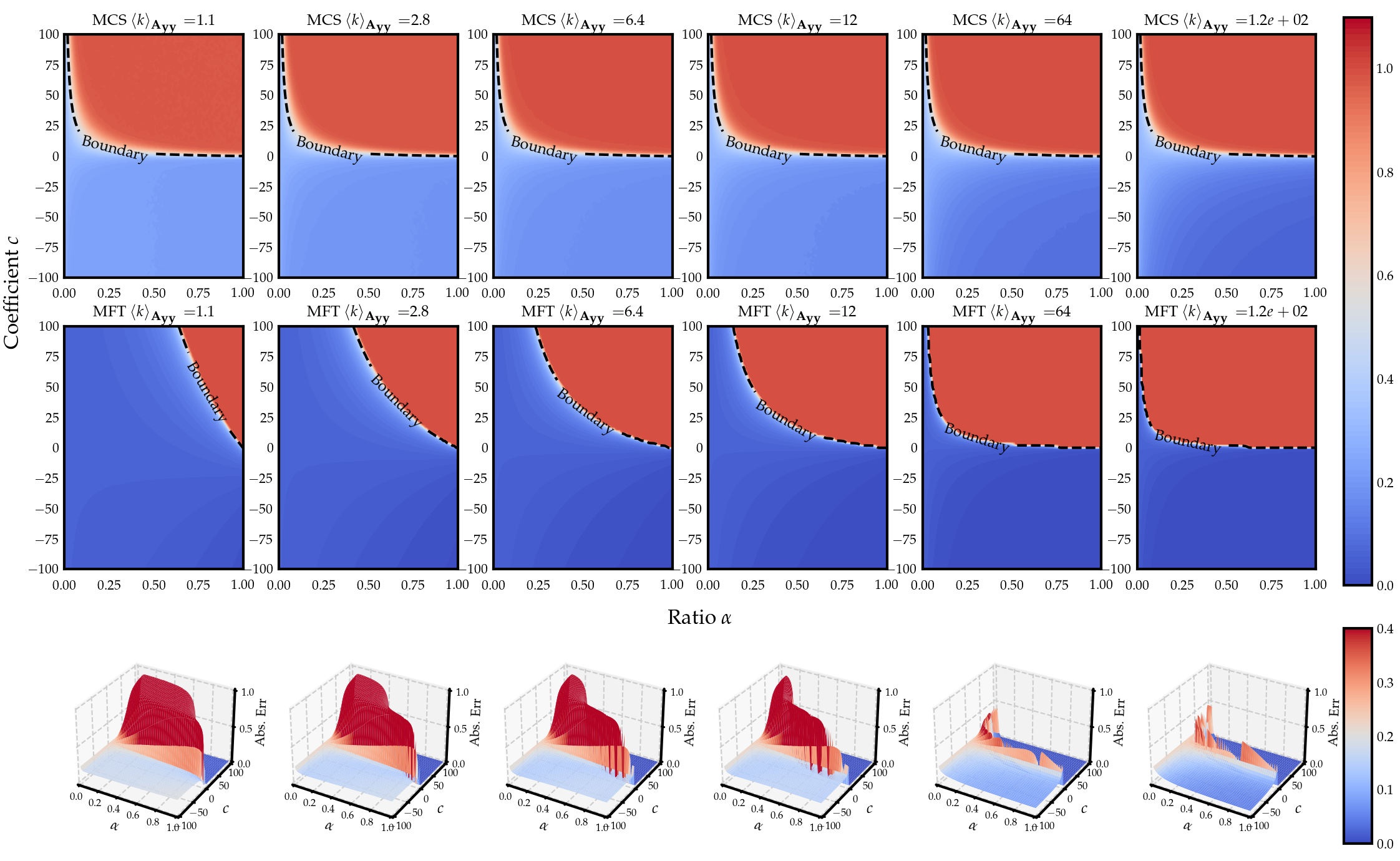}}
\caption[Phase diagrams as a function of $\alpha$ and $c$ with varying average degree of the social networks]{\textbf{Phase diagrams as a function of $\alpha$ and $c$ with varying average degree of the social networks.} The temperature is set as $k_B T=1$. In MCS, the hazard damage-human interaction network is randomly generated with $4000$ nodes, constrained by an average in-degree of $\langle k_{\mathbf y}\rangle_{\mathbf{A}_{\mathbf{xy}}} = 220$. The state variable $\mathbf{X}$, representing physical damage, is randomly assigned values of 0 or 1, ensuring $\langle x\rangle_{\mathbf{A}_{\mathbf{xy}}} = 12$. The social interaction network, also consisting of $4000$ nodes, is randomly generated with an average in-degree $\langle k\rangle_{\mathbf{A}_{\mathbf{yy}}}$ varying from $1$ to $120$. In MFT, the same values of $\langle k_{\mathbf y}\rangle_{\mathbf{A}_{\mathbf{xy}}}$, $\langle k\rangle_{\mathbf{A}_{\mathbf{yy}}}$, and $\langle x\rangle_{\mathbf{A}_{\mathbf{xy}}}$ as those used in MCS are employed.}
    \label{fig:phase_diagram_k_yy}
\end{figure}

We vary the average in-degree of the social interaction network to assess the accuracy of MFT. \cref{fig:phase_diagram_k_yy} illustrates phase diagrams as functions of $\alpha$ and $c$ for different average in-degrees of the social interaction network. The six subfigures in the first row are generated using MCS, serving as the ground truth, while the subfigures in the second row are produced using MFT. The third row presents the absolute error between the MCS results and the MFT approximations. The dashed lines indicate the phase boundary, where $\langle m \rangle = 0.5$. As shown in \cref{fig:phase_diagram_k_yy}, the MCS results reveal that increasing the average degree of the social network does not significantly alter the shape of the phase boundary but intensifies stress prevalence, as indicated by the deepening color. In contrast, the MFT results exhibit noticeable changes in the phase boundary shape. This discrepancy is expected, as MFT tends to break down in low-dimensional systems \citepS{mussardo2010statistical}, particularly when the average in-degree of the social interaction network is small. However, for high-dimensional cases, we observe that MFT provides generally accurate approximations.

\begin{figure}[h]
    \centering
    \makebox[\textwidth][c]{\includegraphics[width=0.93\textwidth]{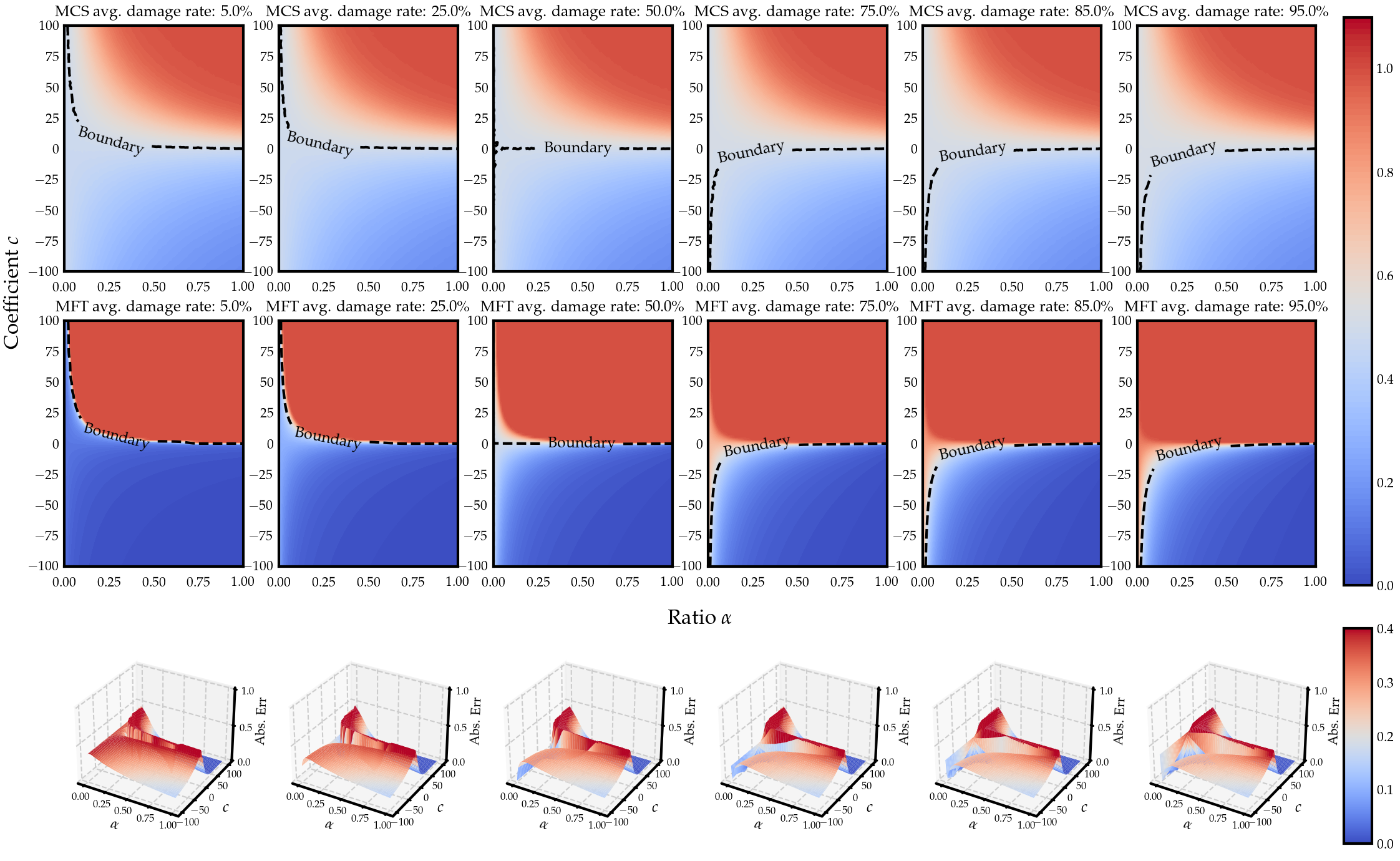}}
    \caption[Phase diagrams over $\alpha$ and $c$ for varying average damage rate $p$]{\textbf{Phase diagrams over $\alpha$ and $c$ for varying average damage rate $p$.} Parameters: $k_B T = 1$, $\langle k\rangle_{\mathbf{A}_{\mathbf{yy}}} = 70$, and $\langle k_{\mathbf y}\rangle_{\mathbf{A}_{\mathbf{xy}}} = 220$ for both MCS and MFT.}
    \label{fig:phase_diagram_beta0}
\end{figure}

\begin{figure}[h]
    \centering
    \makebox[\textwidth][c]{\includegraphics[width=0.93\textwidth]{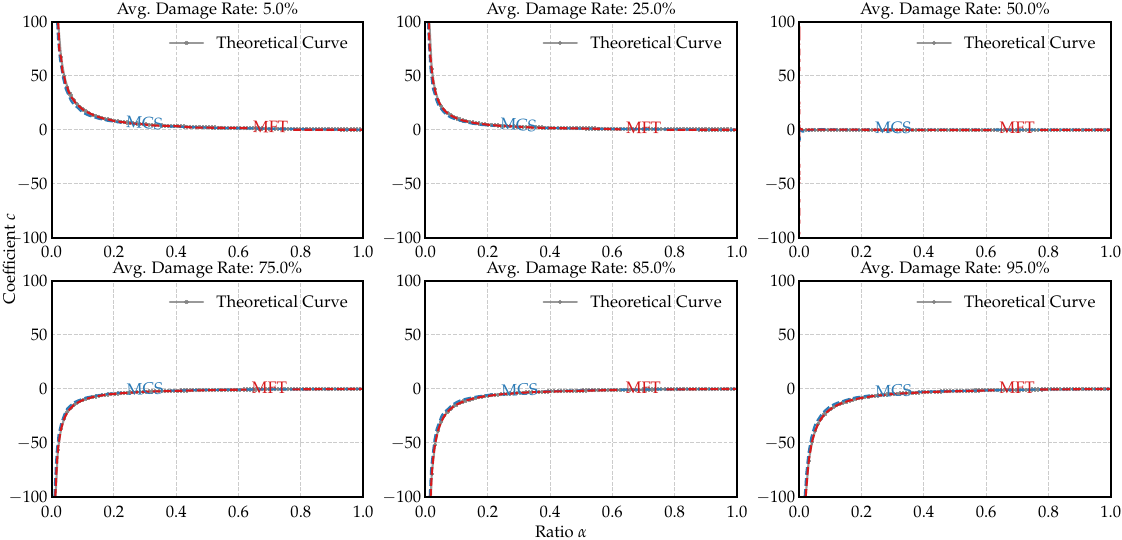}}
    \caption[Comparison of phase-boundary curves from MCS, MFT, and theory]{\textbf{Comparison of phase-boundary curves from MCS, MFT, and theory.} Minor mismatch is due to finite $\alpha$-$c$ grid resolution.}
    \label{fig:criticality_curve}
\end{figure}

\cref{fig:phase_diagram_beta0} shows phase diagrams for varying average damage rates
$p=\frac{\langle x\rangle_{\mathbf{A}_{\mathbf{xy}}}}{\langle k_{\mathbf y}\rangle_{\mathbf{A}_{\mathbf{xy}}}}$.
The boundary shift around $p\approx 0.5$ is consistent with the MFT relation
\begin{equation*}
    c = 2\frac{\langle k_{\mathbf y}\rangle_{\mathbf{A}_{\mathbf{xy}}}}{\langle k\rangle_{\mathbf{A}_{\mathbf{yy}}}} \left(1 - 2p\right)\left(\frac{1}{\alpha}-1\right),
\end{equation*}
from \cref*{eq:limit_function}. \cref{fig:criticality_curve} compares fitted MCS/MFT boundaries with theory.

Overall, the MCS results are in good agreement with the MFT predictions across the tested parameter sweeps and phase-boundary diagnostics. The remaining differences are localized near critical boundaries and are consistent with finite-size and gridding effects. This consistency supports the feasibility of using the mean-field framework as a practical and reliable approximation for regime-level analysis in this study.

\newpage
\section{Identifiability and Inference Diagnostics}
\label{sec:supp_identifiability}
\label{sec:supp_bayes_aies}

\subsection{Posterior formulation and marginalization over temperature}
\label{sec:supp_posterior_T}

Let $\theta \triangleq (\alpha,c)$ denote the community-level parameters assumed to be invariant over the observation period, and let $T$ be a hyperparameter controlling the intrinsic emotional variability (temperature), with $\beta = 1/T$. For each time window $i=1,\dots,I$, the sentiment classifier yields observed counts
\begin{equation}
(N_0^{(i)},N_1^{(i)},N_2^{(i)}), \qquad N^{(i)}=N_0^{(i)}+N_1^{(i)}+N_2^{(i)}.
\end{equation}

The empirical stress prevalence is computed as
\begin{equation}
m^{(i)} = \frac{N_1^{(i)}}{N_0^{(i)}+N_1^{(i)}}.
\end{equation}

In the inference, however, the mean-field (MF) model governs the \emph{latent} emotional prevalence
\begin{equation}
m = \frac{n_1}{n_0+n_1},
\end{equation}
where $(n_0,n_1)$ denote the unknown true numbers of unaffected and affected posts, respectively, and $n_2=N-n_0-n_1$ is the true number of irrelevant posts.

Given the quasi-equilibrium assumption, we treat each time window as independent. The joint likelihood factorizes as
\begin{equation}
f(\{m^{(i)}\}\mid \theta, T) = \prod_{i=1}^I f\!\left(m^{(i)} \mid \theta, T\right).
\end{equation}

We adopt independent uniform priors for the components of $\theta=(\alpha,c)$, with
\[
\alpha \sim \mathcal{U}(0,1),
\qquad
c \sim \mathcal{U}(-10,10),
\]
so that the joint prior density satisfies $f(\theta)\propto \mathcal{U}(0,1)\times\mathcal{U}(-10,10)$ over the admissible parameter domain. In addition, we place a uniform hyperprior on $T$, namely $T \sim \mathcal{U}(10,30)$. This support is chosen to keep the mean-field equilibrium in a non-degenerate regime: for very large $T$ ($\beta\to 0$), the entropic term dominates and the equilibrium tends to $m\approx 0.5$; for very small $T$ ($\beta\to\infty$), entropy becomes negligible and the equilibrium is controlled mainly by the energetic terms $A$ and $B$. Therefore, $[10,30]$ is used as a practical range for emotional variability in this study. Marginalizing over $T$ yields

\begin{align}
f(\theta \mid \{m^{(i)}\})
&\propto \int_{10}^{30} \left[\prod_{i=1}^I f\!\left(m^{(i)} \mid \theta, T\right)\right] f(\theta)\, f(T)\, dT \nonumber\\
&\propto f(\theta)\int_{10}^{30} \left[\prod_{i=1}^I f\!\left(m^{(i)} \mid \theta, T\right)\right]\frac{1}{20}\, dT.
\label{eq:supp_posterior_margT}
\end{align}

The temperature integral in \cref{eq:supp_posterior_margT} is evaluated with the adaptive quadrature routine quad on $[10,30]$. For each proposed parameter value $\theta$ during MCMC sampling, we compute
\begin{equation}
\mathcal{L}(\theta) \triangleq \int_{10}^{30}\left[\prod_{i=1}^{I}f\!\left(m^{(i)}\mid \theta,T\right)\right]\frac{1}{20}\,dT,
\label{eq:supp_quad_integral}
\end{equation}
and evaluate the sampler target as $\log f(\theta)+\log\mathcal{L}(\theta)$.


\subsection{Likelihood construction with aleatory and epistemic uncertainty}
\label{sec:supp_likelihood}

The likelihood $f(m^{(i)}\mid\theta,T)$ combines: (i) aleatory variability governed by the MF model, and (ii) epistemic uncertainty induced by sentiment misclassification. For each time window with total post count $N$, we marginalize over all feasible latent configurations $(n_0,n_1)$ that could have generated the observed counts $(N_0,N_1,N_2)$:
\begin{equation}
f(m^{(i)}\mid\theta,T)
\equiv f(N_0,N_1,N_2\mid\theta,T)
= \sum_{(n_0,n_1)\in\mathcal{D}(N)} 
f(n_0,n_1\mid\theta,T)\;
f(N_0,N_1,N_2\mid n_0,n_1).
\label{eq:supp_total_law}
\end{equation}

The feasible domain is
\begin{equation}
\mathcal{D}(N)=\Big\{(n_0,n_1): n_0\in\{0,\dots,N\},~ n_1\in\{0,\dots,N-n_0\},~ n_0+n_1\ge 1\Big\}.
\end{equation}

The MF-induced distribution on latent configurations is defined by a Boltzmann form with phase-space factor $\Omega(\cdot)$:
\begin{equation}
f(n_0,n_1\mid\theta,T)=\frac{1}{Z_{\mathrm{MF}}(\theta,T)}
\exp\!\left(-\beta\,\mathcal{H}_{\mathrm{MF}}\!\left(\frac{n_1}{n_0+n_1};\theta\right)\right)\;
\Omega\!\left(\frac{n_1}{n_0+n_1}\right),
\label{eq:supp_latent_mf}
\end{equation}
where the partition function is
\begin{equation}
Z_{\mathrm{MF}}(\theta,T)=
\sum_{(n_0,n_1)\in\mathcal{D}(N)}
\exp\!\left(-\beta\,\mathcal{H}_{\mathrm{MF}}\!\left(\frac{n_1}{n_0+n_1};\theta\right)\right)\;
\Omega\!\left(\frac{n_1}{n_0+n_1}\right).
\label{eq:supp_Zmf}
\end{equation}

The classifier observation model is specified via a symmetric confusion structure with average misclassification rate $q$. Let the true class proportions be
\begin{equation}
\pi(n_0,n_1)=\frac{1}{N}\big[\,n_0,\; n_1,\; N-n_0-n_1\,\big],
\end{equation}
and define the induced observed-class probabilities as
\begin{equation}
p_c(n_0,n_1)=\pi(n_0,n_1)\,
\begin{bmatrix}
1-q & q/2 & q/2 \\
q/2 & 1-q & q/2 \\
q/2 & q/2 & 1-q
\end{bmatrix}.
\label{eq:supp_pc}
\end{equation}

Conditional on $p_c(n_0,n_1)$, the observed counts follow a multinomial distribution:
\begin{equation}
f(N_0,N_1,N_2\mid n_0,n_1)
=\mathrm{Multinomial}(N_0,N_1,N_2;\,N,\,p_c(n_0,n_1)).
\label{eq:supp_mult}
\end{equation}

Substituting \cref{eq:supp_latent_mf}--\cref{eq:supp_mult} into \cref{eq:supp_total_law} yields the likelihood expression reported in the main text. In the released code, this finite latent sum is evaluated directly in probability domain for each $(\theta,T)$ query.

\subsection{Affine-invariant ensemble sampler (AIES): algorithmic details}
\label{sec:supp_aies}

To sample from the posterior in \cref{eq:supp_posterior_margT}, we employ the affine-invariant ensemble sampler (AIES)\citeS{si_goodman2010ensemble}, which is particularly effective when the posterior exhibits strong correlations and anisotropic geometry in parameter space. The sampler maintains an ensemble of $W$ walkers $\{X_k\}_{k=1}^W$, where $X_k\in\mathbb{R}^d$ and $d=2$ in our setting ($X_k=(\alpha,c)$). The core proposal mechanism is the \emph{stretch move}, which is invariant under affine transformations of the state space.

\paragraph{Stretch-move proposal.}
Given the current ensemble, for each walker $k$ we randomly select a complementary walker $j\neq k$ uniformly from $\{1,\dots,W\}\setminus\{k\}$. We then draw a stretch factor $z$ from the distribution
\begin{equation}
g(z) \propto \frac{1}{\sqrt{z}},\qquad z\in\left[\frac{1}{a},\,a\right],
\label{eq:supp_stretch_dist}
\end{equation}
where $a>1$ is a tuning parameter (commonly $a=2$). The proposed state for walker $k$ is
\begin{equation}
Y = X_j + z\,(X_k - X_j).
\label{eq:supp_stretch_prop}
\end{equation}

This proposal stretches (or contracts) the vector from $X_j$ to $X_k$ by a factor $z$ around $X_j$.

\paragraph{Metropolis--Hastings acceptance probability.}
Let $\pi(x)$ denote the target density (here, the posterior up to normalization), and let $\log\pi(x)$ be the corresponding log-density. Under the stretch move, the acceptance probability is
\begin{equation}
A(X_k\to Y) = \min\left\{1,\; z^{d-1}\,\frac{\pi(Y)}{\pi(X_k)}\right\}
= \min\left\{1,\;\exp\Big[(d-1)\log z + \log\pi(Y)-\log\pi(X_k)\Big]\right\}.
\label{eq:supp_accept}
\end{equation}
where the factor $z^{d-1}$ arises from the Jacobian of the affine transformation implicit in \cref{eq:supp_stretch_prop} and is essential for detailed balance.

\paragraph{Algorithm summary.}
Let $\log \tilde{\pi}(x)$ denote the log-posterior up to an additive constant:
\begin{equation}
\log \tilde{\pi}(\theta) = \log f(\theta) + \log \mathcal{L}(\theta),
\end{equation}
where $\log \mathcal{L}(\theta)$ is computed via \cref{eq:supp_quad_integral}. The AIES algorithm proceeds as follows:

\begin{algorithm}[H]
\caption{Affine-invariant ensemble sampler (stretch move) for $\theta=(\alpha,c)$}
\label{alg:supp_aies}
\begin{algorithmic}[1]
\Require Number of walkers $W$, dimension $d=2$, stretch parameter $a$, chain length $L$, initial ensemble $\{X_k^{(0)}\}_{k=1}^W$
\For{$t=0$ to $L-1$}
    \State Partition walkers into two disjoint groups $\mathcal{G}_1$ and $\mathcal{G}_2$
    \For{each group $\mathcal{G}\in\{\mathcal{G}_1,\mathcal{G}_2\}$}
        \State Let $\mathcal{G}^c$ denote the complementary group
        \For{each walker index $k\in\mathcal{G}$}
            \State Sample $j$ uniformly from $\mathcal{G}^c$
            \State Draw $z\sim g(z)\propto z^{-1/2}$ on $[1/a,a]$
            \State Propose $Y \leftarrow X_j^{(t)} + z\,(X_k^{(t)}-X_j^{(t)})$
            \State Compute $\Delta \leftarrow (d-1)\log z + \log\tilde{\pi}(Y) - \log\tilde{\pi}(X_k^{(t)})$
            \State Accept with probability $\min\{1,\exp(\Delta)\}$:
            \If{accepted}
                \State $X_k^{(t+1)} \leftarrow Y$
            \Else
                \State $X_k^{(t+1)} \leftarrow X_k^{(t)}$
            \EndIf
        \EndFor
    \EndFor
\EndFor
\State \Return $\{X_k^{(t)}\}_{k=1}^W$ for $t=1,\dots,L$
\end{algorithmic}
\end{algorithm}

\subsection{Posterior Diagnostics and Robustness}

With $W=20$ walkers and $20$ retained samples per walker, each state has $M=400$ retained samples in total. We report integrated autocorrelation time (IACT) and effective sample size (ESS) on these retained samples after applying the physical-domain filter ($0<\alpha<1$, $-10<c<10$). For scalar summary $u_t$ with retained-sample size $M$, ESS is estimated as
\begin{equation}
\mathrm{ESS}(u)\approx \frac{M}{1+2\sum_{\tau=1}^{\infty}\rho_u(\tau)},
\end{equation}
where $\rho_u(\tau)$ is the lag-$\tau$ autocorrelation. In practice, ESS converts autocorrelated retained draws into an equivalent number of independent draws, so it directly quantifies the Monte Carlo precision of posterior summaries and associated uncertainty estimates \citepS{geyer1992practical}.

State-level ranges are: $\mathrm{ESS}_{\alpha}\in[9.2,400.0]$ (median 348.5), $\mathrm{ESS}_{c}\in[147.8,400.0]$ (median 345.0), $\mathrm{IACT}_{\alpha}\in[1.00,41.28]$ (median 1.00), and $\mathrm{IACT}_{c}\in[1.00,2.56]$ (median 1.00). Overall, this indicates good practical sampling quality for most states.

\begin{table}[h]
\centering
\caption[Inference diagnostics across 50 states]{\textbf{Inference diagnostics across 50 states.} Values are reported as min/median/max across states.}
\label{tab:supp_inference_diagnostics}
\begin{tabular}{p{3.9cm} p{4.0cm} p{6.5cm}}
\toprule
\textbf{Diagnostic} & \textbf{Empirical summary} & \textbf{Interpretation} \\
\midrule
Valid posterior fraction & 0.787 / 0.895 / 1.000 & Fraction of samples satisfying $0<\alpha<1$ and $-10<c<10$; high values indicate most posterior mass is physically interpretable. \\
\midrule
\(\mathrm{ESS}_{\alpha}\) & 9.2 / 348.5 / 400.0 & Small values indicate stronger dependence in the retained sequence; most states have large effective sample support for $\alpha$. \\
\midrule
\(\mathrm{ESS}_{c}\) & 147.8 / 345.0 / 400.0 & Effective sample support for $c$ is generally high across states. \\
\midrule
\(\mathrm{IACT}_{\alpha}\) & 1.00 / 1.00 / 41.28 & Larger values correspond to slower decay of autocorrelation for $\alpha$. \\
\midrule
\(\mathrm{IACT}_{c}\) & 1.00 / 1.00 / 2.56 & Autocorrelation for $c$ is short-range in most states. \\
\bottomrule
\end{tabular}
\end{table}

We further assess robustness via two checks with distinct data sources. \textbf{Prior sensitivity} is computed from the same archived state-level posterior draws used above (no new data generation): we apply a narrower admissible window ($0.1<\alpha<0.9$, $-5<c<5$) and recompute posterior means; the resulting shifts are modest for $\alpha$ (median 0.029; max 0.084) and moderate for $c$ (median 0.149; max 0.508). \textbf{Synthetic recovery} is computed from simulation data: we run 200 synthetic experiments generated from known $(\alpha^\star,c^\star,T^\star)$ under the same reduced likelihood family and re-estimate $(\alpha,c)$ with temperature marginalization; empirical 95\% interval coverage is 1.00 for $\alpha$ and 0.98 for $c$, with median absolute errors 0.128 ($\alpha$) and 0.681 ($c$). Together, these diagnostics support that the inferred regime-level conclusions are stable and not dominated by prior-window truncation or sampling pathologies.

\begin{figure}[H]
\centering
\includegraphics[width=\textwidth]{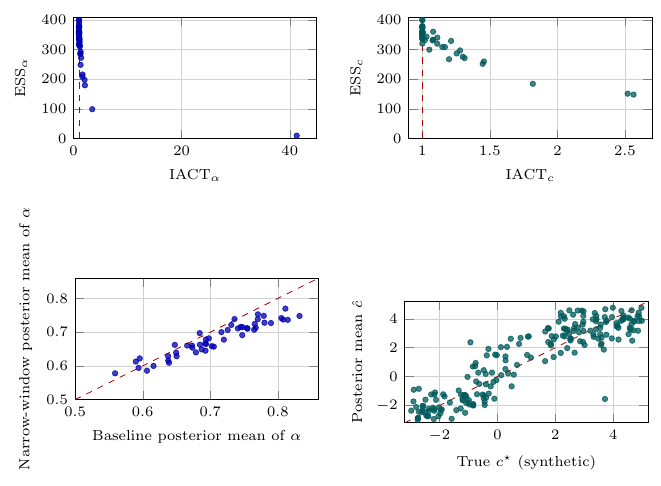}
\caption[Convergence diagnostics across 50 states and synthetic recovery runs]{\textbf{Convergence diagnostics across 50 states and synthetic recovery runs.} (a) Retained-chain mixing for $\alpha$: \((\mathrm{IACT}_{\alpha},\mathrm{ESS}_{\alpha})\). (b) Retained-chain mixing for $c$: \((\mathrm{IACT}_{c},\mathrm{ESS}_{c})\). (c) Prior-window sensitivity for $\alpha$: baseline vs narrowed-window posterior means. (d) Synthetic recovery for $c$: true value $c^\star$ versus recovered posterior mean $\hat c$ across synthetic experiments. Dashed lines mark ideal/reference relations: $\mathrm{IACT}=1$ in (a,b), and equality $y=x$ in (c,d).}
\label{fig:supp_s5_diagnostics}
\end{figure}

\newpage
\section{Data Pipeline}
\label{sec:supp_data_pipeline}
\label{sec:supp_tweet_details}

In this section, we provide detailed information regarding the data collection, preprocessing, dataset construction, and the text classification models used to estimate the emotional prevalence $m$.

\subsection{Data Collection and Preprocessing}

\noindent Twitter/X \citepS{ghani2019social} is selected for this study. Due to restrictions on accessing archived data via the standard API, we utilized open-source datasets from Kaggle containing tweets related to COVID-19. \Cref{tab:dataset_composition} summarizes the datasets used.

\begin{table}[h!]
    \centering
    \caption[Dataset composition]{\textbf{Dataset composition.}}
    \label{tab:dataset_composition}
    \begin{tabular}{l l c c l}
        \toprule
        \textbf{Dataset Name} & \textbf{Hashtag} & \textbf{Rows $\times$ Columns} & \textbf{Truncated} & \textbf{Source} \\
        \midrule
        COVID19 Tweets & \#covid & 179,108$\times$13 & Yes & \href{https://www.kaggle.com/datasets/gpreda/covid19-tweets/data}{Kaggle} \\
        Omicron Rising & \#omicron & 48,168$\times$16 & Yes & \href{https://www.kaggle.com/datasets/gpreda/omicron-rising/data}{Kaggle} \\
        \bottomrule
    \end{tabular}
\end{table}

We retained ``user location,'' ``date,'' and ``text'' columns. Tweets originating from U.S. states were extracted. The text data underwent the following preprocessing steps: (1) removing URLs, (2) removing mentions, (3) removing ``\#'' symbol, (4) converting emojis to text, (5) lowercasing, (6) removing punctuation, (7) normalizing whitespace. \Cref{tab:data_examples} shows examples of raw and preprocessed tweets.

\begin{table}[h]
    \centering
    \caption[Sample tweets of the dataset (all hashtaged with ``\#covid'')]{\textbf{Sample tweets of the dataset (all hashtaged with ``\#covid'').}}
    \label{tab:data_examples}
    \begin{tabular}{l p{1.7cm} p{4.6cm} p{4.3cm} c}
        \toprule
        \textbf{User Location} & \textbf{Date} & \textbf{Raw Text} & \textbf{Preprocessed} & \textbf{Label} \\
        \midrule
        Deer Park, NY & 2020-08-01 16:02:22 & Incase you needed a guide to help you practice social distancing, please inspect the post below \includegraphics[height=1em]{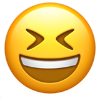}\includegraphics[height=1em]{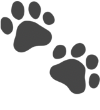}… https://t.co/pn9pisEm9a & incase you needed a guide to help you practice social distancing please inspect the post below grinning\_squinting\_face paw\_prints & \textsc{Unaffected} \\
        \midrule
        St Louis, MO & 2020-08-29 21:12:48 & This was my worst fear. Now it's a reality on a human and systemic level. We are sacrificing people so we can eat… https://t.co/ALbVRTuxNN & This was my worst fear. Now it's a reality on a human and systemic level. We are sacrificing people so we can eat & \textsc{Affected} \\
        \midrule
        New York, NY & 2020-07-29 16:23:26 & My blue jean cut offs are not for partying. They are for official business ONLY. \#summer \#america… https://t.co/AjnFDg24MG & my blue jean cut offs are not for partying they are for official business only summer america & \textsc{Other} \\
        \bottomrule
    \end{tabular}
\end{table}

\subsection{Supervised Dataset Construction}

A supervised dataset was constructed by manually labeling 2,500 randomly sampled tweets, where the valid sample after preprocessing was 2,460. The dataset was then split into training (2,160 tweets) and validation (300 tweets) sets. The labeling criteria are detailed in \Cref{tab:label_criteria}.

\begin{table}[h]
    \centering
    \caption[Labeling criteria]{\textbf{Labeling criteria.}}
    \label{tab:label_criteria}
    \begin{tabular}{p{2.5cm} c p{11.5cm}}
        \toprule
        \textbf{Emotional State} & \textbf{Label} & \textbf{Description} \\
        \midrule
        \textsc{Unaffected} & 0 & Tweets conveying a nonnegative/optimistic attitude, promoting control or recovery efforts. \\
        \midrule
        \textsc{Affected} & 1 & Tweets reflecting a tense/pessimistic attitude, heightening stress. \\
        \midrule
        \textsc{Other} & 2 & Tweets unrelated to COVID-19, advertisements, or non-informative content. \\
        \bottomrule
    \end{tabular}
\end{table}

\begin{figure}[h]
    \centering
    \makebox[\textwidth][c]{%
        \resizebox{0.99\textwidth}{!}{%
            \includegraphics[height=0.20\textheight]{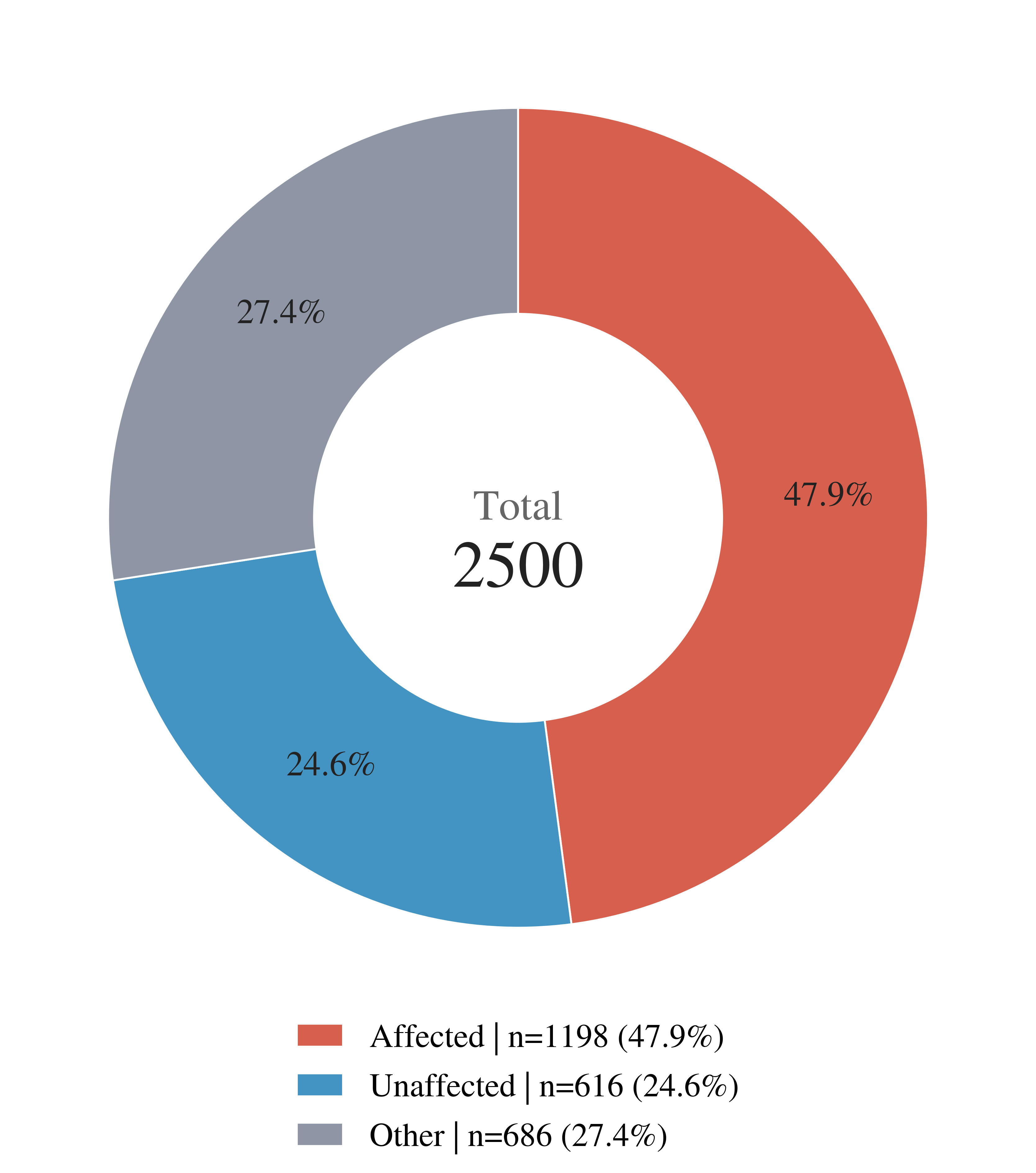}%
            \hspace{0.005\textwidth}%
            \includegraphics[height=0.20\textheight]{pictures/S6/dataset_state_label_pies_5x10.png}%
        }%
    }
    \caption[Label distribution of the raw 2,500 labeled tweets]{\textbf{Label distribution of the raw 2,500 labeled tweets.} (a) Overall distribution across the three classes. (b) State-level distributions for 50 states, where 2,367 out of 2,500 tweets have at least one valid state location, and the remaining 133 tweets are unassigned. For tweets with multiple valid locations, they are assigned to both states, resulting to a total of 2,592 total state assignments.}
    \label{fig:dataset_label_distribution}
\end{figure}

The class distribution in the raw manually labeled dataset across the three labels is shown in \Cref{fig:dataset_label_distribution}, together with the corresponding state-level distributions. Overall, the dataset is relatively balanced between \textsc{Unaffected} and \textsc{Other}, with a larger proportion of \textsc{Affected}. At the state level, some variability is observed, and certain states have limited representation in the manually labeled subset.

Our objective is to distinguish among the three labels, rather than to model state-specific patterns. The 2,460 manually labeled samples used for training and validation contain no valid instances from \textsc{WY}, while the remaining 49 states each have at least one valid sample. However, this does not affect the reliability of the model. The final model is applied to the full dataset, which contains hundreds of thousands of samples, including a sufficient number from \textsc{WY}. Since the task focuses on label discrimination rather than state-level generalization, the absence of \textsc{WY} in the labeled subset does not compromise the overall classification accuracy.

\FloatBarrier
\newpage
\section{Sentiment Model Reliability}
\label{sec:supp_sentiment_reliability}

\subsection{Comparison of Classification Methods}

We evaluated two main strategies for conducting such a text classification task: Prompt Engineering with Large Language Models (LLMs) and Fine-tuning pre-trained Transformer encoder-only models.

\paragraph{Zero-shot, Few-shot, and CoT in-context learning.}
We experimented with prompt engineering techniques to adapt the state-of-the-art OpenAI GPT-5.2 model for our task at hand. With increasingly complex prompting strategies, we tested the following methods:
\begin{itemize}
    \item \textbf{Zero-shot prompting}: Asking the model to classify the tweet without examples.
    \item \textbf{Few-shot prompting}: Providing a small number of labeled examples (2-3) in the context window to guide the model's classification.
    \item \textbf{Chain-of-Thought (CoT)}: Encouraging the model to explicitly follow the dedicated reasoning process before assigning a label.
\end{itemize}
These methods yielded accuracy scores hovering around 62\% on average, which was deemed insufficient for reliable downstream parameter calibration. Another key issue that hindered the adoption of LLM-based classification in our final pipeline was the computational cost and efficiency, as each classification query requires a separate API call, which becomes prohibitive when scaling to the large number of tweets in our final application. The detailed prompts are shown below, each of which is tested three times, and the overall performance is recorded as the average value, summarized in Table~\ref{tab:llm_classification_results}.

\begin{table}[htbp]
\centering
\caption[Experimental Results of GPT-5.2 under Zero-shot, Few-shot and Chain-of-Thought Prompting]{\textbf{Experimental Results of GPT-5.2 under Zero-shot, Few-shot and Chain-of-Thought Prompting}}
\label{tab:llm_classification_results}
\begin{tabular}{llcccccccc}
\toprule
Model & Prompt & run 1: acc & run: 1 F\textsubscript{1} & run 2: acc & run 2: F\textsubscript{1} & run 3: acc & run 3: F\textsubscript{1} & ave acc & ave F\textsubscript{1} \\
\midrule
gpt-5.2 & Zero-shot & 62.33\% & 0.6299 & 62.67\% & 0.6317 & 62.00\% & 0.6260 & 62.33\% & 0.6292 \\
gpt-5.2 & Few-shot  & 63.67\% & 0.6388 & 61.67\% & 0.6223 & 62.00\% & 0.6230 & 62.45\% & 0.6280 \\
gpt-5.2 & CoT       & 65.33\% & 0.6616 & 64.00\% & 0.6490 & 63.67\% & 0.6448 & 64.33\% & 0.6518 \\
\bottomrule
\end{tabular}
\end{table}

\begin{promptbox}[fontupper=\ttfamily\footnotesize]{Zero-shot Prompt}
\RoleSystem\ \par
You are an expert psychologist and social media analyst specializing in understanding how global events affect people's mental states. \\

Your task is to classify tweets posted during the COVID-19 pandemic to determine whether the author's psychological mindset was affected by COVID-19. \\

\#\# Classification Labels

You must classify each tweet into exactly ONE of the following three categories:\\

1. **unaffected**: Tweets in this category convey a nonnegative or optimistic attitude toward COVID-19, promoting control or recovery efforts through encouraging messages. These tweets may express subjective viewpoints that alleviate concerns about the pandemic or share content that positively influences the emotional state of the community.\\

2. **affected**: Tweets in this category reflect a tense or pessimistic attitude toward the pandemic. These texts may either heighten stress about COVID-19 through subjective viewpoints or share factual content that contributes to negative emotional responses in the community.\\

3. **other**: Tweets in this category are unrelated to COVID-19. These include advertisements for profit or promotion, emotionally charged complaints about politicians unrelated to pandemic policies, misuse of COVID-related hashtags, and any content that does not fit the ``Unaffected'' or ``Affected'' categories.\\

\#\# Important Guidelines\\

- Focus on the AUTHOR's psychological state, not the content they're sharing \\
- Neutral reporting of facts without emotional expression -> "other"\\
- When uncertain, prefer "other" over guessing\\
- The tweet text has been preprocessed (URLs, mentions removed, emojis converted to text)\\
- For some reason, the tweets are truncated, so you should infer from such partial contents.\\

\#\# Response Format\\

You must respond with a valid JSON object containing:\\
- "label": one of "unaffected", "affected", or "other"\\
- "confidence": a number between 0 and 1 indicating your confidence\\

\medskip
\RoleUser\ \par
Classify the following tweet:\\

"\{tweet\_text\}"\\

Respond with a JSON object containing "label" and "confidence".
\end{promptbox}

\begin{promptbox}[fontupper=\ttfamily\footnotesize]{Few-shot Prompt}
\RoleSystem\ \par
\dots same to zero-shot \dots

\medskip
\RoleUser\ \par
Here are some examples of how to classify tweet\\

\{examples\}\\

Now classify the following tweet:\\

"\{tweet\_text\}"\\

Respond with a JSON object containing "label" and "confidence".
\end{promptbox}

\begin{promptbox}[fontupper=\ttfamily\footnotesize]{Chain-of-Thought Prompt}
\RoleSystem\ \par
\dots same to zero-shot except "Response Format" \dots

\medskip
\RoleUser\ \par
Now, classify a (possibly truncated) tweet into one of the predefined labels from the system prompt.\\

Internally follow this reasoning procedure:\\
1. Analyze the main topic or message of the tweet. If the tweet is truncated, infer a plausible completion based only on the visible content and typical tweet patterns.\\
2. Identify emotional indicators (positive, negative, neutral) present in the text.\\
3. Determine whether the author is expressing personal feelings or merely sharing information.\\
4. Decide the final label according to the category definitions provided in the system prompt.\\

Do not reveal your intermediate reasoning steps.\\
Only output the final result as a valid JSON object in the following format\\

\{\{\\
  "reasoning": "A brief, high-level summary of why this label was chosen (no step-by-step reasoning).",\\
  "label": "unaffected" | "affected" | "other",\\
  "confidence": 0.0-1.0 \\
\}\}\\

Tweet:\\
"\{tweet\_text\}"
\end{promptbox}

All three prompting strategies, as shown in the prompts design, output a JSON object containing the predicted label and confidence score. The zero-shot prompt relies solely on the model's understanding of the task from the provided instructions, while the few-shot prompt guides the model with some extra examples, which generally leads to a slight improvement in performance. The CoT prompt further encourages the model to internally reason through the classification process by allowing it to follow a structured thought process before arriving at the final label, which typically yields better performance in complex tasks like this truncated tweet classification. Representative CoT examples for all three labels are provided in \Cref{tab:cot_reasoning_examples}.

\begin{table}[htbp]
\centering
\caption[Representative CoT reasoning examples]{\textbf{Representative CoT reasoning examples.}}
\label{tab:cot_reasoning_examples}
\scriptsize
\setlength{\tabcolsep}{4pt}
\begin{tabular}{p{0.08\textwidth}p{0.31\textwidth}p{0.41\textwidth}p{0.12\textwidth}}
\toprule
Target label & Tweet text (preprocessed) & CoT rationale summary & Output \\
\midrule
\textsc{Affected} &
literally in tears knowing we have to close the bar to keep ourselves amp others safe  downcast\_face\_with\_sweati know its in the best inter &
The author expresses sadness and distress about having to close the bar due to COVID-19 safety concerns, indicating a negative emotional impact despite acknowledging it's necessary. &
\texttt{affected} (0.86) \\
\midrule
\textsc{Unaffected} &
incase you needed a guide to help you practice social distancing please inspect the post below  grinning\_squinting\_face paw\_prints &
The tweet encourages practicing social distancing in a light, helpful tone, suggesting a constructive and nonnegative mindset toward COVID-19 precautions. &
\texttt{unaffected} (0.78) \\
\midrule
\textsc{Other} &
the stock markets sure suffered terribly the 8 years biden was vice president oh wait the dow increased 140 i &
The tweet is a political comment about stock market performance during Biden's vice presidency and does not reference COVID-19 or the pandemic's impact on the author's mental state. &
\texttt{other} (0.93) \\
\bottomrule
\end{tabular}
\end{table}

\begin{figure}[h]
    \centering
    \makebox[\textwidth][c]{\includegraphics[width=0.99\textwidth]{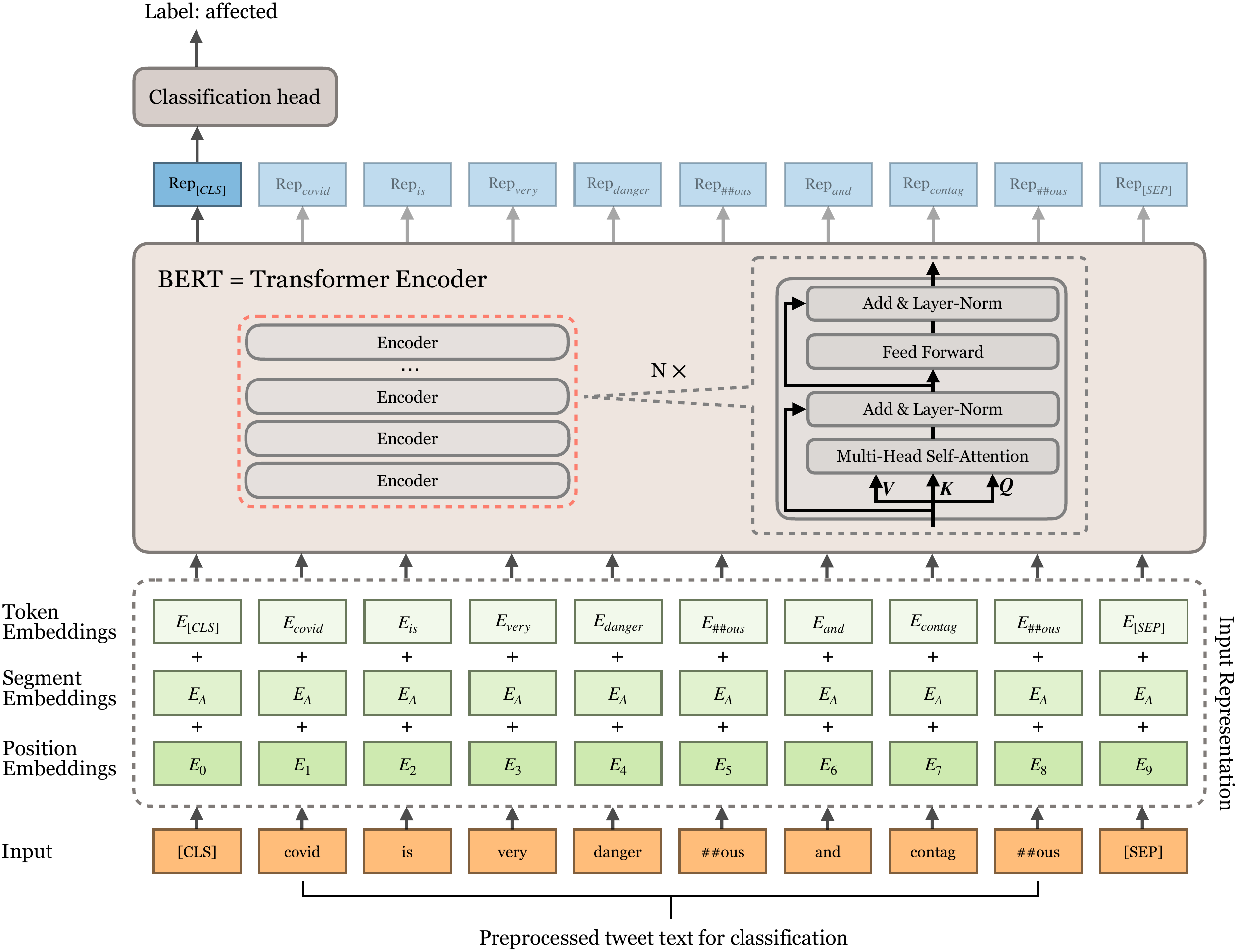}}
    \caption[Embedding, representing, and predicting by BERT]{\textbf{Embedding, representing, and predicting by BERT.}}
		\label{fig:BERT}
\end{figure}

\paragraph{Fine-tuning BERTweet.}
For the Transformer encoder-based approach, we selected BERTweet \citepS{si_nguyen2020bertweet}, a variant of BERT pretrained on English tweets that specifically optimized for social media text analysis tasks. The architecture of BERT and its variants is based on the Transformer encoder (see \Cref{fig:BERT}). Each layer uses multi-head self-attention (MSA). For an input sequence $\mathbf{U}$, query $\mathbf{Q}$, key $\mathbf{K}$, and value $\mathbf{V}$ matrices are computed as $\mathbf{Q} = \mathbf{U}\mathbf{W}^Q, \mathbf{K} = \mathbf{U}\mathbf{W}^K, \mathbf{V} = \mathbf{U}\mathbf{W}^V$. The attention output is $\text{Attention}(\mathbf{Q}, \mathbf{K}, \mathbf{V}) = \text{softmax}(\frac{\mathbf{Q}\mathbf{K}^{\top}}{\sqrt{d_k}}) \mathbf{V}$.
We fine-tuned BERTweet on our supervised dataset by adding a classification head. This approach achieves an accuracy around 76.00\% (see \Cref*{tab:TextClassifier} in the main text).

\begin{figure}[h]
    \centering
    \makebox[\textwidth][c]{%
        \resizebox{0.9\textwidth}{!}{%
            \includegraphics[height=0.20\textheight]{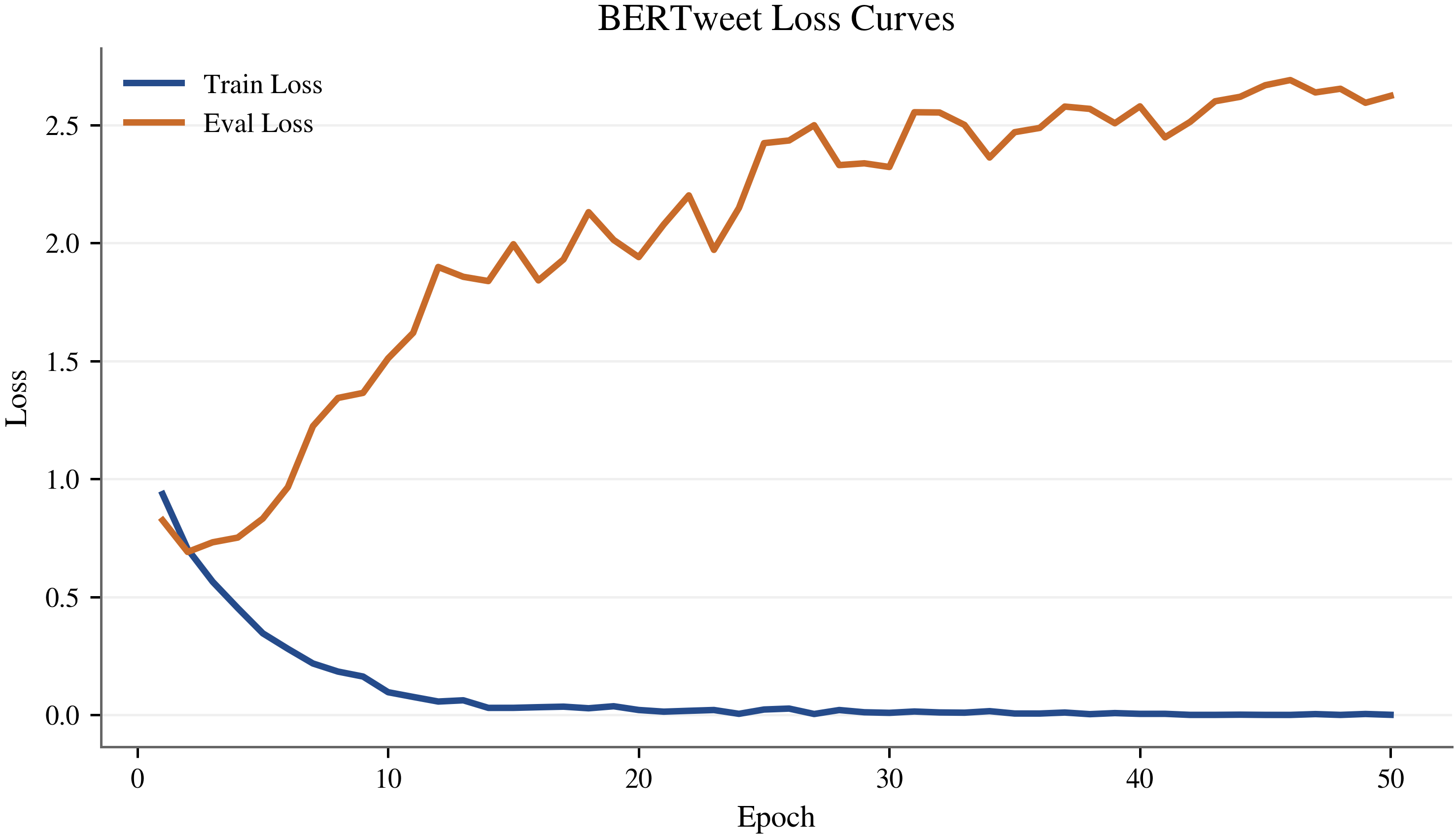}%
            \hspace{0.005\textwidth}%
            \includegraphics[height=0.20\textheight]{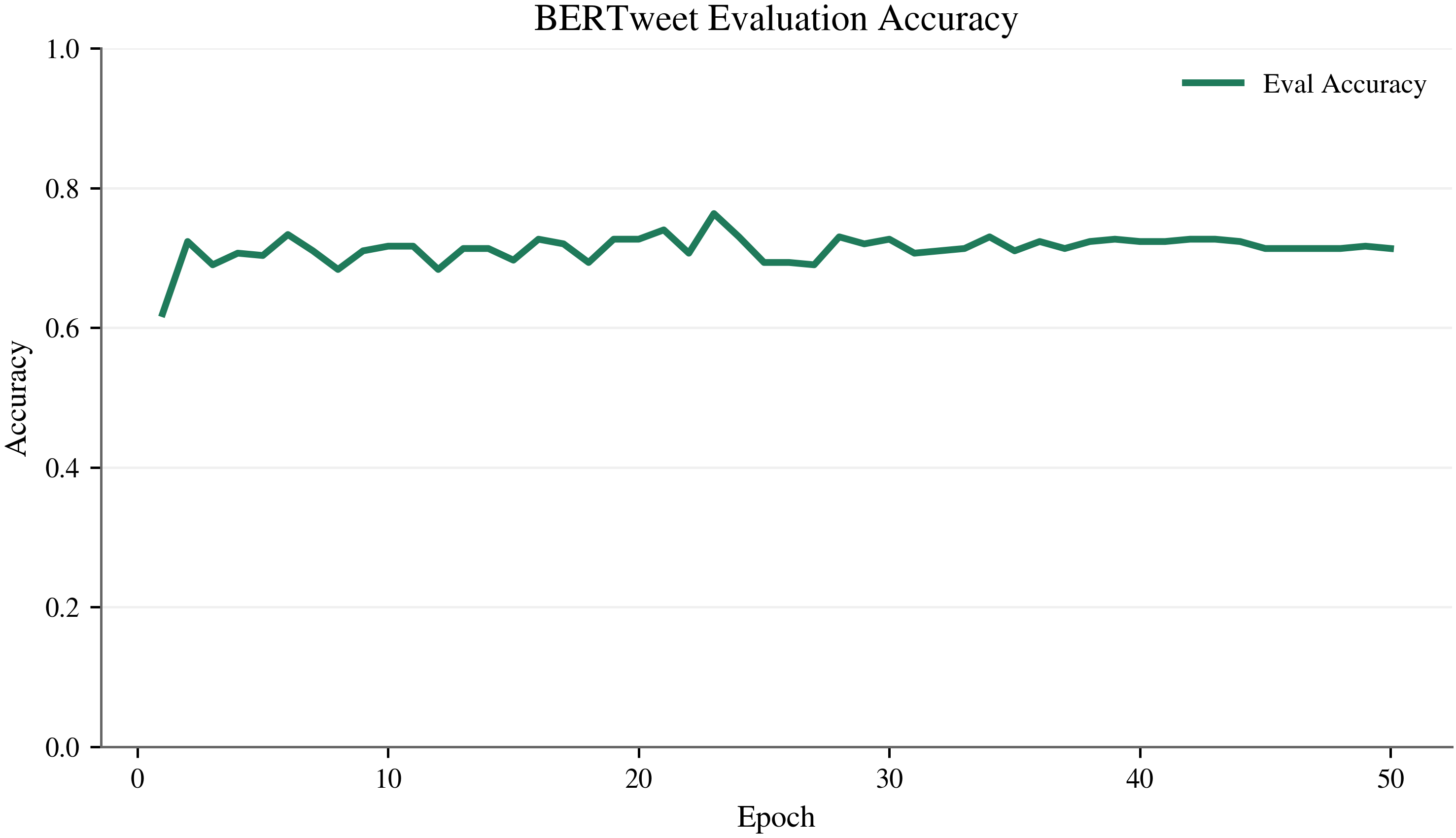}%
        }%
    }
    \caption[Learning curves for fine-tuning BERTweet]{\textbf{Learning curves for fine-tuning BERTweet.} (a) Training and validation loss over epochs. (b) Validation accuracy over epochs.}
    \label{fig:bertweet_learning_curves}
\end{figure}

The learning curves for the fine-tuning process are shown in \Cref{fig:bertweet_learning_curves}. The training loss steadily decreases, indicating that the model is learning from the training set, while the validation loss decreases at the beginning and then increases over epochs, suggesting the onset of overfitting, which is common in fine-tuning such a large model on a small dataset. From the curve of validation accuracy, one can observe that the performance increases initially and then occasionally fluctuates around a certain level, which is consistent with the overfitting pattern observed in the loss curves. Therefore, the checkpoint at epoch with the best validation accuracy is selected for the final model used in downstream inference. The demonstrated learning curves is one representative example, and similar patterns are observed across multiple runs with different random seeds for data shuffling and model initialization, confirming the robustness of the fine-tuning process.

\begin{figure}[H]
    \centering
    \makebox[\textwidth][c]{\includegraphics[width=0.99\textwidth]{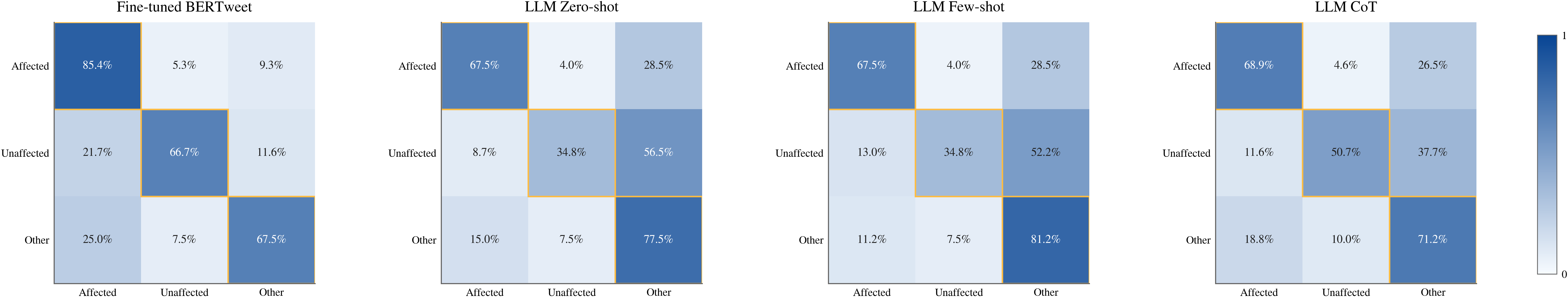}}
    \caption[Confusion matrices for all methods]{\textbf{Confusion matrices for all methods.}}
		\label{fig:confusion_matrices_all_methods}
\end{figure}

As a comparison alongside the accuracy and F1 metrics shown in \Cref*{tab:TextClassifier}, the confusion matrix of each method's best run on the validation set is visualized in \Cref{fig:confusion_matrices_all_methods}. 

It can be observed that while LLMs demonstrate competitive performance on clearly defined categories, they struggle with semantically subtle classes such as \textsc{Unaffected}, often collapsing predictions into broader categories (e.g., \textsc{Other}). In contrast, supervised fine-tuning remains superior for this boundary-sensitive classification task, achieving consistently strong performance across all three classes with relatively balanced precision and recall, and minimal bias toward the majority class (\textsc{Affected} in this case). 

Given its superior performance, robustness, and substantially lower computational cost compared to LLM-based prompting strategies, we adopted the fine-tuned BERTweet classifier for our final pipeline.

\newpage
\section{Intervention Simulations}
\label{sec:supp_intervention}

This section provides implementation details for the numerical control experiments presented in the main text (\cref*{fig:control_strategy}). We simulate six scenarios to examine how different interventions affect the evolution of collective stress prevalence over a $70$-day horizon following the onset of a COVID-19 outbreak.

\subsection{Simulation setup}

\begin{figure}[h]
\centering
\includegraphics[width=0.75\textwidth]{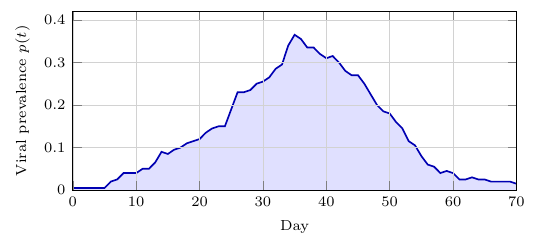}
\caption[Baseline viral prevalence trajectory used in all intervention simulations]{\textbf{Baseline viral prevalence trajectory used in all intervention simulations.} The daily prevalence $p(t)$ follows a bell-shaped epidemic curve over $70$ days, peaking near day $35$ at $p\approx 0.37$. This trajectory serves as the common exogenous forcing for all six scenarios; in the no-control, $\alpha=0$, and positivity-bias scenarios it is applied without modification, while in the active intervention scenarios (Scenarios~2--4) the effective prevalence or network structure is adjusted when the limit-state function $G<0$.}
\label{fig:supp_prevalence_trajectory}
\end{figure}

All simulations share the following configuration:
\begin{itemize}
    \item \textbf{Population size}: $N=200$ agents.
    \item \textbf{Baseline parameters}: $\alpha=0.6672$, $c=0.7049$, representative of many U.S.\ states (see main text).
    \item \textbf{Effective temperature}: $T=0.5$ (i.e., $\beta=1/T=2$), controlling the stochasticity of individual emotional updates.
    \item \textbf{State space}: Both emotional states and damage states are binary ($y_i\in\{0,1\}$, $x_j\in\{0,1\}$), so DimY $=$ DimX $=2$.
    \item \textbf{Prevalence trajectory}: The daily viral prevalence $p(t)$ for $t=0,1,\dots,70$ is generated using the Covasim agent-based epidemic simulator (v3.1.6) \citeS{si_kerr2021covasim}. The simulation is configured with a hybrid population structure (\texttt{pop\_type='hybrid'}) using U.S.\ demographic characteristics (\texttt{location='usa-florida'}), a population of $200$ agents (matching the emotional model), and a transmission rate $\beta_{\text{epi}}=0.012$. From the Covasim output, the daily prevalence (fraction of currently infectious and not yet recovered agents) is extracted and stored. \cref{fig:supp_prevalence_trajectory} shows the resulting baseline trajectory, which produces a bell-shaped outbreak peaking around day~$35$ at approximately $p\approx0.37$ and declining thereafter.
    \item \textbf{Scenario inputs}: For each day and each control scenario, the Monte Carlo solver is driven by the corresponding day-specific hazard state vector and network inputs associated with that scenario. The no-control, $\alpha=0$, and positivity-bias cases inherit the baseline epidemic forcing, whereas the active-control cases use the scenario-specific updates induced by the chosen intervention lever. This matches the workflow used to generate the trajectories reported in \cref{fig:supp_stress_trajectories}.
\end{itemize}

\subsection{Monte Carlo simulation procedure}

\begin{figure}[h]
\centering
\includegraphics[width=0.75\textwidth]{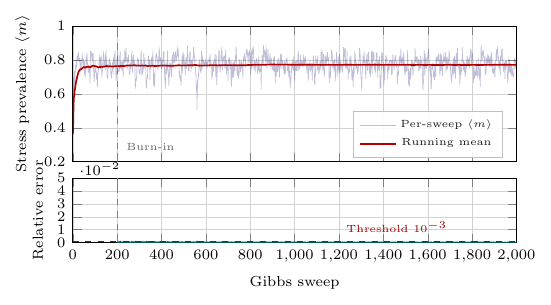}
\caption[Gibbs sampler convergence diagnostic at day~$35$ (peak prevalence)]{\textbf{Gibbs sampler convergence diagnostic at day~$35$ (peak prevalence).} \emph{Top}: per-sweep stress prevalence (light blue) and its running mean (red). The chain is initialized at $\langle m\rangle=p=0.365$ and equilibrates within $\sim\!200$ sweeps to $\langle m\rangle\approx0.78$. The vertical dashed line marks the end of the burn-in period. \emph{Bottom}: relative change in the running mean between successive sweeps. The horizontal dashed line marks the convergence threshold of $10^{-3}$, which is reached at sweep~$200$.}
\label{fig:supp_gibbs_convergence}
\end{figure}

At each day $t$, the stress prevalence $\langle m\rangle$ is computed via the Gibbs simulation model (Algorithm~\ref*{alg:Gibbs} in the main text). The procedure is as follows:

\begin{enumerate}
    \item \textbf{Model assembly}: For the current day, the hazard state vector and the scenario-specific network inputs are assembled into the local Hamiltonian. As in the implementation used for the control-study figures, neighbour contributions are averaged within each agent's current social and hazard neighbourhoods so that local energies are degree-normalized.
    \item \textbf{Initialization}: The emotional configuration is initialized from the current hazard state, i.e., $y_i^{(0)}=\lfloor \text{DimYX}\cdot x_i\rceil$, so the chain starts from the day-specific damage realization before social amplification is applied.
    \item \textbf{Single-site updates}: Each sweep visits all $N$ agents in a random permutation. For each agent $i$:
    \begin{itemize}
        \item A proposal state $y_i'$ is generated: if $y_i=0$, the proposal is $y_i'=1$ with probability $1/2$ (otherwise it stays at $0$); if $y_i=1$ (the maximum), the proposal is $y_i'=0$ with probability $1/2$; otherwise, $y_i'=y_i\pm 1$ with equal probability.
        \item The energy change $\Delta\mathcal{H}_i=\mathcal{H}_i(y_i')-\mathcal{H}_i(y_i)$ is computed using the local Hamiltonian:
        \begin{equation}
        \mathcal{H}_i = \alpha\sum_{j\in\mathcal{N}_i^{(\mathbf{yy})}}b_{j\to i}^{(\mathbf{yy})}\left[(y_j-y_i)^2-c\,y_j\,y_i\right]
        +(1-\alpha)\sum_{j\in\mathcal{N}_i^{(\mathbf{xy})}}b_{j\to i}^{(\mathbf{xy})}\left(x_j-y_i\right)^2.
        \end{equation}
        \item The proposal is accepted with probability $\min\{1,\exp(-\Delta\mathcal{H}_i/T)\}$.
    \end{itemize}
    \item \textbf{Convergence}: The chain is run for up to $10{,}000$ sweeps. After a $200$-sweep burn-in, convergence is monitored through the relative change of the running mean; the sampler is terminated once this diagnostic falls below the prescribed tolerance. In the representative baseline example shown in \cref{fig:supp_gibbs_convergence}, this occurs around sweep~$200$.
    \item \textbf{Output}: The Monte Carlo estimate $\langle m\rangle^{\text{MCS}}$ is obtained from the equilibrated samples and recorded day by day to construct the scenario trajectories in \cref{fig:supp_stress_trajectories}.
\end{enumerate}

\cref{fig:supp_gibbs_convergence} illustrates the convergence behavior of the Gibbs sampler for a representative configuration at day~$35$ (peak prevalence, $p=0.365$) under the baseline parameters ($\alpha=0.6672$, $c=0.7049$). The chain is initialized at $\langle m\rangle=p=0.365$ and rapidly equilibrates within approximately $200$ sweeps to a stationary magnetization of $\langle m\rangle\approx0.78$, well above the viral prevalence---consistent with the social amplification regime. After equilibration, the running mean stabilizes and the relative error drops below $10^{-3}$, confirming convergence. The chain remains in this stationary regime for the remaining sweeps, with per-sweep fluctuations reflecting the stochastic nature of the sampler at finite population size. Although \cref{fig:supp_gibbs_convergence} shows only a single representative example, convergence was verified for all six scenarios across all $71$ time steps: in every case, the sampler satisfied the $10^{-3}$ relative-error criterion well within the $10{,}000$-sweep budget. The close agreement between MCS and MFT estimates across all scenarios in \cref{fig:supp_stress_trajectories} provides additional confirmation that the chains have equilibrated.

\subsection{Mean-field comparison}

Mean-field theory is derived in the main text and analyzed further in \cref{sec:supp_theoretical_properties}. For each day $t$, we substitute the daily prevalence $p(t)$ into the free-energy density (\cref*{eq:free_eng}), i.e., $\langle x\rangle_{\mathbf{A}_{\boldsymbol{xy}}}=p(t)\,\langle k_{\boldsymbol{y}}\rangle_{\mathbf{A}_{\boldsymbol{xy}}}$, and compute
\begin{equation}
m^*(t)=\arg\min_{m\in[0, 1]} f_t(m).
\end{equation}

\subsection{Trigger rule and control variables}

The limit-state function $G$ (\cref*{eq:limit_function} in the main text) determines whether an intervention is activated. Evaluated at day $t$ with the current network degrees and prevalence, the function is
\begin{equation}
G(t) = -\frac{1}{2}\langle k\rangle_{\mathbf{A}_{\boldsymbol{yy}}}(t)\,c
+\langle k_{\boldsymbol{y}}\rangle_{\mathbf{A}_{\boldsymbol{xy}}}(t)\left(1-2p(t)\right)\left(\frac{1}{\alpha}-1\right).
\end{equation}
Under the main-text reliability convention, $G(t)<0$ indicates the majority-arousal (undesirable) regime; therefore this is the trigger condition for activating interventions. The control lever applied depends on the scenario; in the no-control, $\alpha=0$, and positivity-bias scenarios, no active intervention is applied regardless of $G(t)$.

\subsection{Network modification procedure}

For scenarios that involve modifying network degrees, we adjust the adjacency matrix while preserving symmetry:
\begin{itemize}
    \item \textbf{Adding $\Delta k$ edges per node}: The total number of edges to add is $E_{\text{add}}=\lfloor N\cdot\Delta k/2\rfloor$. All non-existing edges $(i,j)$ with $i<j$ are enumerated, and $E_{\text{add}}$ pairs are selected uniformly at random; both $\mathbf{A}(i,j)$ and $\mathbf{A}(j,i)$ are set to $1$.
    \item \textbf{Removing $\Delta k$ edges per node}: The total number of edges to remove is $E_{\text{rem}}=\lfloor N\cdot|\Delta k|/2\rfloor$. All existing edges $(i,j)$ with $i<j$ are enumerated, and $E_{\text{rem}}$ pairs are selected uniformly at random; both $\mathbf{A}(i,j)$ and $\mathbf{A}(j,i)$ are set to $0$.
\end{itemize}
In both cases, the operation preserves the undirected structure and avoids self-loops. The maximum per-step adjustment is $|\Delta k|=2$ (i.e., the average degree changes by at most $2$ per day).

\subsection{Description of different control strategies}

\subsubsection{Scenario 1: No control (baseline)}

No intervention is applied. Both networks $\mathbf{A}_{\boldsymbol{yy}}$ and $\mathbf{A}_{\boldsymbol{xy}}$ remain fixed at $\langle k\rangle=20$ throughout all $70$ days. The parameters $\alpha=0.6672$ and $c=0.7049$ are used. This scenario serves as the reference against which the improvement of other strategies is measured.

\subsubsection{Scenario 2: Increasing $\langle k_{\boldsymbol{y}}\rangle_{\mathbf{A}_{\boldsymbol{xy}}}$}

When $G(t)<0$, the hazard-exposure network $\mathbf{A}_{\boldsymbol{xy}}$ is modified by adding edges so that the average degree increases by $\Delta k=2$ per intervention step. The rationale follows from the partial derivative of the limit-state function:
\begin{equation}
\frac{\partial G}{\partial \langle k_{\boldsymbol{y}}\rangle_{\mathbf{A}_{\boldsymbol{xy}}}}=(1-2p)\left(\frac{1}{\alpha}-1\right).
\end{equation}
When $p<1/2$, this derivative is positive, and increasing $\langle k_{\boldsymbol{y}}\rangle_{\mathbf{A}_{\boldsymbol{xy}}}$ pushes $G$ toward positive values (i.e., toward the safe regime). The social network $\mathbf{A}_{\boldsymbol{yy}}$ remains unchanged, and $\alpha$, $c$ are held fixed. As noted in the main text, this strategy is primarily illustrative: increasing physical contacts may elevate viral prevalence through dynamic feedback, an effect not captured in the static limit-state formulation.

\subsubsection{Scenario 3: Reducing $\langle k\rangle_{\mathbf{A}_{\boldsymbol{yy}}}$}

When $G(t)<0$, the social network $\mathbf{A}_{\boldsymbol{yy}}$ is modified by removing edges so that the average degree decreases by $|\Delta k|=2$ per intervention step (subject to the constraint that the average degree remains non-negative). This corresponds to reducing social media interactions or limiting peer-to-peer emotional contagion channels. The partial derivative
\begin{equation}
\frac{\partial G}{\partial \langle k\rangle_{\mathbf{A}_{\boldsymbol{yy}}}}=-\frac{c}{2}
\end{equation}
is negative when $c>0$, confirming that reducing $\langle k\rangle_{\mathbf{A}_{\boldsymbol{yy}}}$ increases $G$ and steers the system away from the majority-arousal regime. The hazard-exposure network $\mathbf{A}_{\boldsymbol{xy}}$ and parameters $\alpha$, $c$ remain unchanged.

\subsubsection{Scenario 4: Reducing prevalence rate $p$}

When $G(t)<0$, the viral prevalence in all subsequent days is reduced by removing $40\%$ of the currently infected agents. Specifically, for every future day $t'>t$, the infection-state vector $\mathbf{x}(t')$ is updated by randomly selecting $40\%$ of the $N$ agent indices and setting the corresponding infected entries to $x_j=0$. This mimics population-level interventions such as vaccination, mask mandates, or treatment rollout. The upper bound of $40\%$ reduction is motivated by empirical estimates of face-mask efficacy and early vaccine effectiveness. The networks and parameters $\alpha$, $c$ are held fixed.

The partial derivative
\begin{equation}
\frac{\partial G}{\partial p}=-2\langle k_{\boldsymbol{y}}\rangle_{\mathbf{A}_{\boldsymbol{xy}}}\left(\frac{1}{\alpha}-1\right)
\end{equation}
is negative (for $\alpha<1$), so decreasing $p$ increases $G$. Its absolute value is on the order of $10$ for typical parameter values---substantially larger than the partial derivatives with respect to the network degrees ($\sim 0.1$). This explains why reducing $p$ is the most effective single-lever control strategy in the main-text results.

\subsubsection{Scenario 5: No social influence ($\alpha=0$)}

No intervention is applied, and the model parameter is set to $\alpha=0$ (all other parameters remain at their baseline values), eliminating all social influence. In implementation, we keep the same population, networks, and Monte Carlo update procedure, but disable the social-interaction contribution so that updates depend only on hazard-exposure mismatch terms from $\mathbf{A}_{\boldsymbol{xy}}$. Thus, each agent's emotional state is driven solely by the infection status of hazard-exposure neighbours, while $\mathbf{A}_{\boldsymbol{yy}}$ remains present but dynamically inactive. This scenario serves as a counterfactual to quantify the net contribution of social amplification: the difference between the no-control baseline ($\alpha=0.6672$, $c=0.7049$) and this scenario isolates the effect of peer-to-peer emotional contagion on collective stress.

\subsubsection{Scenario 6: Positivity bias ($c<0$)}

No intervention is applied, and the sentiment parameter is set to $c=-0.7049$ (all other parameters remain at their baseline values, $\alpha=0.6672$). Under this setting, the amplification-attenuation term $-c\,y_j\,y_i$ in the local Hamiltonian becomes a penalty on jointly elevated emotional states, favouring emotional attenuation rather than amplification. The networks remain fixed throughout.

This counterfactual scenario illustrates the role of negativity bias: by flipping the sign of $c$, social interactions now encourage individuals to moderate each other's stress rather than amplify it. As a result, the stress prevalence remains noticeably below the viral prevalence even during the outbreak peak, in contrast to the baseline scenario where negativity bias drives stress prevalence above the viral prevalence.

\subsection{Summary of scenario parameters}

\begin{table}[h]
\centering
\caption[Parameter settings for the six intervention scenarios in \cref*{fig:control_strategy}]{\textbf{Parameter settings for the six intervention scenarios in \cref*{fig:control_strategy}.}}
\label{tab:supp_scenario_params}
\begin{tabular}{p{4.5cm} c c c p{4.0cm}}
\toprule
\textbf{Scenario} & $\alpha$ & $c$ & \textbf{Active lever} & \textbf{Control action when $G<0$} \\
\midrule
1. No control & $0.6672$ & $0.7049$ & None & --- \\
\midrule
2. Increasing $\langle k_{\boldsymbol{y}}\rangle_{\mathbf{A}_{\boldsymbol{xy}}}$ & $0.6672$ & $0.7049$ & $\mathbf{A}_{\boldsymbol{xy}}$ & Add $\Delta k=+2$ to $\langle k_{\boldsymbol{y}}\rangle_{\mathbf{A}_{\boldsymbol{xy}}}$ per step \\
\midrule
3. Reducing $\langle k\rangle_{\mathbf{A}_{\boldsymbol{yy}}}$ & $0.6672$ & $0.7049$ & $\mathbf{A}_{\boldsymbol{yy}}$ & Remove $|\Delta k|=2$ from $\langle k\rangle_{\mathbf{A}_{\boldsymbol{yy}}}$ per step \\
\midrule
4. Reducing prevalence $p$ & $0.6672$ & $0.7049$ & $p$ & Remove $40\%$ of infections in future days \\
\midrule
5. No social influence & $0$ & $0.7049$ & None & --- \\
\midrule
6. Positivity bias & $0.6672$ & $-0.7049$ & None & --- \\
\bottomrule
\end{tabular}
\end{table}

\subsection{Stress prevalence trajectories}

\begin{figure}[H]
    \centering
    \makebox[\textwidth][c]{\includegraphics[width=1.08\textwidth]{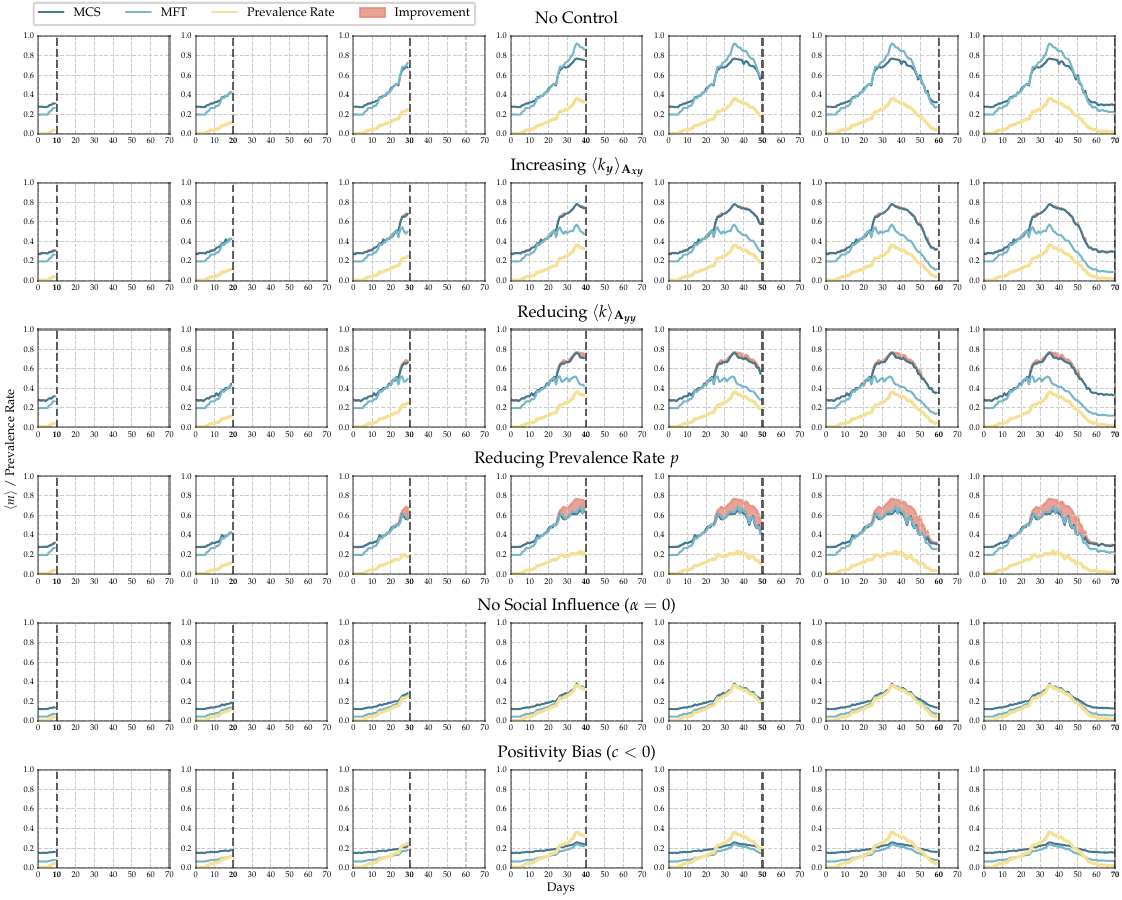}}
    \caption[Stress prevalence trajectories under the six intervention scenarios]{\textbf{Stress prevalence trajectories under the six intervention scenarios.} Each row corresponds to one scenario and each column shows the cumulative trajectory up to a checkpoint day (day~$10$ through day~$70$). In every subplot, the dark blue line represents MCS, the light blue line represents MFT, and the yellow line represents the viral prevalence $p(t)$. The dashed vertical line marks the checkpoint day. The red shaded area (rows~2--4) quantifies the reduction in stress prevalence relative to the no-control baseline. Reducing viral prevalence (row~4) yields the largest improvement, consistent with the dominant magnitude of $\partial G/\partial p$. In the no-social-influence case ($\alpha=0$, row~5), stress closely tracks viral prevalence. Under positivity bias ($c<0$, row~6), stress prevalence remains well below the viral prevalence even at peak infection.}
    \label{fig:supp_stress_trajectories}
\end{figure}

\cref{fig:supp_stress_trajectories} presents the time evolution of stress prevalence $\langle m\rangle$ under all six scenarios. The figure is organized as a $6\times 7$ grid: each row corresponds to one scenario (No Control, Increasing $\langle k_{\boldsymbol{y}}\rangle_{\mathbf{A}_{\boldsymbol{xy}}}$, Reducing $\langle k\rangle_{\mathbf{A}_{\boldsymbol{yy}}}$, Reducing Prevalence Rate~$p$, No Social Influence, and Positivity Bias), and each column shows the trajectory up to a successive checkpoint (days $10,20,\dots,70$). In every subplot, the dark blue line shows the Monte Carlo simulation (MCS) result, the light blue line shows the mean-field theory (MFT) prediction, and the yellow line shows the viral prevalence $p(t)$. The vertical dashed line marks the current checkpoint day. For the three active intervention scenarios (rows~2--4), the shaded red region indicates the improvement in stress prevalence relative to the no-control baseline.

\subsection{Network visualization}

\begin{figure}[H]
    \centering
    \includegraphics[width=0.135\textwidth]{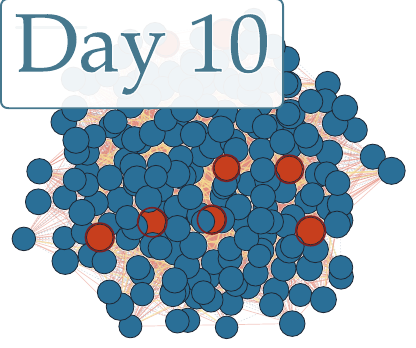}\hfill
    \includegraphics[width=0.135\textwidth]{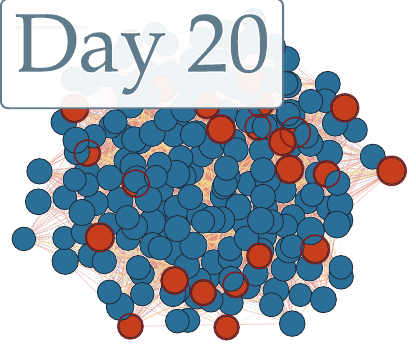}\hfill
    \includegraphics[width=0.135\textwidth]{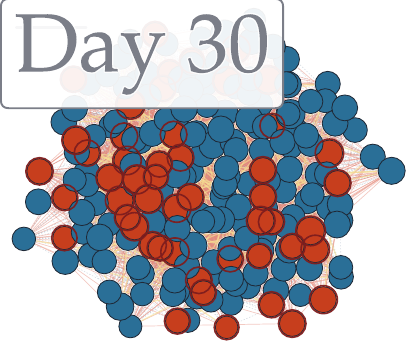}\hfill
    \includegraphics[width=0.135\textwidth]{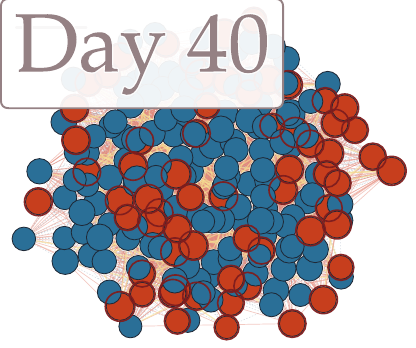}\hfill
    \includegraphics[width=0.135\textwidth]{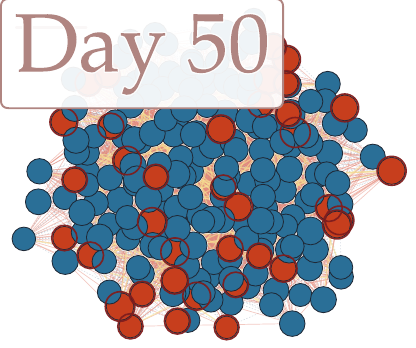}\hfill
    \includegraphics[width=0.135\textwidth]{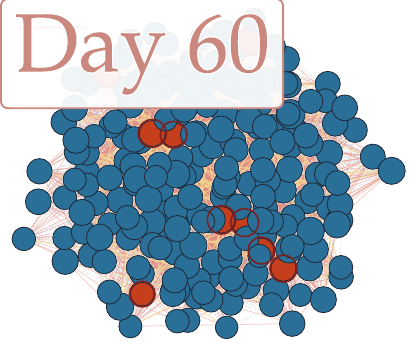}\hfill
    \includegraphics[width=0.135\textwidth]{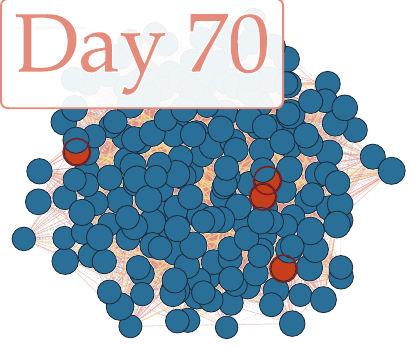}
    \caption[Network snapshots for Scenario~1 (No control)]{\textbf{Network snapshots for Scenario~1 (No control).} Snapshots at days $10$--$70$ showing the evolution of infection status across the population. Both networks $\mathbf{A}_{\boldsymbol{xy}}$ and $\mathbf{A}_{\boldsymbol{yy}}$ remain fixed throughout. Red nodes denote infected agents; blue nodes denote healthy agents. Solid edges represent the hazard-exposure network; dashed edges represent the social network. The infection wave peaks around day $35$ and recedes thereafter.}
    \label{fig:supp_net_none}
\end{figure}

\begin{figure}[H]
    \centering
    \includegraphics[width=0.135\textwidth]{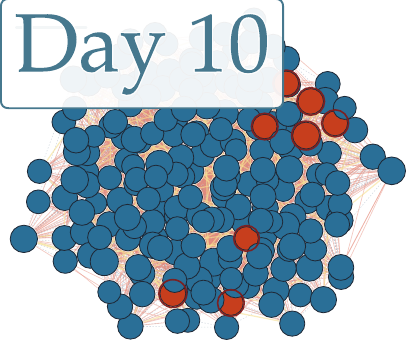}\hfill
    \includegraphics[width=0.135\textwidth]{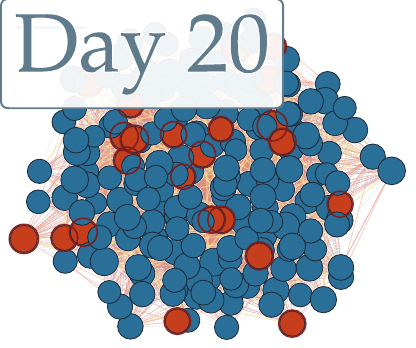}\hfill
    \includegraphics[width=0.135\textwidth]{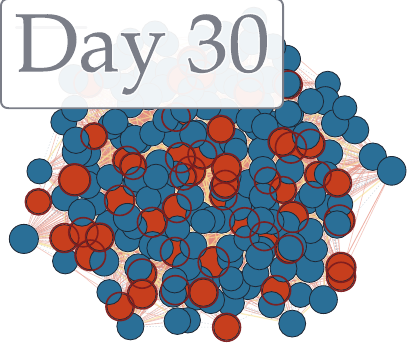}\hfill
    \includegraphics[width=0.135\textwidth]{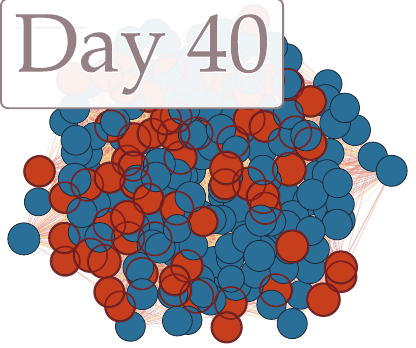}\hfill
    \includegraphics[width=0.135\textwidth]{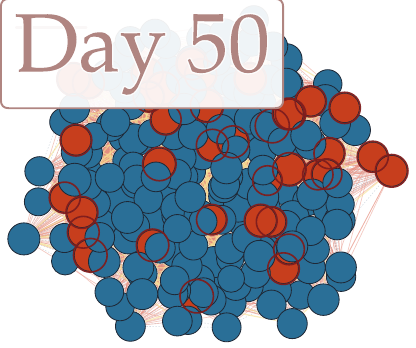}\hfill
    \includegraphics[width=0.135\textwidth]{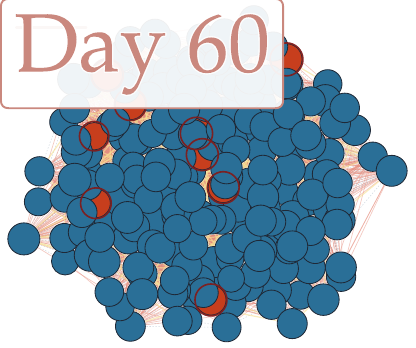}\hfill
    \includegraphics[width=0.135\textwidth]{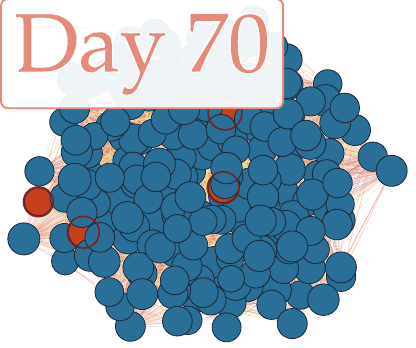}
    \caption[Network snapshots for Scenario~2 (Increasing $\langle k_{\boldsymbol{y}}\rangle_{\mathbf{A}_{\boldsymbol{xy}}}$)]{\textbf{Network snapshots for Scenario~2 (Increasing $\langle k_{\boldsymbol{y}}\rangle_{\mathbf{A}_{\boldsymbol{xy}}}$).} When the limit-state function $G<0$, edges are added to the hazard-exposure network $\mathbf{A}_{\boldsymbol{xy}}$ (solid lines), progressively densifying the network over time. The social network $\mathbf{A}_{\boldsymbol{yy}}$ (dashed lines) remains unchanged. The increasing density of solid edges is visible across successive snapshots.}
    \label{fig:supp_net_kxy}
\end{figure}

\begin{figure}[H]
    \centering
    \includegraphics[width=0.135\textwidth]{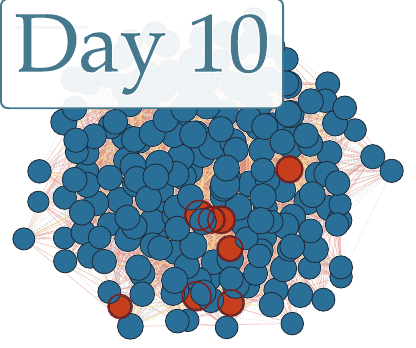}\hfill
    \includegraphics[width=0.135\textwidth]{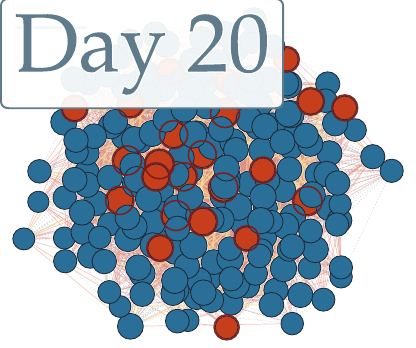}\hfill
    \includegraphics[width=0.135\textwidth]{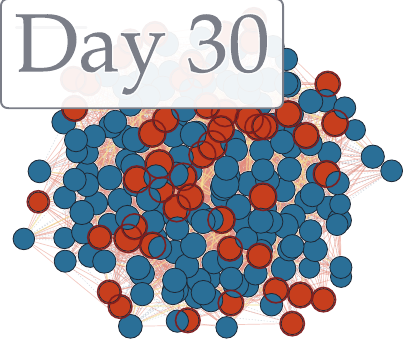}\hfill
    \includegraphics[width=0.135\textwidth]{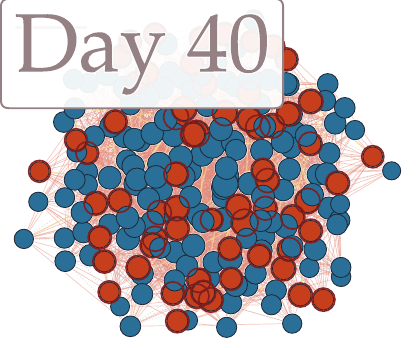}\hfill
    \includegraphics[width=0.135\textwidth]{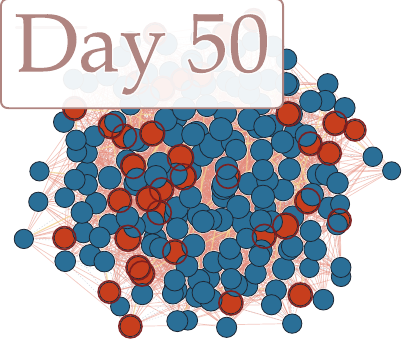}\hfill
    \includegraphics[width=0.135\textwidth]{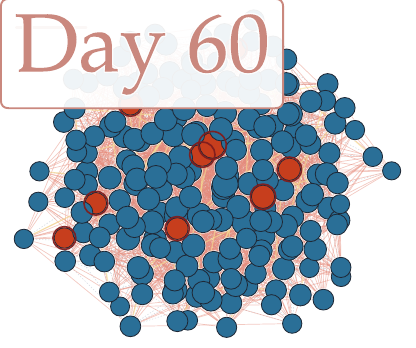}\hfill
    \includegraphics[width=0.135\textwidth]{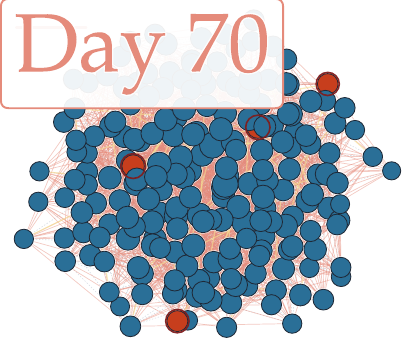}
    \caption[Network snapshots for Scenario~3 (Reducing $\langle k\rangle_{\mathbf{A}_{\boldsymbol{yy}}}$)]{\textbf{Network snapshots for Scenario~3 (Reducing $\langle k\rangle_{\mathbf{A}_{\boldsymbol{yy}}}$).} When $G<0$, edges are removed from the social network $\mathbf{A}_{\boldsymbol{yy}}$ (dashed lines), progressively sparsifying the social connectivity. The hazard-exposure network $\mathbf{A}_{\boldsymbol{xy}}$ (solid lines) remains unchanged. The decreasing density of dashed edges illustrates the reduction in social contagion channels.}
    \label{fig:supp_net_kyy}
\end{figure}

\begin{figure}[H]
    \centering
    \includegraphics[width=0.135\textwidth]{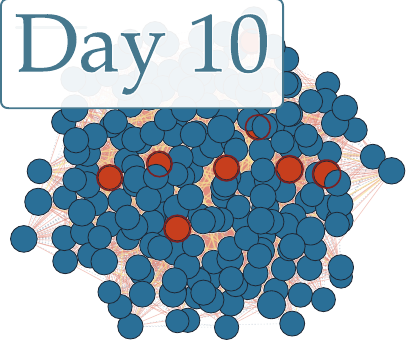}\hfill
    \includegraphics[width=0.135\textwidth]{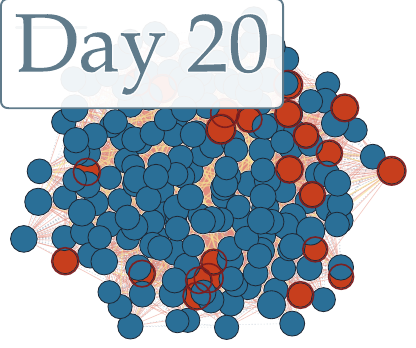}\hfill
    \includegraphics[width=0.135\textwidth]{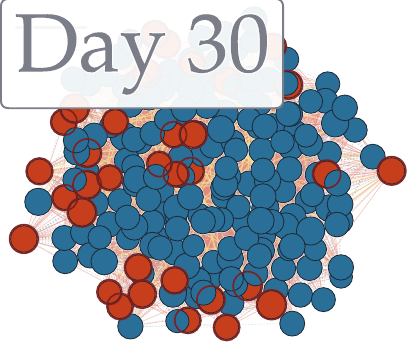}\hfill
    \includegraphics[width=0.135\textwidth]{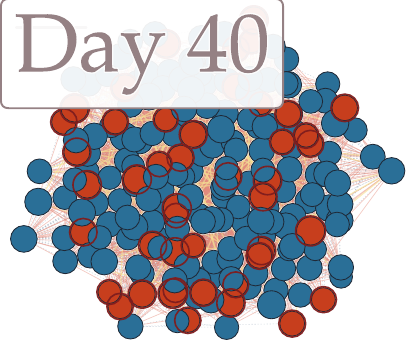}\hfill
    \includegraphics[width=0.135\textwidth]{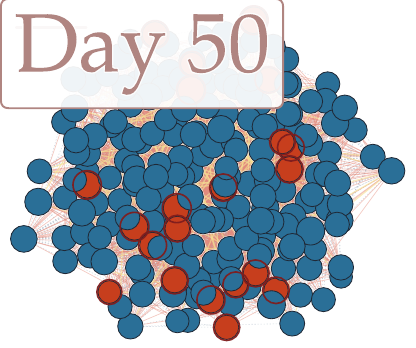}\hfill
    \includegraphics[width=0.135\textwidth]{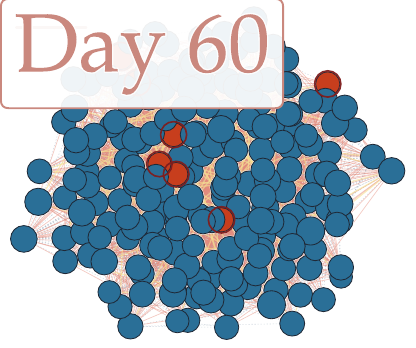}\hfill
    \includegraphics[width=0.135\textwidth]{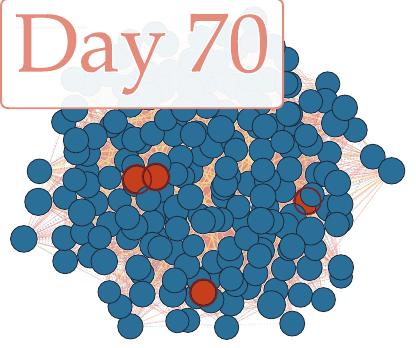}
    \caption[Network snapshots for Scenario~4 (Reducing prevalence $p$)]{\textbf{Network snapshots for Scenario~4 (Reducing prevalence $p$).} Both networks remain structurally fixed. Compared to the no-control baseline (\cref{fig:supp_net_none}), the number of infected nodes (red) is visibly reduced following intervention activation, reflecting the $40\%$ removal of infections applied to all future days once $G<0$.}
    \label{fig:supp_net_p}
\end{figure}

\begin{figure}[H]
    \centering
    \includegraphics[width=0.135\textwidth]{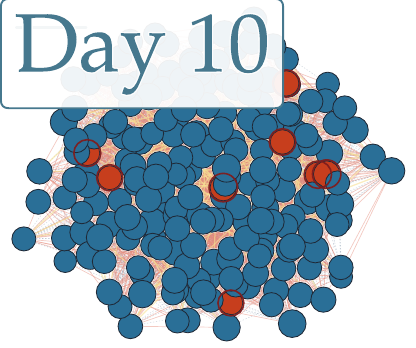}\hfill
    \includegraphics[width=0.135\textwidth]{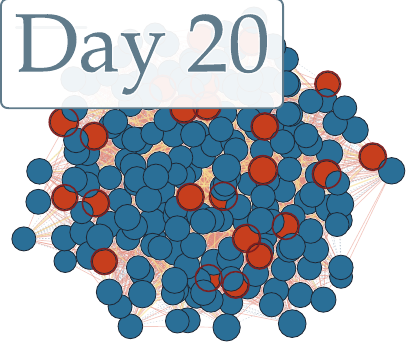}\hfill
    \includegraphics[width=0.135\textwidth]{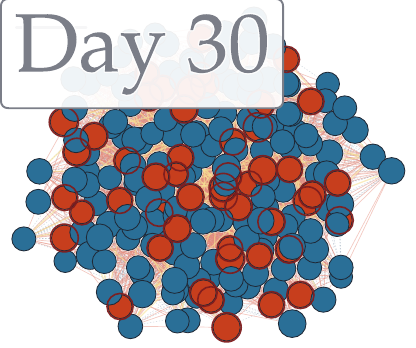}\hfill
    \includegraphics[width=0.135\textwidth]{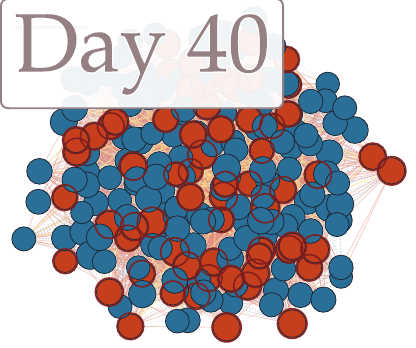}\hfill
    \includegraphics[width=0.135\textwidth]{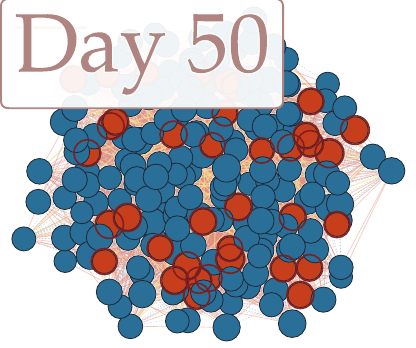}\hfill
    \includegraphics[width=0.135\textwidth]{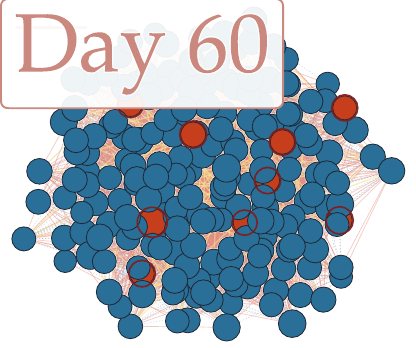}\hfill
    \includegraphics[width=0.135\textwidth]{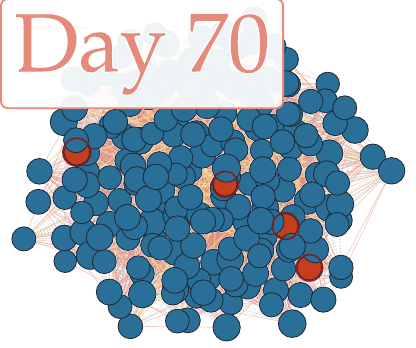}
    \caption[Network snapshots for Scenario~5 (No social influence, $\alpha=0$)]{\textbf{Network snapshots for Scenario~5 (No social influence, $\alpha=0$).} Both networks remain fixed and the infection trajectory is identical to the no-control baseline. The network structure is shown for reference; in this scenario, the social network $\mathbf{A}_{\boldsymbol{yy}}$ exerts no influence on emotional dynamics ($\alpha=0$), so only the hazard-exposure network drives emotional states.}
    \label{fig:supp_net_rational}
\end{figure}

\begin{figure}[H]
    \centering
    \includegraphics[width=0.135\textwidth]{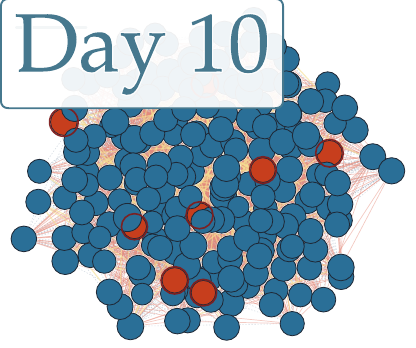}\hfill
    \includegraphics[width=0.135\textwidth]{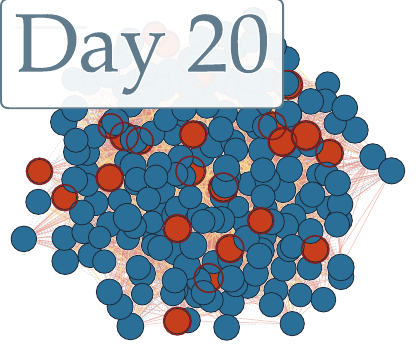}\hfill
    \includegraphics[width=0.135\textwidth]{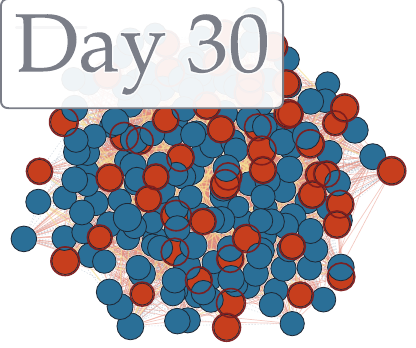}\hfill
    \includegraphics[width=0.135\textwidth]{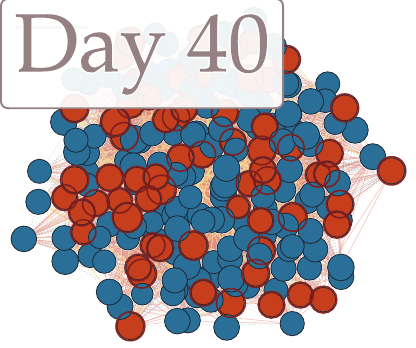}\hfill
    \includegraphics[width=0.135\textwidth]{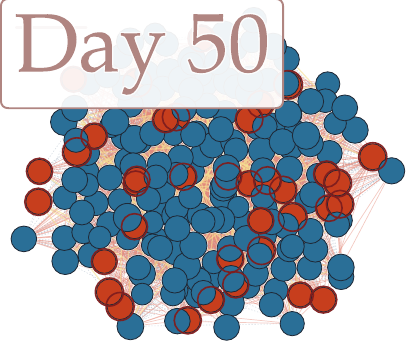}\hfill
    \includegraphics[width=0.135\textwidth]{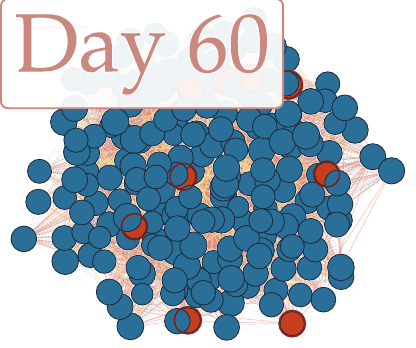}\hfill
    \includegraphics[width=0.135\textwidth]{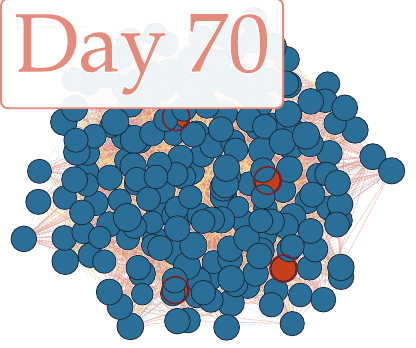}
    \caption[Network snapshots for Scenario~6 (Positivity bias, $c<0$)]{\textbf{Network snapshots for Scenario~6 (Positivity bias, $c<0$).} Both networks remain fixed and the infection trajectory is identical to the no-control baseline. Despite the same epidemic forcing, the stress prevalence under positivity bias is substantially lower (\cref{fig:supp_stress_trajectories}), illustrating the protective role of emotional attenuation through social interactions.}
    \label{fig:supp_net_positive_bias}
\end{figure}

To provide spatial intuition for the co-evolution of infection landscape and network structure under each scenario, we visualize network snapshots at seven checkpoints (days $10,20,\dots,70$). In each snapshot, node positions are fixed across all scenarios and time steps using a common spring layout. Nodes are coloured by infection status: blue for healthy individuals ($x_j=0$) and red for infected individuals ($x_j=1$). Edges of the hazard-exposure network $\mathbf{A}_{\boldsymbol{xy}}$ are drawn as solid lines, edges of the social network $\mathbf{A}_{\boldsymbol{yy}}$ as dashed lines, and edges present in both networks are highlighted. A condensed view of selected snapshots (days $10$, $30$, $50$) is inset in the main-text \cref*{fig:control_strategy}; the full set of snapshots is presented in \cref{fig:supp_net_none,fig:supp_net_kxy,fig:supp_net_kyy,fig:supp_net_p,fig:supp_net_rational,fig:supp_net_positive_bias}.

\newpage
\section{Economic Linkage Robustness}
\label{sec:supp_economic_robustness}
\label{sec:supp_economic_details}
\subsection{Monthly panel construction}
This section documents the monthly state-level panel used for the economic linkage analysis. For each state, daily stress prevalence (Mag) and daily COVID-19 prevalence (IncRate) are aggregated to monthly means and then temporally aligned with monthly macroeconomic indicators. The aligned panel spans 50 U.S.\ states and 36 months (April 2020 to March 2023). Numeric missing values are linearly interpolated after temporal alignment.

\begin{figure}[H]
    \centering
    \makebox[\textwidth][c]{\includegraphics[width=1.04\textwidth]{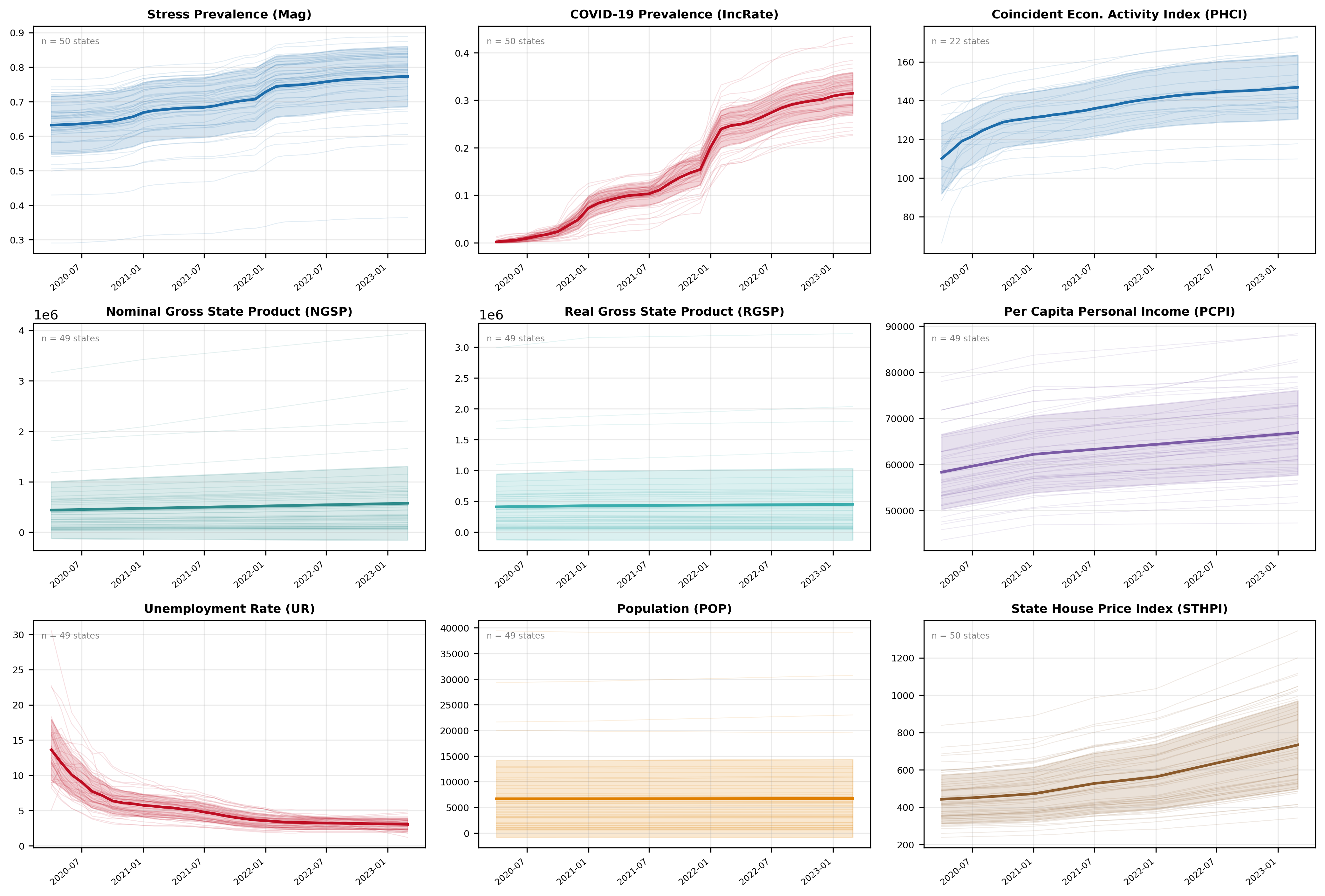}}
    \caption[Monthly state-level data used in the economic linkage analysis]{\textbf{Monthly state-level data used in the economic linkage analysis.} Each panel shows one variable in the aligned monthly panel: stress prevalence (Mag), COVID-19 prevalence (IncRate), and seven macroeconomic indicators. Thin lines denote individual states, thick lines denote cross-state means, and shaded bands indicate $\pm1$ standard deviation across states. The number of states with available data is noted in each panel.}
    \label{fig:supp_s9_econ_timeseries}
\end{figure}

The seven macroeconomic indicators considered in this section are the Coincident Economic Activity Index (PHCI), Nominal Gross State Product (NGSP), Real Gross State Product (RGSP), Per Capita Personal Income (PCPI), Unemployment Rate (UR), Population (POP), and the State House Price Index (STHPI). \Cref{fig:supp_s9_econ_timeseries} summarizes the monthly trajectories of all variables entering the analysis.

\subsection{Sobol index computation}

The two explanatory variables are monthly stress prevalence (Mag) and monthly COVID-19 prevalence (IncRate). For each month, they are related to each economic indicator using a state-level cross-sectional analysis.

First-order Sobol indices are estimated month by month. For each $(X,Y_j)$ pair, we fit a 3-component Gaussian mixture model (GMM) to the bivariate state-level samples:
\begin{equation}
p(x,y)=\sum_{k=1}^{K}\pi_k\,\mathcal{N}\!\left(
\begin{bmatrix}
x\\y
\end{bmatrix}
\middle|
\boldsymbol{\mu}_k,\boldsymbol{\Sigma}_k
\right),\quad
\pi_k>0,\ \sum_{k=1}^{K}\pi_k=1,\quad K=3.
\end{equation}
Model parameters are estimated by EM with covariance regularization ($10^{-6}$) and multiple replicate initializations. With responsibilities
\begin{equation}
\gamma_{ik}=
\frac{\pi_k\,\mathcal{N}(\mathbf{z}_i\mid\boldsymbol{\mu}_k,\boldsymbol{\Sigma}_k)}
{\sum_{j=1}^{K}\pi_j\,\mathcal{N}(\mathbf{z}_i\mid\boldsymbol{\mu}_j,\boldsymbol{\Sigma}_j)},
\end{equation}
the M-step updates are
\begin{equation}
N_k=\sum_{i=1}^{n}\gamma_{ik},\qquad
\pi_k=\frac{N_k}{n},\qquad
\boldsymbol{\mu}_k=\frac{1}{N_k}\sum_{i=1}^{n}\gamma_{ik}\mathbf{z}_i,
\end{equation}
\begin{equation}
\boldsymbol{\Sigma}_k=\frac{1}{N_k}\sum_{i=1}^{n}\gamma_{ik}
\left(\mathbf{z}_i-\boldsymbol{\mu}_k\right)\left(\mathbf{z}_i-\boldsymbol{\mu}_k\right)^\top.
\end{equation}
Writing
\(
\boldsymbol{\mu}_k=
\begin{bmatrix}\mu_{x,k}\\ \mu_{y,k}\end{bmatrix}
\)
and
\(
\boldsymbol{\Sigma}_k=
\begin{bmatrix}
\Sigma_{xx,k} & \Sigma_{xy,k}\\
\Sigma_{yx,k} & \Sigma_{yy,k}
\end{bmatrix},
\)
the conditional mean under component $k$ is
\begin{equation}
m_k(x)=\mu_{y,k}+\Sigma_{yx,k}\Sigma_{xx,k}^{-1}\left(x-\mu_{x,k}\right),
\end{equation}
with posterior weight
\begin{equation}
\omega_k(x)=
\frac{\pi_k\,\mathcal{N}(x\mid\mu_{x,k},\Sigma_{xx,k})}
{\sum_{j=1}^{K}\pi_j\,\mathcal{N}(x\mid\mu_{x,j},\Sigma_{xx,j})}.
\end{equation}
Hence,
\begin{equation}\label{eq:EYX}
\mathbb{E}[Y\mid X=x]=\sum_{k=1}^{K}\omega_k(x)\,m_k(x).
\end{equation}

The unconditional variance $\mathrm{Var}(Y)$ can be computed from the marginal mixture distribution. Using the law of total variance across the $K$ components, we have
\begin{equation}
\mathrm{Var}[Y] = \sum_{k=1}^{K} \pi_k \left( \Sigma_{yy,k} + \mu_{y,k}^2 \right) - \left( \sum_{k=1}^{K} \pi_k \mu_{y,k} \right)^2.
\end{equation}
The term $\mathrm{Var}[\mathbb{E}[Y\mid X]]$ in the Sobol index does not have a closed-form expression; therefore, we estimate it via a 1D Monte Carlo integration over the marginal distribution of $X$, namely $p(x) = \sum_{k=1}^{K} \pi_k\,\mathcal{N}(x \mid \mu_{x,k}, \Sigma_{xx,k})$. Specifically, a large number $M$ of synthetic samples $\{x^{(m)}\}_{m=1}^{M}$ is drawn from $p(x)$. For each sample, the conditional mean \(\mathbb{E}[Y\mid X=x^{(m)}]\) is evaluated using Eq.~\eqref{eq:EYX}, and $\mathrm{Var}[\mathbb{E}[Y\mid X]]$ is then approximated as
\begin{equation}
\widehat{\mathrm{Var}}[\mathbb{E}[Y\mid X]] = \frac{1}{M-1}\sum_{m=1}^{M}\left( \mathbb{E}[Y\mid X=x^{(m)}] - \overline{\mathbb{E}} \right)^2,
\end{equation}
where
\begin{equation}
\overline{\mathbb{E}} = \frac{1}{M}\sum_{m=1}^{M}\mathbb{E}[Y\mid X=x^{(m)}].
\end{equation}
The estimated first-order Sobol index is then computed as
\begin{equation}
\widehat{S}_X = \frac{\widehat{\mathrm{Var}}[\mathbb{E}[Y\mid X]]}{\mathrm{Var}[Y]}.
\end{equation}
This estimation leverages the exact parametric expectation from the GMM to provide a robust, nonlinear, variance-based sensitivity measure, quantifying the proportion of cross-state variability in the economic response that can be attributed to the predictor.

\newpage
\section{Reproducibility Package}
\label{sec:supp_reproducibility}

All codes and data used in this study will be made publicly available upon acceptance, together with an executable package for reproducing all major figures, tables, and posterior summaries in both the main text and Supplementary Information.

\addcontentsline{toc}{section}{References}
\makeatletter
\setcounter{NAT@ctr}{0}
\makeatother
\bibliographystyleS{unsrtnat}
\bibliographyS{references}

\end{document}